\documentclass[aps,prd,11pt,onecolumn,superscriptaddress,preprintnumbers,floatfix,nofootinbib,longbibliography,eqsecnum,showkeys]{revtex4-2}
\usepackage[german,english]{babel}
\usepackage{amsmath,amssymb}
\usepackage{hyperref}
\usepackage{titlesec}
\usepackage{bm}
\usepackage{bbm}
\usepackage{bbold}
\usepackage[normalem]{ulem}
\usepackage{braket}
\usepackage{tablefootnote}
\usepackage{slashed}
\usepackage{multirow}
\usepackage{epsfig}
\usepackage{epstopdf}
\usepackage{diagbox}
\usepackage[table,xcdraw]{xcolor}
\usepackage[mathlines]{lineno}
\usepackage[utf8]{inputenc}
\usepackage{comment}
\usepackage{graphicx}
\usepackage{mathrsfs}
\usepackage{tipa}
\usepackage{mathtools}
\usepackage{ragged2e}
\usepackage{txfonts}
\usepackage{enumitem}

\usepackage{adjustbox}

\usepackage{charter}

\usepackage{setspace}
\usepackage{booktabs} 

\newcommand{\RN}[1]{%
  \textup{\uppercase\expandafter{\romannumeral#1}}%
}

\newcommand{\beq}{\begin{equation}}
\newcommand{\eeq}{\end{equation}}

\newcommand{\be}{\begin{equation}}
\newcommand{\ee}{\end{equation}}
\newcommand{\bea}{\begin{eqnarray}}
\newcommand{\eea}{\end{eqnarray}}

\makeatletter
\let\LN@align\align
\let\LN@endalign\endalign
\renewcommand{\align}{\linenomath\LN@align}
\renewcommand{\endalign}{\LN@endalign\endlinenomath}
\let\LN@gather\gather
\let\LN@endgather\endgather
\renewcommand{\gather}{\linenomath\LN@gather}
\renewcommand{\endgather}{\LN@endgather\endlinenomath}
\makeatother
\allowdisplaybreaks

\begin{document}

\preprint{TUM-EFT 216/26}
\title{Quarkoniumlike states  above open-flavor thresholds in Born-Oppenheimer EFT}
\author{Nora Brambilla}
\email{nora.brambilla@tum.de}
\affiliation{Technical University of Munich,\\
TUM School of Natural Sciences, Physics
Department,\\   
James-Franck-Strasse~1, 85748 Garching, Germany.}
\affiliation{Technical University of Munich, Institute for Advanced Study, \\ 
Lichtenbergstrasse 2 a, 85748 Garching, Germany.}
\affiliation{Technical University of Munich, Munich Data Science Institute, \\ 
Walther-von-Dyck-Strasse 10, 85748 Garching, Germany.}
\author{Roberto Bruschini}
\email{roberto.bruschini@tum.de}
\affiliation{Technical University of Munich,\\
TUM School of Natural Sciences, Physics
Department,\\   
James-Franck-Str.~1, 85748 Garching, Germany.}
\author{Abhishek Mohapatra}
\email{amohapatra@prl.res.in}
\affiliation{
Theoretical Physics Division, Physical Research Laboratory, \\ 
Navrangpura, Ahmedabad 380009, India.}
\author{Fang-Zheng Peng}
\email{fz.peng@tum.de}
\affiliation{Technical University of Munich,\\
TUM School of Natural Sciences, Physics
Department,\\   
James-Franck-Str.~1, 85748 Garching, Germany.}
\author{Tommaso Scirpa}
\email{tommaso.scirpa@tum.de}
\affiliation{Technical University of Munich,\\
TUM School of Natural Sciences, Physics
Department,\\   
James-Franck-Str.~1, 85748 Garching, Germany.}

\begin{abstract}

Many quarkoniumlike states have been observed above open-flavor
thresholds, but their organization and internal structure remain
unsettled. We study the isoscalar hidden-charm and hidden-bottom sectors
in Born--Oppenheimer effective field theory (BOEFT), between the
spin--isospin averaged $S+S$ and $S+P$ thresholds. At leading order,
heavy-quark spin decouples, and the quarkonium static potential mixes
through string breaking with the lowest tetraquark/open-flavor BO
potentials of the same quantum numbers. These potentials are constrained
by QCD symmetries, their short- and long-distance behavior, and
lattice-QCD data. The only calibrated parameter is the lowest $1^{--}$
adjoint meson mass, fixed from the shallow multiplet associated with the
$\chi_{c1}(3872)$.

Using $T$-matrix, $K$-matrix, and complex-scaling methods, we determine
bound states and resonance poles, their masses, pole widths from the
included nonstrange $S+S$ channels, normalized pole couplings, and
prescription-dependent quarkonium and open-flavor composition measures.
Hybrid BOEFT multiplets are also included.

The spectrum exhibits a common heavy-quark-spin-symmetry multiplet organization.
Most poles are predominantly quarkonium resonances localized at relatively short distances, with the largest open-flavor components closest to threshold.
The same coupled equations generate shallow, spatially extended, open-flavor-dominated bound states with molecular long-distance characteristics, associated with the $\chi_{c1}(3872)$ and its bottomoniumlike counterpart $X_b$.
Their binding energies, radii, and small quarkonium components are highly sensitive to the adjoint meson mass, whereas the higher predominantly quarkonium spectrum is more stable.
In addition, the equations predict a more compact open-flavor-dominated resonance in the $l = 0$ bottomoniumlike sector, approximately $50$~MeV above the common spin–isospin averaged threshold, with a $P$-wave open-flavor component and a $0^{-+}$, $1^{--}$ BOEFT multiplet structure.

Together with the hybrid levels, the spectrum provides multiplet assignments for most experimental candidates. States not naturally accommodated point to the need for hidden-strange and $S+P$ tetraquark/open-flavor BO sectors and for hybrid--tetraquark and hybrid--quarkonium mixings.

\end{abstract}


\maketitle

\clearpage

\section{Introduction}\label{sec:intro}

The discovery of the $\chi_{c1}(3872)$, originally denoted
$X(3872)$~\cite{Belle:2003nnu}, showed that the hidden-heavy meson spectrum above open-flavor threshold cannot be understood as a simple
continuation of the below-threshold quarkonium spectrum. Since then, a large number of quarkoniumlike candidates have been observed in the
hidden-charm and hidden-bottom sectors by Belle, BESIII, LHCb, and
other experiments
\cite{BESIII:2013ouc,Belle:2013yex,Belle:2011aa,
Belle:2013urd,BESIII:2015cld,Belle:2008qeq,Belle:2014nuw,
BESIII:2020qkh,LHCb:2021uow,PhysRevLett.131.131901,
LHCb:2021vvq,Brambilla:2010cs,QuarkoniumWorkingGroup:2004kpm}.
Many of these states lie close to heavy-light meson--antimeson
thresholds and do not fit straightforwardly into the conventional
quarkonium spectrum. The central challenge is therefore not merely to
interpret each candidate separately, but to determine whether
conventional quarkonia, threshold-dominated exotic states, and hybrids
can be organized within a common QCD framework controlled by
symmetries and coupled-channel dynamics.
In particular, one would like to understand whether the proliferation of states above threshold reflects a collection of unrelated dynamical mechanisms or an underlying
spectroscopic organization controlled by QCD symmetries.

Potential models laid the foundations for the successful description
of heavy quarkonium as a nonrelativistic system. In particular, the
Cornell model and its coupled-channel extensions played a central role in the study of quarkonium spectroscopy, threshold effects, and strong
decays
\cite{Bhanot:1978mj,Richardson:1978bt,Martin:1980jx,
LeYaouanc:1972vsx,Eichten:1974af,Eichten:1978tg,Eichten:1979ms,
Ono:1980js,Heikkila:1983wd,Eichten:2005ga}. Below the open-flavor
threshold, the potential description has been placed on a systematic
QCD foundation through nonrelativistic effective field theories, in
particular potential NRQCD, in which the potentials are matching
coefficients whose nonperturbative content can be accessed through
lattice QCD~\cite{Brambilla:2004jw}. Above threshold, the confining
quarkonium potential has also been continued phenomenologically and
coupled to heavy-light meson--antimeson channels.

A broad range of descriptions has been developed for states that do
not resemble conventional quarkonia, including hadronic molecules,
compact tetraquarks, diquark--antidiquark configurations,
hadro-quarkonia, and hybrids; see Ref.~\cite{Brambilla:2019esw} and
references therein. These approaches have illuminated complementary
dynamical regimes, but they generally begin by selecting a particular
clustering or organization of the relevant degrees of freedom, and are therefore not always naturally suited to treating conventional
quarkonia, threshold-dominated states, and hybrids within the same
description. Direct lattice QCD calculations provide the most
fundamental route to the spectrum, and have achieved important results
in several quantum-number sectors
\cite{HadronSpectrum:2012gic,Cheung:2016bym,Ryan:2020iog,
Prelovsek:2013cra,Prelovsek:2020eiw,Wilson:2023anv}, but a calculation
including all relevant channels and quantum numbers throughout the above-threshold spectrum is not yet available.

Born--Oppenheimer effective field theory (BOEFT) provides a
QCD-constrained and systematically improvable framework for addressing
this problem. It exploits the separation between the slow motion of
the heavy quarks and the faster dynamics of the light quarks and
gluons
\cite{Brambilla:2017uyf,Berwein:2015vca,Oncala:2017hop,
Soto:2020xpm,Berwein:2024ztx}. In the static limit, the light degrees
of freedom generate static energies classified by Born--Oppenheimer
(BO) quantum numbers. The motion of the heavy quark--antiquark pair is
then governed by Schr\"odinger equations whose potentials and mixing
terms encode this nonperturbative light-field dynamics. At leading
order in the heavy-quark mass expansion, the heavy-quark spin decouples, and the resulting states form heavy-quark-spin-symmetry (HQSS) multiplets. 
Conservation of the BO quantum numbers also determines which static
configurations can mix. The coupled-channel structure is therefore
constrained by QCD rather than chosen separately for each observed
state. In this sense, BOEFT provides an effective-field-theory
generalization of phenomenological coupled-channel potential
descriptions.

This structure is particularly important near open-flavor thresholds.
The quarkonium static potential can mix only with
tetraquark/open-flavor BO potentials carrying the same BO quantum
numbers. These potentials should not be identified with constant
meson--antimeson threshold lines. At short heavy-quark separation, the
lowest isoscalar tetraquark static energies approach a repulsive
color-octet $Q\bar Q$ potential offset by the energy of an adjoint
meson. At large separation, the same static energies approach the
corresponding static heavy-light meson--antimeson thresholds
\cite{TarrusCastella:2022rxb,TarrusCastella:2024zps,
Berwein:2024ztx,Braaten:2024tbm}. A tetraquark/open-flavor BO channel therefore interpolates continuously between short-distance
adjoint-hadron dynamics and a long-distance heavy-light
meson--antimeson configuration. Its mixing with quarkonium is generated
by string breaking and is concentrated in the region where the
corresponding static energies approach one another
\cite{Bali:2005fu,Bulava:2019iut,Bulava:2024jpj}.

This interpolation relates the compact-tetraquark and
hadronic-molecule descriptions without treating them as mutually exclusive microscopic hypotheses or as separate channels in the
Schr\"odinger equation. A state localized mainly at short or
intermediate distances in a tetraquark/open-flavor BO potential has a
compact character, whereas a shallow state dominated by the large-distance region of the same BO channel has molecular
long-distance characteristics. More generally, a physical state may
sample both spatial regimes and may also contain a quarkonium
component through string-breaking mixing. BOEFT therefore determines
the spatial and channel structure dynamically, without assuming a
specific clustering of the heavy and light constituents from the
outset.

The lowest isoscalar quarkonium--tetraquark/open-flavor mixing potential has been constrained by lattice QCD in the string-breaking
region~\cite{Bali:2005fu,Bulava:2019iut,Bulava:2024jpj}. This input
has already been used within BOEFT to study threshold effects below
the lowest spin--isospin averaged open-flavor threshold, including the
$\chi_{c1}(3872)$, the $T_{cc}^{+}(3875)$ in the double-charm sector,
and the modifications of the conventional quarkonium spectrum induced
by nearby thresholds
\cite{Brambilla:2024imu,Brambilla:2026zfw}.

In the present work, we extend this program to the bound-state and
resonance spectrum in the isoscalar hidden-charm and hidden-bottom
sectors between the lowest spin--isospin averaged nonstrange $S+S$
and $S+P$ heavy-meson thresholds. This is the lowest energy window in
which open-flavor dynamics is unavoidable, while the nonstrange $S+S$
BO channels can still be studied before the explicit inclusion of the higher $S+P$ channels becomes necessary. Hidden-strange channels,
which may be important for some physical states in this interval, are
not included in the present calculation.

We solve the leading-order coupled Schr\"odinger equations in which the quarkonium static potential mixes with the lowest isoscalar
tetraquark/open-flavor BO potentials carrying the same quantum
numbers. Their forms are constrained by QCD symmetries, available
lattice QCD data, and their known short- and long-distance behavior.
The only parameter calibrated to experimental spectroscopy is the
lowest $1^{--}$ adjoint meson mass. In the mass and threshold
convention adopted here, its central value,
$\bar\Lambda_{1^{--}}^{\rm HL}= -0.113~\mathrm{GeV}$, is fixed so that the
spin-averaged $2P$ multiplet associated with the $\chi_{c1}(3872)$
lies $96~\mathrm{keV}$ below the common spin--isospin averaged
$D^{(\ast)}\bar D^{(\ast)}$ threshold. Once this input is fixed, the
remaining bound states and resonance poles in both heavy-flavor
sectors follow without further calibration to the observed spectrum.

At leading order, the heavy-quark spin decouples, and the predicted
levels form spin-averaged HQSS multiplets. Spin-dependent corrections,
whose structure has been formulated in BOEFT
\cite{Oncala:2017hop,Soto:2020xpm,Brambilla:2018pyn,
Brambilla:2019jfi,Soto:2023lbh}, lift these degeneracies. For
predominantly quarkonium states, the expected relative corrections are
of order $v^2$. For tetraquark/open-flavor configurations, spin and
nonstatic effects enter already at order $1/m_Q$ and are naturally
related to the splittings among the physical heavy-light meson
thresholds. They may shift individual multiplet members between
different thresholds and induce additional mixings and decay channels.
These effects are not included here; the present calculation isolates
the leading nonperturbative quarkonium--tetraquark/open-flavor mixing
already present in the static limit.

We determine the bound states and resonance poles through three
complementary methods: analytic continuation of the $T$ matrix to the
unphysical Riemann sheet, analysis of $K$ matrix poles on the real
axis, and the complex scaling method. We extract the pole masses, the pole widths generated by the included spin-averaged nonstrange $S+S$ channels, and the normalized pole couplings to the asymptotically open channels; we also calculate phase shifts and effective range parameters. The normalized pole couplings characterize relative residue strengths and, for narrow isolated resonances, may approximate the branching fraction ratios only within the truncated channel space. They are not
predictions for physical branching fractions.

We also characterize the quarkonium--open-flavor structure of the
states. For bound states, the components of the normalized wavefunction
have the ordinary probability interpretation. For resonances, no
unique probabilistic compositeness exists
\cite{Weinberg:1965zz,Baru:2010ww,Sekihara:2016xnq,
Kinugawa:2024crb}. We therefore employ the complementary complex scaling method
and $K$ matrix prescriptions and present their results as
prescription-dependent composition measures rather than as
resonance probabilities. These measures characterize the internal
channel structure and should not be identified with unique physical probabilities above threshold.

The principal result is a global HQSS-multiplet organization of the
spectrum rather than a set of independent assignments for individual states. Most resonance poles in the energy region considered emerge as
predominantly quarkonium states that remain localized mainly at
comparatively short distances, with the largest open-flavor
contributions occurring closest to the threshold. Strikingly, the same
coupled equations also generate exceptional shallow, spatially
extended, open-flavor-dominated states. In particular, the multiplet
associated with the $\chi_{c1}(3872)$ samples predominantly the
large-distance meson--antimeson regime of the complete
tetraquark/open-flavor BO channel and therefore has molecular
long-distance characteristics.

The shallow states are highly sensitive to the adjoint meson mass
because they lie close to the critical condition for binding. To illustrate this behavior, we also repeat the calculation using
$\bar\Lambda_{1^{--}}^{\rm HL}=-0.134~\mathrm{GeV}$, the value employed in the
previous below-threshold study~\cite{Brambilla:2026zfw}. In the
equations and spin-isospin averaged threshold convention of the present work, this second value is only a sensitivity test and not an alternative central calibration or an uncertainty endpoint. The comparison shows that the binding energies, radii, and small quarkonium components of the shallow multiplets vary strongly, whereas the ordering, multiplet organization, and dominant channel content of the higher predominantly quarkonium spectrum are considerably more stable.

To extend the spectroscopic organization beyond the coupled
quarkonium--open-flavor sector, we also calculate the corresponding
hybrid BOEFT multiplets
\cite{Berwein:2015vca,Oncala:2017hop,Brambilla:2022hhi}. The hybrid
potentials are not coupled here to the tetraquark/open-flavor channels,
and the $1/m_Q$-suppressed hybrid--quarkonium mixing is also omitted.
These multiplets are therefore included only as spectroscopic levels, not as predictions for physical hybrid resonance pole masses, pole widths, pole couplings, or mixed-state compositions.

We compare the coupled-channel multiplets and the hybrid 
levels with the observed hidden-charm and hidden-bottom spectra. The purpose of this comparison is to investigate whether the candidates
can be organized into spin-averaged HQSS multiplets and to identify
where omitted dynamics may be important. The proposed associations
should therefore be understood as possible multiplet assignments
within the accuracy and channel content of the leading-order
calculation, rather than as unique state-by-state identifications.

We further compare our results with related coupled-channel
Born--Oppenheimer studies
\cite{Bicudo:2019ymo,Bicudo:2020qhp,Bicudo:2022ihz}
and with existing \textit{ab initio} lattice QCD calculations
\cite{HadronSpectrum:2012gic,Cheung:2016bym,Ryan:2020iog,Prelovsek:2013cra,Prelovsek:2020eiw,Wilson:2023anv}. These comparisons help isolate the consequences of the QCD constraints implemented here, in particular, the short-distance repulsive color-octet behavior of the
tetraquark static energies, their approach to open-flavor thresholds at large distance, and the lattice-constrained string-breaking
interaction.

Taken together, the coupled quarkonium--open-flavor multiplets and the  hybrid levels provide possible HQSS associations for most observed candidates in the energy window considered.
Candidates not naturally accommodated by the included channels point
toward the additional hidden-strange and $S+P$
tetraquark/open-flavor BO sectors, as well as hybrid--tetraquark and
hybrid--quarkonium mixings. The purpose of the present analysis is therefore both to organize the spectrum within the channel space that
can presently be treated, and to identify the additional BO sectors and
nonperturbative mixing potentials required for a more complete
description.

The paper is organized as follows. Section~\ref{sec:BOEFT} presents the BOEFT framework, the BO quantum numbers, the static potentials, and the
coupled Schr\"odinger equations. Sections~\ref{sec:spectrum} and \ref{sec:probabilities} describe the methods used to determine the resonance poles and to characterize their channel content. Section~\ref{sec:results} presents the numerical results for
the bound states and resonance poles, including their masses, widths,
pole couplings, and, for resonances, prescription-dependent
channel-composition measures. It also analyzes the phase shifts and
effective-range parameters, discusses how the weak-binding analysis is
related to the notions of compositeness used in the literature, and
examines the sensitivity of the shallow states to the adjoint meson
mass. Section~\ref{sec:hybrids} presents the hybrid spectrum.
Sections~\ref{sec:Comparison with experiments} and \ref{sec:Comparison with literature} compare the resulting multiplet organization
with experiment, lattice QCD, and related Born--Oppenheimer studies.
Finally, Section~\ref{sec:conclusions} summarizes the main results and the extensions required for a more complete treatment.

Appendix~\ref{app:changeofbasis} derives the unitary transformation connecting the original Born--Oppenheimer basis with the diabatic basis used in the coupled Schr\"odinger equations. Appendix~\ref{app:bottom-charm} presents the corresponding resonance spectrum and channel-composition results in the bottom-charmed sector. Appendix~\ref{app:Hybrids} summarizes the hybrid Schr\"odinger equations, the matching and parametrization of the
hybrid static potentials, and the resulting hybrid spectrum.

\section{The BOEFT description of the system}\label{sec:BOEFT}

In this work, we apply BOEFT~
\cite{Berwein:2015vca,Oncala:2017hop,Soto:2020xpm,Berwein:2024ztx,Brambilla:2018pyn,Brambilla:2019jfi,Brambilla:2024imu,Brambilla:2026zfw,Brambilla:2017uyf, TarrusCastella:2022rxb, TarrusCastella:2024zps, Brambilla:2025xma}
to the isoscalar hidden-charm and hidden-bottom sectors in the
energy region extending from the lowest spin--isospin averaged
$S+S$ open-flavor threshold to the first $S+P$ threshold.
Following the lattice QCD calculations of
Refs.~\cite{Bali:2005fu,Bulava:2019iut,Bulava:2024jpj},
we work with two degenerate light-quark flavors in the isospin
limit and neglect heavy-strange meson--antimeson channels.
The relevant BO configurations are the quarkonium channel and
the two lowest isoscalar tetraquark channels. The quarkonium
configuration and one of the tetraquark configurations carry the
same BO quantum numbers and therefore mix through string breaking,
producing the narrow avoided crossing visible in the corresponding
adiabatic static energies. Previous BOEFT studies investigated the
effects of this mixing below the lowest open-flavor
threshold~\cite{TarrusCastella:2022rxb,Brambilla:2026zfw}.
Here we extend the analysis to bound states and resonance poles up
to the first spin--isospin averaged $S+P$ threshold.

\subsection{BO quantum numbers and static potentials}\label{subsec:quantum_numbers}

The BO static potentials are the QCD static energies of the light degrees of freedom (LDF), namely light quarks and gluons, in the presence of a static $Q\bar Q$ pair separated by a distance
$\bm r$, where $Q$ denotes a heavy quark and $\bar Q$ a heavy
antiquark. The relevant symmetry group is $D_{\infty h}$, generated
by rotations around the $Q\bar Q$ axis, reflections through planes
containing that axis, and the combined action of charge conjugation
and parity on the LDF. The static potentials are therefore
classified by the \emph{BO quantum numbers} $\Lambda_\eta^\sigma$.
The rotational quantum number is defined as
$\Lambda\equiv|\bm K\cdot\hat{\bm r}|$, where $\bm K$ is the total angular momentum of the LDF and
$\hat{\bm r}=\bm r/r$. Integer values
$\Lambda=0,1,2,\ldots$ are conventionally denoted by
$\Sigma,\Pi,\Delta,\ldots$. The eigenvalue of the combined $CP$
transformation is denoted by $\eta=g,u$, while the eigenvalue under reflection through a plane containing the $Q\bar Q$ axis is denoted
by $\sigma=+,-$.
For $\Lambda>0$, rotational symmetry implies a degeneracy between
the $\sigma=+$ and $\sigma=-$ components, and the superscript
$\sigma$ is therefore omitted. The static potentials also carry
the flavor quantum numbers of the LDF. Since we restrict the
present analysis to the isoscalar sector with two degenerate light flavors, these labels will be suppressed.

\begin{figure}
\centering 
\includegraphics*[width=10.5cm,clip=true]{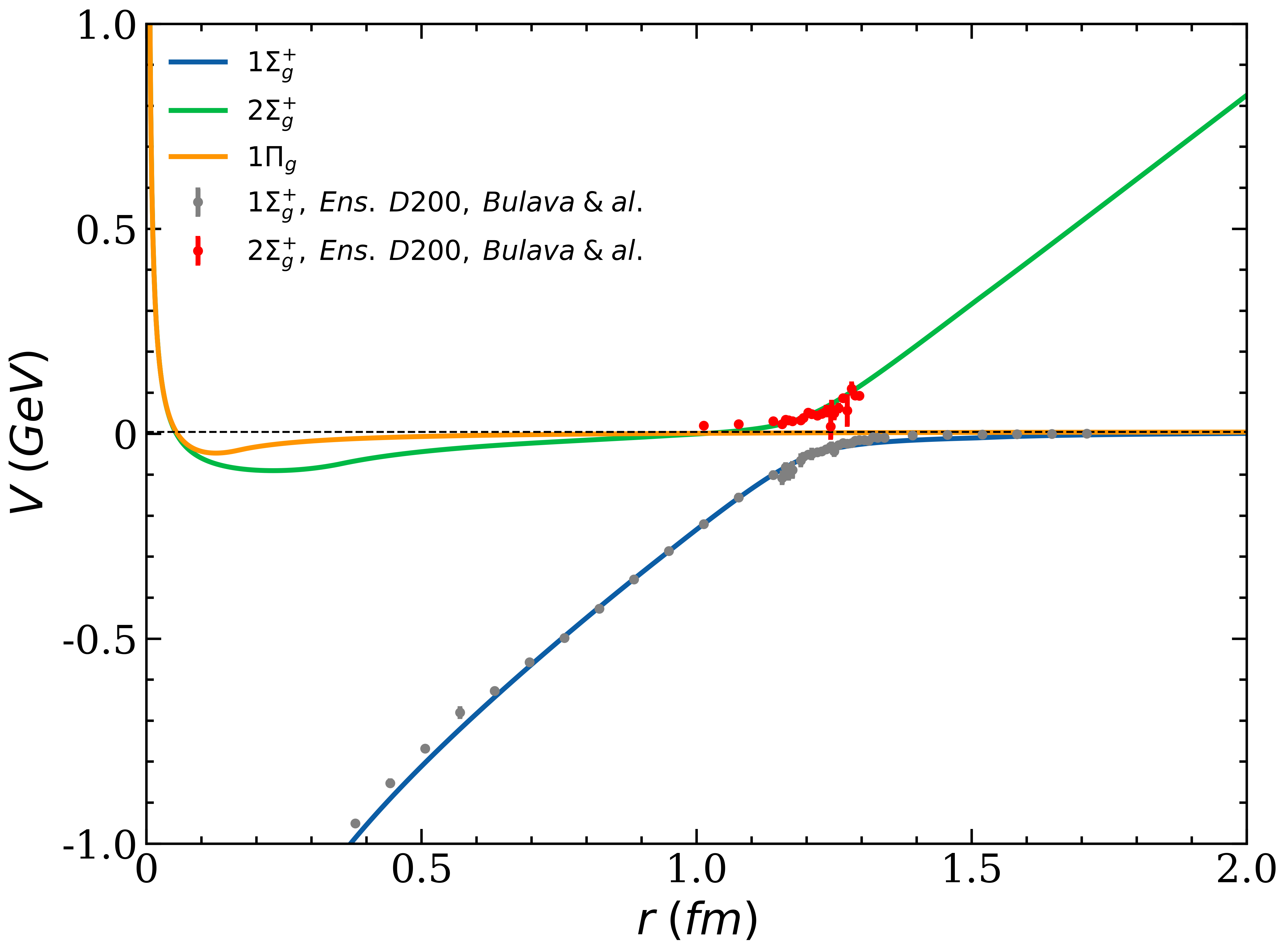}
\caption{The adiabatic QCD static energies relevant to this work.
Note the short-distance degeneracy between the $2\Sigma_g^+$ and $1 \Pi_g$ static energy, due to spherical symmetry restoration, and the narrow avoided crossing between the $1\Sigma_g^+$ and $2\Sigma_g^+$ energies, due to the quarkonium-tetraquark mixing via string breaking.
The lattice QCD data points are taken from Ref.~\cite{Bulava:2024jpj}.
}
\label{fig:isospin0}
\end{figure}

When two BO static potentials sharing the same BO quantum numbers become close at some distance $r$, they mix, and the corresponding adiabatic static energies give rise to the avoided level crossing phenomenon~\cite{Berwein:2024ztx}, displayed in Fig.~\ref{fig:isospin0} for the potentials relevant to this study.

In the $r\to0$ limit, the cylindrical symmetry group
$D_{\infty h}$ enlarges to the spherical symmetry group $O(3)$.
The BO static potentials, therefore, form degenerate multiplets with
multiplicity $2k+1$, where $k(k+1)$ is the eigenvalue of
$\bm K^2$. In this limit, parity and charge conjugation are
separately restored for the LDF, and the multiplets may be
classified by $k^{PC}$. A $k^{PC}$ multiplet contains the static
potentials with
\begin{equation}
  \Lambda=0,\ldots,k,\qquad
  \eta=CP,\qquad
  \sigma=P(-1)^k\quad\text{for }\Lambda=0.
\end{equation}
At vanishing separation, the pointlike $Q\bar Q$ pair acts as a
static color source in either the singlet or octet representation.
For a singlet source, whose static energy approaches an attractive
Coulomb potential, the only LDF configuration is the
vacuum with $k^{PC}=0^{++}$. For an octet source, whose static
energy approaches a repulsive Coulomb potential, the LDF
excitations are \emph{adjoint hadrons}: light-quark and gluonic
configurations labelled by $k^{PC}$ quantum numbers
bound to the static color-octet source so that the complete state is a color singlet.

At leading order in the $1/m_Q$ expansion, the adiabatic BO potentials are given by the static energies~\cite{Berwein:2024ztx,Brambilla:2026zfw}. The quarkonium static potential carries
$\Sigma_g^+$ quantum numbers and is associated at short distance
with the vacuum configuration $k^{PC}=0^{++}$. The lowest
isoscalar tetraquark static potentials relevant to the present
work carry $\Sigma_g^+$ and $\Pi_g$ quantum numbers and belong at
short distance to the $k^{PC}=1^{--}$ adjoint meson multiplet.
A further tetraquark potential with $\Sigma_u^-$ quantum numbers
is associated with $k^{PC}=0^{-+}$, but it does not mix with the
quarkonium channel at leading order and will not be included here.

The same BO quantum numbers occur in the decomposition of a pair
of $S$-wave static-light mesons. Exact conservation of
$\Lambda_\eta^\sigma$, therefore, requires the tetraquark static
energies to approach at large separation the corresponding
static-light meson--antimeson thresholds with the same BO quantum
numbers~\cite{Berwein:2024ztx,Braaten:2024tbm,Brambilla:2026zfw}. Thus, the tetraquark BO channel should not be
identified with a meson--antimeson configuration at every
separation: that interpretation becomes literal only in its
large-$r$ asymptotic region.

It is useful to distinguish the static energies
from the diabatic potentials entering the coupled Schr\"odinger
equation. We denote the static energies by
$n\Lambda_\eta^\sigma$, with $n=1,2,\ldots$. In particular,
$1\Sigma_g^+$ and $2\Sigma_g^+$ are the two static energies
involved in the string-breaking avoided crossing. By contrast,
$V_{\Sigma_g^+}$ and $V_{\Sigma_g^{+\prime}}$ denote,
respectively, the quarkonium and tetraquark diabatic potentials,
while $V_{\Sigma_g^+-\Sigma_g^{+\prime}}$ denotes their mixing.
The prime in $\Sigma_g^{+\prime}$ therefore labels the tetraquark
diabatic channel and not an excitation number.

The static energies $1\Sigma_g^+$, $2\Sigma_g^+$, and $1\Pi_g$ are displayed in Fig.~\ref{fig:isospin0}.
At short distances, the $1\Sigma_g^+$ static energy approaches an attractive Coulomb potential associated with the vacuum $k^{PC}=0^{++}$, while the $2\Sigma_g^+$ and $1\Pi_g$ static energies approach a repulsive Coulomb potential offset by the energy of the lowest isoscalar $1^{--}$ adjoint meson (see Eq.~\eqref{eq:QQbar_V})~\cite{Berwein:2024ztx,Braaten:2024tbm,Brambilla:2024imu,Brambilla:2026zfw}.
The $1\Sigma_g^+$ and $2\Sigma_g^+$ curves have a narrow avoided crossing for distances near $1.2~\mathrm{fm}$.
At large distances, the $1\Sigma_g^+$ and $1\Pi_g$  approach twice the energy of an $S$-wave static-light meson.
At large distances, the $2\Sigma_g^+$  increases linearly.

The two diabatic potentials and their mixing are related to the
adiabatic static energies through the $r$-dependent rotation

\begin{equation}
    \begin{pmatrix}
        V_{\Sigma_g^+} & V_{\Sigma_g^+ \text{-} \Sigma_g^{+\prime}} \\
        V_{\Sigma_g^+ \text{-} \Sigma_g^{+\prime}} & V_{\Sigma_g^{+\prime}} \\
    \end{pmatrix}
    =
    \begin{pmatrix}
        \cos \theta & -\sin\theta \\
        \sin\theta & \cos\theta \\
    \end{pmatrix}
    \begin{pmatrix}
        V_{1 \Sigma_g^+} & 0 \\
        0 & V_{2 \Sigma_g^+} \\
    \end{pmatrix}
    \begin{pmatrix}
        \cos \theta & \sin\theta \\
        -\sin\theta & \cos\theta \\
    \end{pmatrix}.
    \label{eq: transformation between diabatic and adiabatic potential}
\end{equation}

The mixing angle satisfies $\theta(r)\to0$ as $r\to0$ and
$\theta(r)\to\pi/2$ as $r\to\infty$. Consequently,
$V_{\Sigma_g^+}$ follows the adiabatic static energy $1\Sigma_g^+$ at
short distance and $2\Sigma_g^+$ at large distance, whereas
$V_{\Sigma_g^{+\prime}}$ follows $2\Sigma_g^+$ at short distance
and $1\Sigma_g^+$ at large distance. The mixing potential
$V_{\Sigma_g^+-\Sigma_g^{+\prime}}$ vanishes in both limits and is
localized in the string-breaking region. The diabatic
quarkonium and tetraquark potentials cross near $r\simeq1.2$~fm,
where the corresponding adiabatic static energies exhibit their narrow
avoided crossing.

We emphasize again that the prime in
$\Sigma_g^{+\prime}$ does not denote an excitation number.
The adiabatic static energies are denoted by $1\Sigma_g^+$ and
$2\Sigma_g^+$, whereas $\Sigma_g^+$ and
$\Sigma_g^{+\prime}$ label the quarkonium and tetraquark diabatic
channels. Their ordering is reversed across the avoided crossing: below it, the quarkonium channel is the lower diabatic configuration, while above it, the tetraquark channel is the lower one.

\subsection{Multiplets and Coupled Schr\"odinger equations}\label{subsec:coupledequation}

Rotations, parity, and charge conjugation are exact symmetries of BOEFT inherited from QCD. We denote by $\bm L_{Q\bar Q}$ the orbital angular momentum of the heavy pair and by $\bm K$ the total angular momentum of the LDF. Their sum,
\begin{equation}
  \bm L=\bm L_{Q\bar Q}+\bm K,
\end{equation}
is the BO angular momentum. The heavy-quark spin is
denoted by $\bm S_{Q\bar Q}$, and the total angular momentum is
\begin{equation}
  \bm J=\bm L+\bm S_{Q\bar Q}.
\end{equation}
The quantum numbers $\bm{J}^2$ and $J_3$ are conserved at all orders in BOEFT. At leading order, $\bm{L}^2$ and $\bm{S}^2_{Q\bar{Q}}$ are also separately conserved. By contrast,
$\bm K^2$ and $\bm L_{Q\bar Q}^2$ need not be conserved
separately, although their quantum numbers $k$ and $l_{Q\bar Q}$ remain useful for organizing the channel content of
the wavefunctions and the BOEFT multiplets. Further details on the definition of the channels and their quantum numbers can be
found in Ref.~\cite{Brambilla:2026zfw}.

At leading order in BOEFT, the heavy-quark spin decouples from the
LDF dynamics. The energy levels, therefore, depend on $l$ but not
on $s_{Q\bar Q}$ or $J$. Each level forms an HQSS degenerate multiplet containing one spin-singlet state with $s_{Q\bar Q}=0$ and $J=l$, together with one spin-triplet state for $l=0$ or three spin-triplet states with $J=l-1,l,l+1$ for $l>0$
\footnote{For quarkonium, the expected size of relativistic and spin corrections is $v^2$, with $v$ the relative velocity of the heavy quarks, related to operators entering at order $1/m_Q^2$. For tetraquarks, the expected size of those corrections is the mass splitting between vector and pseudoscalar heavy-light mesons related to operators entering at order $1/m_Q$~\cite{Soto:2020xpm,Oncala:2017hop,Brambilla:2018pyn,Brambilla:2019jfi,Soto:2023lbh}.}.
For the channels considered here,
\begin{equation}
  P=(-1)^{l+1},
  \qquad
  C=P(-1)^{s_{Q\bar Q}+1}.
\end{equation}
The quarkonium and isoscalar tetraquark channel multiplets with
$l\leq2$ relevant to the present calculation are listed in
Table~\ref{tab:multiplets for tetraquark}.
Table~\ref{tab:multiplets for tetraquark}
lists the quantum numbers carried by
the separate quarkonium and tetraquark channels. Once channels
with identical $J^{PC}$ are coupled, the physical bound states and resonances are in general mixtures of the corresponding channel
configurations.

\begin{table}
\centering
\renewcommand{\arraystretch}{1.2}
\setlength{\tabcolsep}{7pt}
\begin{tabular}{c c c c c c}  
\toprule[1.5pt]
\multirow{2}{*}{System} & \multirow{2}{*}{$k^{PC}$} & \multirow{2}{*}{Potential(s)} & \multirow{2}{*}{$l$} & \multicolumn{2}{c}{$J^{PC}$} \\
\cmidrule(lr){5-6}
 & & & & $s_{Q \bar{Q}} = 0$ & $s_{Q \bar{Q}} = 1$ \\
\midrule[1.5pt]
\multirow{3}{*}{$Q \bar{Q}$}  &$0^{++}$ & $\Sigma_g^+$ & $0$ & $0^{-+}$ & $1^{--}$ \\
& $0^{++}$ & $\Sigma_g^+$ & $1$ & $1^{+-}$ & $(0,1,2)^{++}$ \\
& $0^{++}$ & $\Sigma_g^+$ & $2$ & $2^{-+}$ & $(1,2,3)^{--}$ \\
\midrule[1.5pt]
\multirow{3}{*}{$Q \bar{Q} q \bar{q}$} & $1^{--}$ & $\Sigma_g^{+'}$, $\Pi_g$ & $1$ & $1^{+-}$ & $(0,1,2)^{++}$ \\
& $1^{--}$ & $\Sigma_g^{+'}$ & $0$ & $0^{-+}$ & $1^{--}$ \\
& $1^{--}$ & $\Sigma_g^{+'}$, $\Pi_g$ & $2$ & $2^{-+}$ & $(1,2,3)^{--}$ \\
\bottomrule[1.5pt]
\end{tabular}
\caption{
HQSS quantum numbers carried by the quarkonium
($Q\bar Q$) and the lowest isoscalar tetraquark
($Q\bar Q\bar q q$) channels with $l\leq2$ relevant to this work.
The corresponding short-distance $k^{PC}$ quantum numbers and
diabatic BO potentials are also shown. Physical states with common
$J^{PC}$ are in general mixtures of the listed channel configurations. The multiplets are ordered by increasing energies of the relative states.}
\label{tab:multiplets for tetraquark}
\end{table}

In the BO diabatic basis, the coupled-channel Schr\"odinger equation for states with $l>0$ and $P=(-1)^{l+1}$ reads\footnote{There is also a different Schr\"odinger equation for tetraquark states with $l>0$ and $P=(-1)^l$, but it is irrelevant to this work since it does not involve a quarkonium channel~\cite{Berwein:2024ztx}.}
\begin{align}\label{eq:diabatic}
\hspace{+0cm}\left[
-\frac{1}{m_Qr^2}\,\partial_r r^2 \partial_r + \frac{1}{m_Qr^2}
{\begin{pmatrix}
l (l + 1) & 0 & 0 \\[4pt]
0 & (l - 1) l & 0 \\[4pt]
0 & 0 & (l + 1) (l  + 2) \\[4pt]
\end{pmatrix}}\right. \hspace{0cm} \left. \right. \nonumber \\
&\hspace{-8cm}\left. + \begin{pmatrix} 
V_{\Sigma_{g}^+} & \sqrt{\frac{l}{2l + 1}} V_{\Sigma_g^+ \text{-} \Sigma_g^{+\prime}} & \sqrt{\frac{l + 1}{2l + 1}} V_{\Sigma_g^+ \text{-} \Sigma_g^{+\prime}} \\[6pt]
\sqrt{\frac{l}{2l + 1}} V_{\Sigma_g^+ \text{-} \Sigma_g^{+\prime}} & \frac{l}{2l + 1} V_{\Sigma_g^{+'}} + \frac{l + 1}{2l + 1} V_{\Pi_g}  & \frac{\sqrt{l(l + 1)}}{2l + 1}(V_{\Sigma_g^{+'}} - V_{\Pi_g}) \\[6pt]
\sqrt{\frac{l + 1}{2l + 1}} V_{\Sigma_g^+ \text{-} \Sigma_g^{+\prime}} & \frac{\sqrt{l(l + 1)}}{2l + 1}(V_{\Sigma_g^{+'}} - V_{\Pi_g}) & \frac{l + 1}{2l + 1} V_{\Sigma_g^{+'}} + \frac{l}{2l + 1}V_{\Pi_g}  \\
\end{pmatrix} \right]
\begin{pmatrix} \psi_{Q \bar{Q}(l)}^{l} \\[4pt] \psi_{M\bar{M}(l-1)}^{l}\\[4pt] \psi_{M\bar{M}(l+1)}^{l} \end{pmatrix}  
=  \mathcal{E}_{l} \begin{pmatrix} \psi_{Q \bar{Q}(l)}^{l} \\[4pt] \psi_{M\bar{M}(l-1)}^{l} \\[4pt] \psi_{M\bar{M}(l+1)}^{l} \end{pmatrix},
\end{align}
where the dependence of the potentials and the wavefunctions on the distance $r$ has been suppressed for ease of notation.
We use a subscript $Q \bar{Q}(l_{Q\bar{Q}})$ or $M\bar{M}(l_{Q\bar{Q}})$ to denote a quarkonium or tetraquark channel, respectively, in the partial wave $l_{Q\bar{Q}}$, this last expressed in terms of the $l$ quantum number.
The notation $M\bar M$ is used for convenience to label the
tetraquark BO channel in a definite heavy-pair partial wave. It
should be interpreted literally as a heavy-light
meson--antimeson configuration only in the large-distance
asymptotic region of that channel.
Note that for the quarkonium channel $l_{Q\bar{Q}}$ coincides with $l$, whereas for the tetraquark channels $l_{Q\bar{Q}}$ differs from $l$ by one unit.
For states with $l=0$ and $P=-1$, the Schr\"odinger equation reduces to
\begin{equation}\label{eq:diabaticl0}
\hspace{+0cm}\left[
-\frac{1}{m_Qr^2}\,\partial_r r^2 \partial_r + \frac{1}{m_Qr^2}
{\begin{pmatrix}
0 & 0 \\[4pt]
0 & 2 \\[4pt]
\end{pmatrix}}
+ \begin{pmatrix} 
V_{\Sigma_{g}^+} & V_{\Sigma_g^+ \text{-} \Sigma_g^{+\prime}} \\[6pt]
V_{\Sigma_g^+ \text{-} \Sigma_g^{+\prime}} & V_{\Sigma_g^{+'}}  \\
\end{pmatrix} \right]
\begin{pmatrix} \psi_{Q \bar{Q}(0)}^{0} \\[4pt] \psi_{M\bar{M}(1)}^{0} \end{pmatrix}  
=  \mathcal{E}_{l} \begin{pmatrix} \psi_{Q \bar{Q}(0)}^{0} \\[4pt] \psi_{M\bar{M}(1)}^{0} \end{pmatrix}.
\end{equation}
The details about the unitary transformation connecting the representation of the Schr\"odinger equation in Eq.~\eqref{eq:diabatic} with the one used in our previous studies \cite{Brambilla:2024imu,Brambilla:2026zfw} can be found in Appendix~\ref{app:changeofbasis}.

If the mixing potential
$V_{\Sigma_g^+-\Sigma_g^{+\prime}}$ is set to zero, the
quarkonium and tetraquark sectors decouple. The tetraquark sector
contains bound states below threshold and open-flavor scattering
states above it, whereas the confining quarkonium potential
supports a discrete tower of levels. Once the mixing is switched
on, quarkonium levels above the open-flavor threshold acquire pole widths and become resonances, while the bound states below threshold generally contain both quarkonium and tetraquark components.  
In the present work, we retain only the states below the first spin--isospin averaged $S+P$ threshold, where the chosen channel basis is applicable.

\subsection{Quark masses and potential parameterizations}\label{subsec:parametrization}

We now specify the diabatic-potential parametrizations and the
heavy-quark mass convention used in Eqs.~\eqref{eq:diabatic},~\eqref{eq:diabaticl0}.
The lowest $1^{--}$ adjoint meson mass is the only parameter
adjusted to experimental spectroscopy. The remaining numerical
inputs are taken from lattice QCD, from the known short- and
long-distance BOEFT constraints, or from the modeling assumptions
specified below.

In this work, we consider spin--isospin averaged charmed and
bottom heavy-light mesons, denoted by $D^{(\ast)}$ and $B^{(\ast)}$,
respectively. Following
Refs.~\cite{Brambilla:2024imu,Brambilla:2026zfw}, we adopt a
heavy-light scheme for the mass parameter entering
the kinetic terms of Eqs.~\eqref{eq:diabatic} and
\eqref{eq:diabaticl0}. We define
\begin{equation}
 m_c^{\rm HL}
 =
 m_{D^{(\ast)}}
 =
 \frac{m_D+3m_{D^\ast}}{4}
 =
 1.973~{\rm GeV},
 \label{eq:Dmeson}
\end{equation}
where
$m_D=(m_{D^0}+m_{D^+})/2$ and
$m_{D^\ast}=(m_{D^{*0}}+m_{D^{*+}})/2$, and
\begin{equation}
 m_b^{\rm HL}
 =
 m_{B^{(\ast)}}
 =
 \frac{m_B+3m_{B^\ast}}{4}
 =
 5.313~{\rm GeV},
 \label{eq:Bmeson}
\end{equation}
where
$m_B=(m_{B^0}+m_{B^+})/2$ and
$m_{B^\ast}=(m_{B^{*0}}+m_{B^{*+}})/2$.
These are the mass parameters used in the kinetic operators of
the coupled Schr\"odinger equations.

Among the potentials entering Eqs.~\eqref{eq:diabatic} and~\eqref{eq:diabaticl0}, the
quarkonium potential $V_{\Sigma_g^+}$ is well constrained by lattice QCD over a
broad range of distances. The tetraquark potential
$V_{\Sigma_g^{+\prime}}$ and the string-breaking mixing potential
$V_{\Sigma_g^+-\Sigma_g^{+\prime}}$ have been determined in
lattice QCD only in the vicinity of the string-breaking region,
whereas the tetraquark potential $V_{\Pi_g}$ is presently
unknown. We therefore construct parametrizations of
$V_{\Sigma_g^{+\prime}}$, $V_{\Pi_g}$, and
$V_{\Sigma_g^+-\Sigma_g^{+\prime}}$ that satisfy the known
short- and long-distance BOEFT constraints 
\cite{Berwein:2024ztx,Braaten:2024tbm}, and reproduce the
available lattice QCD information around
$1~\mathrm{fm}\lesssim r\lesssim1.5~\mathrm{fm}$.
As in Ref.~\cite{Brambilla:2026zfw}, heavy-strange tetraquark channels are not included.

At short distance, we model $V_{\Sigma_g^{+\prime}}$ and
$V_{\Pi_g}$ using the functional form employed for the quenched
hybrid BO potentials in Ref.~\cite{Alasiri:2024nue}, with the
gluelump mass replaced by the adjoint meson mass. At large
distance, we adopt a minimal Yukawa ansatz with range
$m_\pi^{-1}$, motivated by the range of pion exchange between the two static-light mesons.

\begin{align}
&V_{\Sigma_g^+}(r) = V_0 + \frac{\gamma}{r} + \sigma r,\label{eq:Vcornell}\\
&V_{\Lambda^{\sigma}_{\eta}}(r) = 
\begin{cases}
\kappa_8/r + \bar{\Lambda}_{1^{--}}^{\rm HL} + A_{\Lambda_\eta^\sigma}\, r^2 + B_{\Lambda_\eta^\sigma}\, r^4, & r < R_{\Lambda_\eta^\sigma} \\
F_{\Lambda_\eta^\sigma}\,e^{-r/d}/r + E_1, & r > R_{\Lambda_\eta^\sigma}
\end{cases},
\label{eq:QQbar_V}
\end{align}
where $\Lambda^{\sigma}_{\eta} \in \{\Sigma_g^{+\prime}, \Pi_g\}$, $V_0 = -1.142~\mathrm{GeV}$, $\gamma = -0.434$, $\sigma = 0.198\,\mathrm{GeV}^2$, $\kappa_8 = 0.037$, $A_{\Sigma_g^{+\prime}} = 0.0065\,\mathrm{GeV}^3$, $B_{\Sigma_g^{+\prime}} = 0.0018\,\mathrm{GeV}^5$, $A_{\Pi_g} = 0.0726\,\mathrm{GeV}^3$, $B_{\Pi_g} = -0.0051\,\mathrm{GeV}^5$, $d = 1/m_\pi \sim 1/0.15\,\mathrm{GeV}^{-1} = 1.31\,\mathrm{fm}$, $E_1 = 0.005~\mathrm{GeV}$.

The quantity $\bar\Lambda_{1^{--}}^{\rm HL}$ is the adjoint meson mass appearing in the heavy-light scheme used for the masses in Eqs.~\eqref{eq:Dmeson} and \eqref{eq:Bmeson}.
The parameters $F_{\Lambda_\eta^\sigma}$ and $R_{\Lambda_\eta^\sigma}$ are fixed by imposing continuity of the potentials and their first derivatives, giving
$F_{\Sigma_g^{+\prime}}=-0.1376$,
$R_{\Sigma_g^{+\prime}}=0.3384\,\mathrm{fm}$,
$F_{\Pi_g}=-0.0248$, and
$R_{\Pi_g}=0.1527\,\mathrm{fm}$.
As for the mixing potential $V_{\Sigma_g^+ \text{-}\Sigma_g^{+\prime}}$, it must vanish linearly at short distances \cite{TarrusCastella:2022rxb}. It is also expected that the mixing potential vanishes asymptotically at $r \to \infty$, since the quarkonium and tetraquark potentials $V_{\Sigma_g^+}$, $V_{\Sigma_g^{+ \prime}}$ get progressively further away from each other. 
Hence, we parameterize $V_{\Sigma_g^+ \text{-}\Sigma_g^{+\prime}}$ as
\begin{align}
  V_{\Sigma_g^+ \text{-}\Sigma_g^{+\prime}} = 
  \begin{cases}
g\,r/r_1 \quad &r < r_1
\\
g\, 
\quad  &r_1\le r \le r_2
\\
g \, e^{-(r-r_2)/r_0} & r > r_2
\end{cases}, 
\label{eq:Vmix}
\end{align}
where $r_0=0.5$~fm is the Sommer scale, $g=0.05$~GeV is taken from Ref.~\cite{Bulava:2024jpj}, and $r_1=0.95$~fm and $r_2=1.51$~fm\footnote{Another suitable scheme to use for this problem is the Cornell scheme~\cite{Mateu:2018zym,Brambilla:2026zfw}.
The Cornell mass scheme can be viewed as a proxy for the pole mass scheme, with the only difference being that, in the pole mass scheme, the octet potential is given by the perturbative expression for the repulsive Coulomb potential, whereas in the Cornell scheme it is parametrized phenomenologically.
Indicating with $m_Q^{\rm C}$ the masses in the Cornell scheme, in the static limit, we have 
\begin{equation}
 2m_Q^{\rm HL}
 =
 2m_Q^{\rm C}-V_0,
\end{equation}
where $-V_0/2$ equates the static-light mass of the quarkonium heavy-light meson in the Cornell scheme. Hence
\begin{equation}
 m_Q^{\rm C}
 =
 m_Q^{\rm HL}
 +V_0/2.
 \label{eq:HL-Cornell-mass-relation}
\end{equation}
With the parameters used here, this gives
$m_c^{\rm C}=1.402~{\rm GeV}$ and
$m_b^{\rm C}=4.742~{\rm GeV}$. These mass parameters are quoted only to specify the relation between the two different schemes; the masses used in the kinetic operators
of the present calculation remain those of
Eqs.~\eqref{eq:Dmeson} and \eqref{eq:Bmeson}.
The corresponding adjoint meson mass in the Cornell scheme is
\begin{equation}
  \Lambda_{1^{--}}^{\rm C}
  =\bar\Lambda_{1^{--}}^{\rm HL}-V_0.
  \label{eq:adjoint-energy-conventions}
\end{equation}}.

The tetraquark potentials in Eq.~\eqref{eq:QQbar_V} approach
$E_1=0.005~{\rm GeV}$ at large separation. Since the masses in
Eqs.~\eqref{eq:Dmeson} and \eqref{eq:Bmeson} are defined in the
heavy-light threshold scheme, the physical spin--isospin averaged open-flavor threshold is
\begin{equation}
 m_{\rm thresh}=2m_Q^{\rm HL}.
 \label{eq:threshold-general}
\end{equation}
Numerically,
\begin{align}
 m_{D^{(\ast)}\bar D^{(\ast)}}
 &=2m_c^{\rm HL}=3.946~{\rm GeV},
 \\
 m_{B^{(\ast)}\bar B^{(\ast)}}
 &=2m_b^{\rm HL}=10.626~{\rm GeV}.
\end{align}

Consequently, the physical mass associated with an eigenvalue
${\cal E}_l$ of the potentials as written in
Eqs.~\eqref{eq:Vcornell} and \eqref{eq:QQbar_V} is
\begin{equation}
 M_l
 =
 {\cal E}_l+2m_Q^{\rm HL}-E_1.
 \label{eq:quarkoniumlike-masses}
\end{equation}
\newpage

The lowest $1^{--}$ adjoint meson mass has not yet been
determined in lattice QCD and is therefore calibrated
phenomenologically. At leading order, both the heavy-light
thresholds and the members of an HQSS multiplet are spin-averaged.
We therefore require the spin-averaged $2P$ multiplet associated
with the $\chi_{c1}(3872)$ to lie $96~\mathrm{keV}$ below the
static spin--isospin averaged $D^{(\ast)}\bar D^{(\ast)}$ threshold. 
This condition uses the observed shallow nature of the
$\chi_{c1}(3872)$ as input, but it should not be interpreted as a
direct fit to the physical $D^0\bar D^{*0}$ threshold. 
We obtain\footnote{For comparison, Ref.~\cite{Brambilla:2026zfw} used
$\Lambda_{1^{--}}^{\rm C}=1.008~\mathrm{GeV}$ which corresponds to $\bar\Lambda_{1^{--}}^{\rm HL}=-0.134~\mathrm{GeV}$.
When inserted into the present equations, this value produces a binding energy of
$764~\mathrm{keV}$ for the spin-averaged multiplet associated with
the $\chi_{c1}(3872)$. The sensitivity of the near-threshold
states to this change is discussed in Section~\ref{sec:adjoint meson fine tuning}.}
\begin{equation}
  \bar\Lambda_{1^{--}}^{\rm HL}=-0.113~\mathrm{GeV},
  \qquad
  \Lambda_{1^{--}}^{\rm C}
  =\bar\Lambda_{1^{--}}^{\rm HL}-V_0
  =1.029~\mathrm{GeV},
  \label{eq:adjoint-energy-fit}
\end{equation}
where $\bar\Lambda_{1^{--}}^{\rm HL}$, $\Lambda_{1^{--}}^{\rm C}$ indicate respectively the value of the adjoint meson mass in the heavy-light and Cornell schemes.  

The adiabatic energies shown in Fig.~\ref{fig:isospin0} are obtained by diagonalizing the potential matrix entering
Eq.~\eqref{eq: transformation between diabatic and adiabatic potential}. In the string-breaking region, the
$1\Sigma_g^+$ and $2\Sigma_g^+$ adiabatic potentials reproduce the lattice QCD data from the D200 ensemble of Ref.~\cite{Bulava:2024jpj}, which constrain $V_{\Sigma_g^+}$,
$V_{\Sigma_g^{+\prime}}$, and
$V_{\Sigma_g^+-\Sigma_g^{+\prime}}$ in that region.
The $\Pi_g$ potential is not presently constrained by lattice data and is therefore model-dependent.

The parametrization of the presently unknown $V_{\Pi_g}$ potential
and of the short-distance part of
$V_{\Sigma_g^{+\prime}}$ introduces a model uncertainty. Its
effect is small for quarkonium-dominated states. For the calibrated
shallow charmoniumlike state, part of the variation of its mass can
be compensated by retuning the adjoint meson mass. This
compensation does not remove the model dependence on other observables. 
This may indeed significantly affect the mass, the decay width, and the composition 
of resonances with a significant tetraquark component, as discussed in Sections~\ref{sec:spectrum},~\ref{sec:probabilities}.

\section{Determination of resonance poles above threshold}
\label{sec:spectrum}

We determine the resonance poles \footnote{We also determine the bound state multiplets lying in the vicinity of $S + S$ threshold; see Section~\ref{sec:results}.} of Eqs.~\eqref{eq:diabatic} and~\eqref{eq:diabaticl0} using three complementary methods. First, we analytically continue the $T$ matrix to the unphysical Riemann sheet and locate its complex poles. Second, we study the poles of the real $K$ matrix on the physical energy axis. Finally, we use the complex scaling method (CSM) to obtain the resonance energies as discrete complex eigenvalues of the rotated Hamiltonian. The complex-scaled eigenvalues provide an independent determination of the same poles obtained from the
analytically continued $T$ matrix. For narrow and isolated
resonances, far away from the threshold, the $T$ and $K$ matrix pole parameters coincide, whereas appreciable differences may arise for broad resonances or in the presence of a significant nonresonant background.

\subsection{\texorpdfstring{$T$}{T} matrix from
the emergent-wave method}
\label{subsec:emergent}

Bound states and resonances are associated with nonanalytic
structures of the $S$ matrix~\cite{Taylor:1972pty,Oller:1997ti,Bedaque&Van-Colck:2002EFT-for-few-necleon-system,Ceci:2006jj,Friedrich:2015obs,PavonValderrama:2019lsu,Oller:2019opk,Peng:2023lfw}. 
Resonances above the relevant open-flavor threshold are identified
with poles of the analytically continued scattering amplitude on
an unphysical Riemann sheet. The corresponding $T$ matrix can be
extracted from the asymptotic behavior of the scattering solutions in the open channels~\cite{Friedrich:2015obs,Bicudo:2019ymo,Bicudo:2017szl,Hoffmann:2024hbz}.

Throughout this section, $E$ denotes the physical
center-of-mass energy. It is related to the Schr\"odinger
eigenvalue $\mathcal{E}_l$ used in Section~\ref{sec:BOEFT} by  
\begin{equation}
  E= M_l = \mathcal{E}_l+2m_Q^{\rm HL}-E_1.
  \label{eq:physical-scattering-energy}
\end{equation}

The Schr\"odinger equations~\eqref{eq:diabatic}
and~\eqref{eq:diabaticl0} contain one quarkonium channel
$Q\bar{Q}(l_{Q\bar{Q}})$
and one or two tetraquark channels $M\bar{M}(l_{Q\bar{Q}})$. 
 In scattering terminology, the
quarkonium channel is always closed because
$V_{\Sigma_g^+}(r)$ increases without bound at large separation.
The tetraquark channels are closed below the spin-averaged
open-flavor threshold and open above it. The scattering matrix is
therefore defined entirely in the space of asymptotically open
tetraquark channels. At leading order, these channels share the
same threshold and hence the same scattering momentum.

When investigating quarkoniumlike resonances above threshold, the $M\bar{M}(l_{Q\bar{Q}})$ channels are always open.
The corresponding $S$ matrix for the scattering process can therefore be parametrized in terms of these tetraquark channels.

For $l>0$, unitarity and time-reversal symmetry imply that the $S$ matrix can be parametrized as
\begin{eqnarray}
S(E)=
\begin{pmatrix}
\eta(E)\, e^{2i\delta_{l-1}(E)} & i\sqrt{1-\eta^2(E)}\,e^{i(\delta_{l-1}(E)+\delta_{l+1}(E))} \\
i\sqrt{1-\eta^2(E)}\,e^{i(\delta_{l-1}(E)+\delta_{l+1}(E))} & \eta(E)\, e^{2i\delta_{l+1}(E)}
\end{pmatrix},
\label{eq:S-matrix}
\end{eqnarray}
where $\eta$ denotes the inelasticity and $\delta_{l\pm1}$ are the phase shifts of the two open channels $M\bar{M}(l_{Q\bar{Q}})$ with orbital angular momenta $l_{Q\bar{Q}}=l\pm1$.
For $l=0$, the $S$ matrix can be parametrized simply as
\begin{equation}
    S(E) = 
        \begin{pmatrix}
            e^{2i\delta_1(E)}
        \end{pmatrix},
    \label{eq:S-matrix0}
\end{equation}
where $\delta_1$ is the phase shift of the single open channel $M\bar{M}(l_{Q\bar{Q}})$ with $l_{Q\bar{Q}}=1$.

Within our normalization convention, the partial-wave $T$ matrix
is defined by
\begin{eqnarray}
S(E) = I+ 2 i \,T(E)\,,
\label{eq: S and T relation}
\end{eqnarray}
where $I$ is the identity matrix in the space of open channels.

It is convenient to denote the number of open channels by
\begin{equation}
  N_{\rm open}=
  \begin{cases}
    1, & l=0,\\
    2, & l>0.
  \end{cases}
  \label{eq:number-open-channels}
\end{equation}
Indices $i,j=1,\ldots,N_{\rm open}$ will denote asymptotically
open tetraquark channels in the $S$ matrix.
The index $0$ will be reserved for the closed quarkonium component
of the coupled wavefunction which has $l_{Q\bar Q}=l$. 
For $l>0$, the open-channel indices $i,j=1,2$ refer to the
tetraquark channels with heavy-pair orbital angular momenta
$l_{Q\bar Q}=l-1$ and $l_{Q\bar Q}=l+1$, respectively. For
$l=0$, the single open channel $i=j=1$ has
$l_{Q\bar Q}=1$. For each energy above threshold,
there is one linearly independent scattering solution for each
open incoming channel.

Using energy-normalized scattering states, the asymptotic radial
wavefunction in an open channel $i$ for an incoming wave in
channel $j$ takes the form
\begin{eqnarray}
\psi_{ij}(r)\xrightarrow{r\to\infty}
\sqrt{\mu k} \sqrt{\frac{2}{\pi}}\left(\delta_{ij} j_{l_{Q\bar{Q}}}(kr)
+ i\, T_{ij}(E)\, h_{l_{Q\bar{Q}}}^{(1)}(kr)\right)\,,
\label{eq:asymptotic behaviour of u(r)}
\end{eqnarray}
Here $\mu=m_Q^{\rm HL}/2$ is the reduced mass of the heavy pair,
$j_{l_{Q\bar Q}}$ and $h^{(1)}_{l_{Q\bar Q}}$ are the spherical
Bessel and outgoing spherical Hankel functions, respectively, and
\begin{eqnarray}
k=\sqrt{2\mu(E - m_\mathrm{thresh})}
\label{eq:momenta for open channels}
\end{eqnarray}
is the common scattering momentum of the degenerate open channels.
On the physical sheet, $k$ is chosen real and positive above
threshold. To access resonance poles on the adjacent unphysical
sheet, we analytically continue the momentum with
$\operatorname{Im}k<0$. We display the energy dependence
explicitly only for the scattering matrices and suppress it in
the wavefunctions and momenta when no ambiguity arises.

For each incoming open channel $j$, the wavefunction in the closed quarkonium channel is written as
\begin{eqnarray}
\psi_{0j}(r)=\dfrac{u_{0j}(r)}{kr}\,,
\label{eq:radial wave function 1}
\end{eqnarray}
where $u_{0j}(r)$ satisfies the boundary conditions
\begin{eqnarray}
\begin{cases}
u_{0j}(r)\sim r^{l_{Q\bar{Q}}+1} & \text{$r\to 0$,}\\
u_{0j}(r)\rightarrow 0 & \text{$r\to \infty$.}
\end{cases}
\label{eq:boundary conditions QQ}
\end{eqnarray}

The wavefunctions in the open tetraquark channels are decomposed
into an incoming regular wave and an emergent scattered wave,
\begin{eqnarray}
\psi_{ij}(r)=\dfrac{u_{ij}(r)}{kr} =\sqrt{\mu k} \sqrt{\frac{2}{\pi}} \left( \delta_{ij} j_{l_{Q\bar{Q}}}(kr) + \dfrac{\lambda_{ij}(r)}{kr}\right),
\label{eq:radial wave function 2}
\end{eqnarray}
where $\lambda_{ij}(r)$ satisfies the boundary conditions
\begin{eqnarray}
\begin{cases}
\lambda_{ij}(r)\sim r^{l_{Q\bar{Q}}+1} & \text{$r\to 0$,}\\
\lambda_{ij}(r)\rightarrow i\, k r\, T_{ij}(E) \, h_{l_{Q\bar{Q}}}^{(1)}(kr) & \text{$r\to \infty$,}
\end{cases}
\label{eq:boundary conditions MM}
\end{eqnarray}
for all $j$.

We determine the $T$ matrix elements by matching the numerical
solutions of the coupled radial equations to the asymptotic
conditions in Eqs.~\eqref{eq:boundary conditions QQ}
and~\eqref{eq:boundary conditions MM} at a sufficiently large
matching radius $r_{\rm match}$. Resonances are identified with
poles of the analytically continued $T$ matrix on the unphysical
Riemann sheet.
We parametrize a pole position as
\begin{equation}
  E_{\rm pole}=M-\frac{i}{2}\Gamma,
  \label{eq:T-pole-position}
\end{equation}
where $M$ is the pole mass and $\Gamma$ is the pole width generated
by the spin-averaged, nonstrange $S+S$ open-flavor channels
included in the calculation. It is not a prediction for the
physical total width.

Finally, we use the residues of the $T$ matrix at the complex pole to characterize the relative strengths with which the pole couples to the asymptotically open channels.
The residues of the $T$ matrix elements are defined by
\begin{equation}
\mathcal{R}_{ij} = -\frac{1}{2\pi i} \oint \mathrm{d} E \, T_{ij}(E),
\end{equation}
where the integral is over a closed contour enclosing the simple pole at $E_{\mathrm{pole}}$ but no other singularity.
The residues $\mathcal{R}_{ij}$ obey the factorization theorem \cite{Gribov:2009}
\begin{equation}
\mathcal{R}_{ij}^2 = \mathcal{R}_{ii} \mathcal{R}_{jj}
\end{equation}
for any $i$ and $j$.
We can therefore define the complex pole couplings $\tilde{g}_i^2$ as
\begin{equation}
\tilde{g}_i^2 = \mathcal{R}_{ii}.
\end{equation}
Each complex pole coupling $\tilde{g}_i^2$ expresses the strength of the transition of the resonance into the channel $i$.
By taking the absolute value of the complex pole couplings and normalizing them over the available decay channels, we define real, normalized pole couplings $g_i^2$,
\begin{equation}
g_i^2
=
\frac{\left|\widetilde g_i\right|^2}
{\sum_{j=1}^{N_{\rm open}}
 \left|\widetilde g_j\right|^2}
=
\frac{\left|\mathcal{R}_{ii}\right|}
{\sum_{j=1}^{N_{\rm open}}
 \left|\mathcal{R}_{jj}\right|},
\qquad
i=1,\ldots,N_{\rm open}.
\label{eq:normalized-pole-couplings}
\end{equation}
The normalized pole couplings characterize the relative residue
strengths with which the pole couples to the asymptotically open
channels. For a narrow and isolated resonance, and only within the
truncated channel space included here, they may approximate
branching fraction ratios, $\mathrm{Br}_i\approx g_i^2$. 
This is not strictly true for broad or overlapping resonances, but in this case, the pole couplings $g_i^2$ can nonetheless be used to translate the residue information into an intuitive measure of the transition strength of the resonance into the different channels.
In general, these normalized pole couplings should not be identified
with physical branching fractions. Indeed, in our model, they
characterize only the relative residue strengths in the included
$S+S$ open-flavor channels.

\subsection{\texorpdfstring{$K$}{K} matrix method}
\label{subsec:k-matrix}

On the physical energy axis, the $K$ matrix is related to the
$T$ matrix by~\cite{Aitchison:1972ay,Svarc:2012wr}

\begin{equation}
K^{-1}(E) = T^{-1}(E) + i I.
\label{eq:k-matrix_def}
\end{equation}
Or, equivalently,
\begin{equation}
  T(E)=K(E)\left[I-iK(E)\right]^{-1}.
  \label{eq:T-from-K-general}
\end{equation}
Unitarity and time-reversal symmetry imply that $K(E)$ is real and
symmetric for real energies above threshold.
Proper resonances (i.e., resonances whose pole is located above threshold) may correspond to poles of the $K$ matrix for real energies above threshold on the physical Riemann sheet.
To see this, consider a $K$ matrix with a real pole at an energy $M$ and write its Laurent expansion around the pole as,
\begin{equation}
    K(E) = \bm{g} \otimes \bm{g} \frac{\Gamma/2}{M - E} + K^\text{bg}(E),
    \label{eq:k-matrix_Laurent}
\end{equation}
where the ``background'' term $K^\text{bg}(E)$ contains the sum of all the regular terms of the Laurent expansion.
In the pole term, $\bm{g}$ is a unit vector containing the normalized, real pole couplings to each channel, $\sum_{i=1}^{N_{\rm open}} g_i^2=1$, and $\otimes$ denotes the outer product, $(\bm{g}\otimes \bm{g})_{ij}=g_i g_j$.

For a narrow isolated resonance, the mass, width, and pole couplings calculated using the $K$ matrix coincide with those obtained from the $T$ matrix.
To see this, plug Eq.~\eqref{eq:k-matrix_Laurent} into Eq.~\eqref{eq:k-matrix_def} while neglecting the background term $ K^\text{bg}$ and solve for $T$,
\begin{equation}
    T(E) = - \frac{\bm{g} \otimes \bm{g} \, \Gamma/2}{E - M + i \bm{g} \otimes \bm{g} \, \Gamma / 2 }.
    \label{eq:T from K}
\end{equation}
Note that the denominator on the right side of Eq.~\eqref{eq:T from K} can be rewritten as
\begin{equation}
    E - M + i \bm{g} \otimes \bm{g} \, \Gamma / 2 = i \left(\bm{g} \otimes \bm{g} - I\right) \Gamma /2 + (E - M + i \Gamma /2)
\end{equation}
and that the matrix $\left(\bm{g} \otimes \bm{g} - I\right)$ is singular since it maps the nonvanishing vector $\bm{g}$ to zero, $\left(\bm{g} \otimes \bm{g} - I\right)\bm{g} = 0$.
Therefore, we see that the $T$ matrix in Eq.~\eqref{eq:T from K} has a simple pole at $E=M - i \Gamma /2$, and we can further expand it around the pole as
\begin{equation}
    T(E) = - \bm{g} \otimes \bm{g} \frac{\Gamma/2}{E - M + i \Gamma / 2 } +\dotsc,
\end{equation}
omitting regular terms.
So, an isolated pole of the $K$ matrix with a real energy $M$ corresponds to a resonance with complex energy $M - i \Gamma/2$, width $\Gamma$, and pole couplings $\tilde g_i^2= g_i^2 \Gamma / 2$.

When the regular terms in the Laurent expansion are not
negligible, the real-axis $K$ matrix parameters need not coincide
with the complex $T$ matrix pole parameters.
The difference is particularly relevant for broad resonances, overlapping
structures, and poles close to a threshold, where the scattering
background can vary rapidly.
In those cases, it may even happen that a $T$ matrix pole has no correspondent $K$ matrix pole, or vice versa.
We therefore refer to the quantities
extracted from Eq.~\eqref{eq:k-matrix_Laurent} as the
$K$ matrix mass, width parameter, and normalized pole couplings.
They coincide with the pole mass, pole width, and
normalized residue couplings extracted from the $T$ matrix only in the narrow, isolated-resonance limit. More generally, the real-axis
$K$ matrix poles, when applicable, can provide a useful diagnostic of the resonance
structures found through analytic continuation of the $T$ matrix.

The $K$ matrix is obtained by solving the coupled Schr\"odinger
equations at real energies above threshold. The closed
quarkonium-channel boundary conditions are those of
Eqs.~\eqref{eq:radial wave function 1}
and~\eqref{eq:boundary conditions QQ}. In the open tetraquark
channels, the outgoing-wave condition in
Eq.~\eqref{eq:boundary conditions MM} is replaced by the
real standing-wave condition
\begin{equation}
\begin{cases}
\lambda_{ij}(r)\sim r^{l_{Q\bar{Q}}+1} & \text{$r\to 0$,}\\
\lambda_{ij}(r)\rightarrow{} - kr \, K_{ij}(E) \, y_{l_{Q\bar{Q}}}(kr) & \text{$r\to \infty$,}
\end{cases}
\label{eq:real boundary conditions MM}
\end{equation}
where $i,j=1,\ldots,N_{\rm open}$ 
and $y_{l_{Q\bar{Q}}}(kr)$ is the spherical Bessel function of the second kind.
Matching a complete set of linearly independent real solutions to these asymptotic forms determines the $K$ matrix
elements. In practice, we solve the radial equations using the
SPARSE algorithm~\cite{Bruschini:2025xhf}.

\subsection{Complex scaling method}
\label{subsec:complex}
The complex scaling method provides an independent determination
of the resonance energies and a cross-check of the poles obtained
from the analytically continued $T$ matrix. A resonance wavefunction satisfying outgoing boundary conditions is not
square-integrable on the real coordinate axis
\cite{Ashida:2020dkc,Myo:2014ypa,Romo:1968tcz,Myo:2020rni}.
Formally, a resonance is an eigenstate $\ket{\Psi_R}$ of the
Hamiltonian with complex eigenvalue,
\begin{eqnarray}
H\ket{\Psi_R}=(M - i \Gamma / 2)\ket{\Psi_R}\,,
\label{eq: resonance hamiltonian}
\end{eqnarray}
where the complex energy $M - i \Gamma / 2$ is the position of the associated $T$ matrix pole.

The usual hermitian inner product is not suitable for normalizing
a resonance. Taking the Hermitian conjugate of
Eq.~\eqref{eq: resonance hamiltonian} gives
\begin{equation}
  \bra{\Psi_R}H
  =
  \bra{\Psi_R}
  \left(M+\frac{i}{2}\Gamma\right),
  \label{eq:resonant_bra}
\end{equation}
so the Hermitian-conjugate state is associated with the
complex-conjugate energy.\footnote{This reflects the non-Hermitian
spectral structure required for states with complex energies; see
Ref.~\cite{Berggren:1968zz}.}
One therefore introduces the biorthogonal Gamow state
$\bra{\widetilde\Psi_R}$ satisfying
\begin{equation}
  \bra{\widetilde\Psi_R}H
  =
  \bra{\widetilde\Psi_R}
  \left(M-\frac{i}{2}\Gamma\right).
  \label{eq:gamov_state}
\end{equation}

Note that Eqs.~\eqref{eq:resonant_bra} and \eqref{eq:gamov_state}, together with time-reversal symmetry imply that the Gamow state $\bra{\widetilde{\Psi}_R}$ can be expressed as the complex conjugate of the dual vector $\bra{\Psi_R}$, that is, $\bra{\widetilde{\Psi}_R} = \bra{\Psi_R}^\ast$.
The resulting bilinear form is the $c$ product,
\begin{equation}
  \langle\widetilde\Psi_R|\Psi_R\rangle
  =
  \sum_{i=0}^{N_{\rm open}}
  \int_0^\infty dr\,r^2\psi_i(r)^2.
  \label{eq: inner product for CSM}
\end{equation}
For resonance normalization and channel decomposition within the
complex scaling framework, the $c$ product replaces the usual
Hermitian inner product. For bound states, and for real-energy
standing-wave solutions that may be chosen real, it reduces to
the standard norm.

For a proper resonance, the $c$ product $\langle\widetilde{\Psi}_R|\Psi_R\rangle$ can be regularized using, for instance, the Zel'dovich prescription~\cite{Berggren:1968zz}, and the state can be normalized as
\begin{eqnarray}
\langle \widetilde{\Psi}_R|\Psi_R\rangle = N^2 \lim_{\epsilon\to 0}
\sum_{i=0}^{N_{\rm open}}\int \mathrm{d}r\, r^2\psi_i(r)^2\,
e^{-\epsilon r^2}
=1\,,
\label{eq: wave function normalization for Gamov states}
\end{eqnarray}
where $N$ is a complex normalization constant.
More generally, any regulator of the form $e^{-\epsilon r^n}$, with $1 < n < \pi / \bigl[2 \arctan(\Gamma/(M - m_{\mathrm{thresh}}))\bigr]$, leads to the same norm for the resonant state~\cite{Rakityansky:2022,GYARMATI1971523}.

An equivalent and practically convenient framework is provided by the complex scaling method, originally developed by Aguilar, Balslev, and Combes~\cite{Myo:2020rni,10.1007/3-540-50994-1_56,Aguilar:1971ve,Balslev:1971vb}. In this approach, the spatial coordinate is rotated into the complex plane, rendering the resonance wavefunctions square-integrable. The transformation is implemented via a transformation operator $U(\theta)$,
\begin{eqnarray}\label{eq:r rotated}
U(\theta)\,r\,U^{-1}(\theta)=r\,e^{i\theta}\,,
\end{eqnarray}
under which the momentum transforms as%
\begin{eqnarray}\label{eq:k rotated}
U(\theta)\,k\,U^{-1}(\theta)=k\,e^{-i\theta}\,.
\end{eqnarray}
Note the above global complex scaling given in Eqs.~\eqref{eq:r rotated},~\eqref{eq:k rotated}, applied to the piecewise potentials in Eqs.~\eqref{eq:QQbar_V}-\eqref{eq:Vmix} formally violates the assumption of dilation analyticity at the matching points $r_1$, $r_2$, and $R_{\Lambda^{\sigma}_{\eta}}$. To circumvent this problem, we employ the smooth exterior complex scaling method (SECSM), in which the complex rotation of the spatial coordinate is applied only beyond a radius $R_0$, whose value is larger than those of all matching points $r_1$, $r_2$, and $R_{\Lambda^{\sigma}_{\eta}}$. 
In this case, the spatial coordinate is mapped into the complex plane as 
\begin{equation}\label{eq:r rotated gen}
z_{\theta}(r)\equiv U^\prime(\theta)\,r\,U^{^\prime\,-1}(\theta)=r+\left(e^{i\theta}-1\right)f(r),
\end{equation}
where
\begin{equation}\label{eq:mapping}
f(r)=
\begin{cases}
0, & r\le R_0,\\
L\left(-4x^3+7x^4-3x^5\right), & R_0<r<R_1,\\[2mm]
r-R_1, & r\ge R_1,
\end{cases}
\end{equation}
with
\begin{equation}
L=R_1-R_0,
\qquad
x=\frac{r-R_0}{L}.
\end{equation}
In the present calculation, we choose
\begin{equation}
R_0=1.6~\mathrm{fm},
\qquad
R_1=2.1~\mathrm{fm},
\end{equation}
and verify the independence of the results from the specific values of $R_0$, $R_1$ within a $1$~fm range.
With the above mapping function of Eq.~\eqref{eq:mapping}, the complex contour for the radial coordinate is smoothly connected to the physical
real axis and approaches the standard global scaling form of Eq.\eqref{eq:r rotated} at large $r$,
\begin{equation}
U^\prime(\theta)\,r\,U^{^\prime\,-1}(\theta) \approx re^{i\theta},
\qquad r \gg R_1.
\end{equation}
The Schr\"odinger equation is thus mapped to
\begin{eqnarray}
H^{\theta}\ket{\Psi^{\theta}}
=E\ket{\Psi^{\theta}}\,,
\label{eq:schrodinger eq rotated}
\end{eqnarray}
with $H^{\theta}=U^\prime(\theta)HU^{^\prime-1}(\theta)$\footnote{Under the coordinate transformation of Eq.~\eqref{eq:r rotated gen}, the derivatives term transform as
 $
\frac{\partial}{\partial z_{\theta}}
=
\frac{1}{z_{\theta}'(r)}\frac{\partial}{\partial r},
$ 
and
 $
\frac{\partial^2}{\partial z_{\theta}^2}
=
\frac{1}{z_{\theta}'(r)^2}\frac{\partial^2}{\partial r^2}
-
\frac{z_{\theta}''(r)}{z_{\theta}'(r)^3}\frac{\partial}{\partial r}.
$ 
Thus, the SECSM Hamiltonian in Eq.~\eqref{eq:schrodinger eq rotated} is obtained by replacing
$r\rightarrow z_{\theta}(r)$ in the potential and centrifugal terms, and by using the above chain rule relations for the kinetic operator term.}.

Under the complex rotation, the continuum spectrum is rotated
away from the real axis by an angle $2\theta$, whereas the bound-state
energies and exposed resonance poles remain invariant under
changes of $\theta$ within the stability region
\cite{BERGGREN197323,Myo:2020rni}. A pole at
$E_{\rm pole}=M-i\Gamma/2$ is exposed when

\begin{eqnarray}
\theta>
  \frac{1}{2}
  \tan^{-1} \biggl[
    \frac{\Gamma}{2(M-m_{\rm thresh})}
  \biggr].
  \label{eq:complex-scaling-angle}
\end{eqnarray}
The stability of the complex eigenvalue under variations of
$\theta$ distinguishes a resonance or a bound state from the rotated continuum.

Under SECSM, the resonance wavefunction in each channel $i$ transforms as\footnote{
For the global complex scaling map,
$z_\theta(r)=r e^{i\theta}$ and
$z_\theta'(r)=e^{i\theta}$, the present convention gives
$\psi_i^\theta(r)=\psi_i(r e^{i\theta})$.
The conventional global CSM wavefunction includes a Jacobian
prefactor and is given by
$\psi_{i,\mathrm{CSM}}^\theta(r)
=e^{3i\theta/2}\psi_i(r e^{i\theta})$.
In the present convention, the $c$ product is evaluated with
the radial contour measure $z_\theta(r)^2z_\theta'(r)\,dr$,
which reduces to $e^{3i\theta}r^2\,dr$ for global complex scaling.}

\begin{equation}
\psi_i^\theta(r)
\equiv [U'(\theta)\psi_i](r)
=\psi_i(z_\theta(r))
=\frac{\phi_i(r)}{z_\theta(r)},
\end{equation}
where $\phi_i(r)\equiv u_i(z_\theta(r))$ denotes the reduced
radial wavefunction evaluated along the complex contour.
When the scaling angle exposes the resonance pole, the transformed
wavefunction becomes square-integrable as a function of the real
path $r$.

\subsection{Numerical implementation and convergence}
\label{subsec:numerical}

The radial equations~\eqref{eq:diabatic} and~\eqref{eq:diabaticl0} are solved numerically for the $T$ matrix, $K$ matrix, and complex scaling methods.

For the $T$ matrix method, these equations are solved over the interval 
$r\in[r_{\rm min},r_{\rm max}]$, where
$r_{\rm min}=10^{-5}~\mathrm{fm}$ and
$r_{\rm max}=10~\mathrm{fm}$, using an explicit Runge-Kutta method of order 8 with a relative tolerance of $10^{-7}$ and an absolute one of $10^{-10}$.
The numerical radial wavefunctions relative to the open-flavor tetraquark channels are then matched to their asymptotic expression given in Eq.~\eqref{eq:asymptotic behaviour of u(r)} 
at a sufficiently large distance $r_\text{match}$, where the boundary conditions given in Eqs.~\eqref{eq:boundary conditions QQ},~\eqref{eq:boundary conditions MM} for $r \to \infty$ are numerically well satisfied to extract the $T$ matrix entries. The typical values of $r_\text{match}$ are of order $2$-$3~\mathrm{fm}$ for charmoniumlike resonances and $1.5$-$2.5~\mathrm{fm}$ for bottomoniumlike resonances.
In general, the numerical stability of the results can be checked by considering small variations of $r_\text{match}$ within a $0.5~\mathrm{fm}$ range from the optimal value and verifying the very weak dependence of the results for the pole masses, widths, and so on from a specific value of $r_\text{match}$.
The $T$ matrix poles on the complex energy plane are identified by looking for poles of its determinant.

For the $K$ matrix method, Eqs.~\eqref{eq:diabatic},~\eqref{eq:diabaticl0} are solved within the interval $r\in[r_{\rm min},r_{\rm max}]$, with
$r_{\rm min}=10^{-4}~\mathrm{fm}$, $r_{\rm max}=200~\mathrm{fm}$ on an equally spaced grid of spacing $10^{-4}~\mathrm{fm}$.
For this method, an increase in the radial resolution by a factor of $10$ would change the pole masses and widths by less than $1$~eV.

Finally, in the complex scaling calculation, Eqs.~\eqref{eq:diabatic},~\eqref{eq:diabaticl0} are solved within the interval $r\in[r_{\rm min},r_{\rm max}]$ , with
$r_{\rm min}=10^{-3}~\mathrm{fm}$, $r_{\rm max}=30~\mathrm{fm}$ on an equally spaced grid of spacing $10^{-3}~\mathrm{fm}$.
We use rotation angles in the interval $\theta\in[\theta_{\rm min},\theta_{\rm max}]$, with
$\theta_{\rm min}=0^\circ$, $\theta_{\rm max}=40^\circ$.
In the present calculation, we vary the rotation angles between $5^\circ$ and $40^\circ$. The stability of our results has been checked across a range of around $20^\circ$, giving a variation of the pole masses and widths of within $1~{\rm MeV}$.

\section{ Characterization of the channel composition}
\label{sec:probabilities}

For a bound state, the channel content is defined
unambiguously from its normalized wavefunction. The integral of the squared modulus of each channel component gives the
probability of finding the state in the corresponding
configuration, and the sum of all channel probabilities is equal
to one.

The situation is different for a resonance. A resonance wavefunction satisfying the outgoing boundary conditions is not
square-integrable with respect to the usual hermitian norm, and there is no unique probabilistic definition of its channel
composition. A quantitative characterization, therefore, requires
a prescription that reduces to the ordinary channel
probabilities in the bound-state limit. Depending on the
prescription, the resulting resonance components may be complex
or may take real values outside the interval $[0,1]$. These
features do not affect the physical scattering amplitudes, which
are independent of how the resonance channel content is
parametrized.

In this section, we employ two complementary prescriptions. The
first is based on the complex-scaled Gamow state at the complex
resonance energy, while the second is based on the standing-wave
solution associated with a real $K$ matrix pole. To avoid
proliferating notation, we use the same symbols $Z$ and $X_i$ in
both subsections. Their definitions are nevertheless
prescription dependent, and values obtained with the two methods
should not be identified with one another. Their comparison
instead provides an indication of the intrinsic prescription
dependence of the resonance channel content.

\subsection{Complex scaling prescription}
\label{subsec:emergent_complex_composition}

Within the complex scaling method, a resonance is represented by
a square-integrable eigenstate $\ket{\Psi^\theta}$ of the rotated
Hamiltonian. Its normalization is defined through the $c$
product introduced in Section~\ref{subsec:complex},
\begin{equation}
  \langle\widetilde\Psi^\theta|\Psi^\theta\rangle
  =
  N^2
  \sum_{i=0}^{N_{\rm open}}
  \int_0^\infty dr\,z_{\theta}^{'}(r)\,z_{\theta}(r)^2\,
  \psi_i^\theta(r)^2
  =1.
  \label{eq: wave function normalization for CSM}
\end{equation}

Here $i=0$ denotes the quarkonium channel and
$i=1,\ldots,N_{\rm open}$ denote the
open-flavor/tetraquark channels, while $N$ is a constant factor. The integrand contains the square of the complex-scaled wavefunction rather than its modulus squared. Consequently, both the individual channel
integrals and the normalization constant $N$ are in general complex.

We define the quarkonium component $Z$ and the
open-flavor/tetraquark components $X_i$ by
\begin{align}
  Z
  &=
  N^2
  \int_0^\infty dr\,z_{\theta}^{'}(r)\,z_{\theta}(r)^2\,
  \psi_0^\theta(r)^2,
  \\
  X_i
  &=
  N^2
  \int_0^\infty dr\,z_{\theta}^{'}(r)\,z_{\theta}(r)^2\,
  \psi_i^\theta(r)^2,
  \qquad
  i=1,\ldots,N_{\rm open}.
   \label{eq: elementarity and compositeness defined with CSM}
\end{align}
They satisfy
\begin{equation}
  Z+\sum_{i=1}^{N_{\rm open}}X_i=1.
\end{equation}

For a bound state with real wavefunctions, the
components $Z$ and $X_i$ can then be
identified with the quarkonium probability and the
open-flavor/tetraquark-channel probabilities, respectively. For
a resonance, by contrast, $Z$ and $X_i$ are generally complex
and do not admit a direct probabilistic interpretation. They
should instead be understood as the analytic continuation of the
bound-state channel components to the complex resonance energy.

The complex channel components $X_i$ and the pole couplings
introduced in Section~\ref{subsec:emergent} encode related but
distinct information about the coupling of the resonance to the
open channels. Their precise relation depends on the energy
dependence of the interaction and on the normalization
prescription~\cite{Baru:2010ww,Aceti:2012dd,
Aceti:2014ala}. In particular, no direct equality between $X_i$
and the normalized pole couplings $g_i^2$ is implied in the
general coupled-channel case considered here.

In weak-binding analyses, a quantity conventionally denoted by
$Z$ is often associated with the ``elementary'' component of a
state in the sense of Weinberg~\cite{Weinberg:1965zz}. In the
present BOEFT description, however, $Z$ has the more specific
meaning of the quarkonium-channel component, while the $X_i$
characterize the open-flavor/tetraquark channels. We therefore use the terms quarkonium component and open-flavor component
throughout, reserving a probabilistic interpretation for bound
states.

For comparison among different resonances, it is useful to map the complex components onto real normalized quantities. We therefore define the normalized composition measures
\begin{equation}
  P_0
  =
  \frac{|Z|}
       {|Z|+\sum_{j=1}^{N_{\rm open}}|X_j|},
   \label{eq:probabilities def Z}     
\end{equation}

and
\begin{equation}
  P_i
  =
  \frac{|X_i|}
       {|Z|+\sum_{j=1}^{N_{\rm open}}|X_j|},
  \qquad
  i=1,\ldots,N_{\rm open}.
  \label{eq:probabilities def X}
\end{equation}
By construction,
\begin{equation}
  P_0+\sum_{i=1}^{N_{\rm open}}P_i=1.
\end{equation}
For a bound state, these quantities coincide with the ordinary
channel probabilities. For a resonance, they provide a
convenient real normalization of the complex components, but do
not acquire a unique probabilistic interpretation. Taking the absolute values is itself part of the adopted prescription.

\subsection{\texorpdfstring{$K$}{K} matrix prescription}
 \label{subsec:K-matrix_composition}

We now construct a real-axis characterization of the channel
content from the standing-wave solutions associated with a
$K$ matrix pole. In this subsection, $M$ and $\Gamma$ denote the
mass and width parameters appearing in the Laurent expansion of
the $K$ matrix. For a broad resonance or in the presence of a
significant background, they need not coincide with the
corresponding complex $T$ matrix pole parameters.

For a real energy above threshold, the asymptotic wavefunction
in an open channel is
\begin{equation}
  \psi_{ij}(r)
  \xrightarrow{r\to\infty}
  \sqrt{\mu k}\sqrt{\frac{2}{\pi}}
  \left[
    \delta_{ij}j_{l_{Q\bar Q}}(kr)
    -
    K_{ij}(E)y_{l_{Q\bar Q}}(kr)
  \right],
  \qquad
  i,j=1,\ldots,N_{\rm open}.
  \label{eq:asymptotic behaviour real}
\end{equation}
Eq.~\eqref{eq:asymptotic behaviour real} 
applies only to the asymptotically open channels.
The closed quarkonium component vanishes at large separation but
is determined simultaneously as part of the full coupled
solution.

At a generic real energy, there is one linearly independent
standing-wave solution for each open incoming channel. At a real
$K$ matrix pole $E=M$, only one linear combination of these solutions is associated with the pole. It is selected by the
normalized pole-coupling vector $\bm g$ introduced in
Section~\ref{subsec:k-matrix}. Multiplying the solutions by $g_j$
and summing over the incoming channels gives, for an open channel
$i$,
\begin{align}
  \sum_{j=1}^{N_{\rm open}}
  \psi_{ij}(r)g_j
  \xrightarrow{r\to\infty}
  \sqrt{\mu k}\sqrt{\frac{2}{\pi}}
  \Bigg[
  g_i j_{l_{Q\bar Q}}(kr)
  \nonumber
  -
  \left(
    g_i\frac{\Gamma/2}{M-E}
    +
    \sum_{j=1}^{N_{\rm open}}
    K^{\rm bg}_{ij}(E)g_j
  \right)
  y_{l_{Q\bar Q}}(kr)
  \Bigg].
\end{align}

The pole contribution can be isolated by multiplying this
combination by $(E-M)/(\Gamma/2)$ and taking the limit
$E\to M$,
\begin{equation}
  \lim_{E\to M}
  \frac{E-M}{\Gamma/2}
  \sum_{j=1}^{N_{\rm open}}
  \psi_{ij}(r)g_j
  =
  \sqrt{\mu k}\sqrt{\frac{2}{\pi}}\,
  g_i y_{l_{Q\bar Q}}(kr),
  \qquad i=1,\ldots,N_{\rm open}.
  \label{eq:solpolelarger}
\end{equation}
The regular background drops out in this limit.

We therefore define the pole-projected standing-wave profile in
every channel, including the closed quarkonium channel, by
\begin{equation}
  \widetilde\psi_i(r)
  =
  \lim_{E\to M}
  \frac{E-M}{\Gamma/2}
  \sum_{j=1}^{N_{\rm open}}
  \psi_{ij}(r)g_j.
  \label{eq:polewfuncdef}
\end{equation}
For the open channels, this profile has the asymptotic behavior
shown above and is not square-integrable. The same linear combination of the complete coupled solutions defines
$\widetilde\psi_0(r)$ in the quarkonium channel. Since the
quarkonium channel is closed, $\widetilde\psi_0(r)$ is spatially
localized and its squared integral is finite.

The normalized $K$ matrix pole couplings $g_i^2$ characterize the
relative strengths with which the pole couples to the
open-flavor channels. The corresponding open-flavor components
will therefore be distributed according to
\begin{equation}
  X_i\propto g_i^2,
  \qquad
  i=1,\ldots,N_{\rm open}.
   \label{eq:bbbarprob}
\end{equation}

The quarkonium contribution can instead be characterized through
the localized pole-projected profile,
\begin{equation}
  Z
  \propto
  \int_0^\infty dr\,r^2
  \widetilde\psi_0(r)^2.
   \label{eq:qqbarprob}
\end{equation}

Because $\widetilde\psi_0$ inherits the normalization of a
scattering solution, the integral on the right-hand side is
dimensionful. The overall normalization is therefore part of the
prescription.

We fix it by requiring that the definition give $Z=1$ for the
limiting case of a single discrete level coupled to a continuum
whose threshold lies well below the resonance
\cite{Friedrich:2015obs,Fano:1961zz}. This condition leads to
\begin{equation}
  Z
  =
  \frac{\pi\Gamma}{2}
  \int_0^\infty dr\,r^2
  \widetilde\psi_0(r)^2.
   \label{eq:QQbar probability from K-matrix}
\end{equation}
Here $\Gamma$ is the width parameter extracted from the
$K$ matrix Laurent expansion. Equation~(\ref{eq:QQbar probability from K-matrix}) defines a
prescription-dependent quarkonium component; it does not
establish a unique probability for a resonance.

Having defined the total non-quarkonium component as $1-Z$, we
distribute it among the open channels according to the normalized
$K$ matrix pole couplings,
\begin{equation}
  X_i
  =
  (1-Z)g_i^2,
  \qquad
  i=1,\ldots,N_{\rm open}.
   \label{eq:open channel probability from K-matrix}
\end{equation}
This relation is a prescription for allocating the total
open-flavor component; it is not obtained by integrating the
non-normalizable open-channel standing-wave profiles. The
components satisfy
\begin{equation}
  Z+\sum_{i=1}^{N_{\rm open}}X_i=1.
\end{equation}
So it is tempting to interpret them as probabilities.
The $K$ matrix components $Z$ and $X_i$ are real, but they need
not lie in the interval $[0,1]$. In particular, the prescription
may yield $Z>1$, in which case the components $X_i$ defined in
Eq.~(\ref{eq:open channel probability from K-matrix}) are negative. Such values do not signal an
inconsistency of the scattering amplitude. They instead reflect
the fact that the resonance components are not probabilities.

For the real $K$ matrix solution, $\widetilde\psi_0(r)$ may be chosen real. Since the width parameter is positive, Eq.~\eqref{eq:QQbar probability from K-matrix} implies $Z\geq0$. Using the same normalized absolute-value measures defined in Eqs.~(\ref{eq:probabilities def Z}) and (\ref{eq:probabilities def X}), with the normalization condition $\sum_{i=1}^{N_{\rm open}} g_i^2=1$, in the $K$ matrix method the denominator on the right sides of Eqs.~\eqref{eq:probabilities def Z} and \eqref{eq:probabilities def X} can be expressed explicitly as
\begin{equation} 
|Z|+\sum_{j=1}^{N_{\rm open}}|X_j| =Z+|1-Z| = \begin{cases} 1, & 0\leq Z\leq1,\\[1mm] 2Z-1, & Z>1. 
\end{cases}
\end{equation}

Consequently, when $0\leq Z\leq1$, the normalized composition
measures reduce to
\begin{equation}
  P_0=Z,
  \qquad
  P_i=X_i.
\end{equation}
When $Z>1$, the absolute-value normalization maps the real
components onto positive quantities summing to one, but these
quantities remain prescription-dependent measures rather than
physical probabilities.

The complex scaling and $K$ matrix prescriptions probe related
aspects of the resonance channel content, but are not expected to
give identical numerical values, particularly for broad
resonances or poles strongly affected by a nearby threshold or
nonresonant background. Agreement on the dominant channel assignment provides a useful qualitative characterization,
whereas differences between the two prescriptions quantify part
of the intrinsic ambiguity in assigning a composition to a
resonance. Neither set of resonance components should be
interpreted as a unique compositeness.

\section{Numerical results}
\label{sec:results}

In this section, we present the pole masses, pole widths, normalized pole
couplings, and channel-composition measures of the charmoniumlike and
bottomoniumlike states between the spin--isospin averaged $S+S$ and
$S+P$ heavy-meson thresholds. The quoted widths are generated only by
the spin-averaged, nonstrange $S+S$ open-flavor channels included in
the coupled equations and should not be identified with physical total
widths. The corresponding results for the bottom-charmed sector are
given in Appendix~\ref{app:bottom-charm}.

\subsection{Resonance spectrum, pole widths, and pole couplings}
\label{subsec:spectrum}

The bound-state levels and resonance poles obtained from
Eqs.~\eqref{eq:diabatic} and \eqref{eq:diabaticl0} are labeled by the
conserved BO angular-momentum quantum number $l$ and by an excitation
number $n$. We denote them by $nl$, with $l=0,1,2,\ldots$ represented
by $S,P,D,\ldots$. The corresponding physical energy convention is
given in Eq.~\eqref{eq:quarkoniumlike-masses}.

The letters $S$, $P$, $D$, \ldots\ in this notation refer to the BO
angular momentum $l$ and should not be confused with the orbital
angular momentum of every component of the coupled wavefunction. The
quarkonium component has $l_{Q\bar Q}=l$, whereas the tetraquark/open-
flavor components have $l_{Q\bar Q}=l-1$ and $l_{Q\bar Q}=l+1$ when
allowed. Thus, for example, the BOEFT $2P$ multiplet contains a $P$-wave
quarkonium component together with $S$- and $D$-wave
tetraquark/open-flavor components. In the following, we restrict the
analysis to $l\leq 2$.

For clarity, we briefly recall the bound state multiplets that determine the radial numbering of the above-threshold poles. In the charmoniumlike
sector we obtain the bound state multiplets $1S$, $1P$, $2S$, $1D$, and
$2P$. The last of these is the shallow, tetraquark-dominated multiplet
associated with the $\chi_{c1}(3872)$. The adjoint meson mass was
calibrated so that this spin-averaged multiplet lies $96$~keV below
the spin--isospin averaged $D^{(\ast)}\bar D^{(\ast)}$ threshold, as
described in Section~\ref{subsec:parametrization}. The BOEFT label $2P$
should therefore not be identified with the conventional quark-model
$2P$ charmonium multiplet.

In the bottomoniumlike sector, we obtain the bound state multiplets $1S$,
$1P$, $2S$, $1D$, $2P$, $3S$, $2D$, $3P$, $4S$, and $4P$. The
BOEFT $4P$ multiplet is a shallow, tetraquark-dominated state below
the spin--isospin averaged $B^{(\ast)}\bar B^{(\ast)}$ threshold, associated with the bottom counterpart state of $\chi_{c1}(3872)$, $X_b$. It
should not be confused with the conventional quark-model $4P$
bottomonium multiplet. A detailed discussion of the below-threshold spectrum is given in
Refs.~\cite{TarrusCastella:2022rxb,Brambilla:2026zfw}.

Between the spin--isospin averaged $S+S$ and $S+P$ thresholds, we find
four charmoniumlike resonance multiplets and six bottomoniumlike
resonance multiplets. The $T$ matrix complex-pole parameters and the corresponding real-axis $K$ matrix parameters are collected in
Tables~\ref{Table: c c-bar masses and decay widths} and
\ref{Table: b b-bar masses and decay widths}.
Note that the bottomoniumlike resonance multiplet $5S$ has no corresponding $K$ matrix pole. It is a tetraquark-dominated state, and it should not be confused with the conventional quark model $5S$ bottomonium multiplet, as discussed in more detail in Section~\ref{subsec:probabilities results}.
The complex-scaled
eigenvalues are not listed in the tables since they 
agree with the analytically continued $T$ matrix poles over the adopted range of rotation angles.
More details about the numerical implementation of the $T$ matrix, $K$ matrix, and complex scaling methods are given in Section~\ref{subsec:numerical}.
We present the extension to the bottom-charmed sector in Appendix~\ref{app:bottom-charm}, which yields four quarkoniumlike resonances between the corresponding $S+S$ and $S+P$ thresholds, while no shallow tetraquark/open-flavor bound state is found with the present adjoint-meson calibration.

\begin{table}
    \renewcommand{\arraystretch}{1.2}
    \centering
    \begin{tabular}{cccrrccrr}
        \toprule [1.5pt]
        \multirow{2}{*}{$nl$} & \multicolumn{4}{c}{$T$ matrix} & \multicolumn{4}{c}{$K$ matrix}  \\
         \cmidrule(lr){2-5} \cmidrule(lr){6-9} 
       & $M$ (MeV) & $\Gamma$ (MeV) & $g^2_{l-1}$ & $g^2_{l+1}$ & $M$ (MeV) &  $\Gamma$ (MeV) & $g^2_{l-1}$ & $g^2_{l+1}$ \\
     \midrule [1.5pt]
        $3P$ & $3988$ & $28$  & $86\%$  & $14\%$   &  $3979$  & $50$   & $94\%$ & $6\%$  \\
        $3S$ & $4145$ & $1$  &   & $100\%$   & $4145$ & $1$  & & $100\%$  \\
        $2D$ & $4217$ & $22$  &  $4\%$ & $96\%$   & $4216$  & $20$  & $5\%$ & $95\%$ \\
        $4P$ & $4364$ &  $8$ &  $8\%$ & $92\%$  & $4363$ & $8$  & $8\%$ & $92\%$ \\
\bottomrule  [1.5pt]
    \end{tabular}
   \caption{Masses $M$ and widths $\Gamma$ of the charmoniumlike multiplets obtained from the $T$ matrix and $K$ matrix poles. The normalized pole couplings $g^2_{l-1}$ and $g^2_{l+1}$ characterize the relative residue strengths in the $D^{(\ast)}\bar{D}^{(\ast)}$ open-flavor channels with orbital angular momentum $l-1$ and $l+1$, respectively. For narrow isolated resonances, they may
   approximate branching fraction ratios within this truncated channel
   space.}
     \label{Table: c c-bar masses and decay widths}
\end{table}

\begin{table}
    \setlength{\tabcolsep}{.5em}
    \renewcommand{\arraystretch}{1.2}
    \centering
    \begin{tabular}{rrrrrrrrr}
        \toprule [1.5pt]
        \multirow{2}{*}{$nl$} & \multicolumn{4}{c}{$T$ matrix} & \multicolumn{4}{c}{$K$ matrix}  \\
         \cmidrule(lr){2-5} \cmidrule(lr){6-9} 
       & $M$ (MeV) & $\Gamma$ (MeV) & $g^2_{l-1}$ & $g^2_{l+1}$ & $M$ (MeV) &  $\Gamma$ (MeV) & $g^2_{l-1}$ & $g^2_{l+1}$ \\
     \midrule [1.5pt]
        $5S$ & $10676$ & $53$ &  & $100\%$ & N/A & N/A &  & N/A \\
        $3D$ & $10681$ &$32$   & $53\%$ & $47\%$   & $10672$   & $58$  & $76\%$ & $24\%$  \\
        $5P$ &  $10778$ & $2$  & $12\%$ & $88\%$   & $10777$ & $2$ & $14\%$ & $86\%$ \\
        $6S$ & $10858$ & $2$  &   &  $100\%$   & $10858$ & $3$  & & $100\%$ \\
     $4D$ & $10919$ & $28$  & $27\%$ & $73\%$   & $10916$ & $28$ & $33\%$ & $67\%$ \\
     $6P$ & $11007$ & $23$  & $28\%$ & $72\%$   & $11004$  & $25$ & $27\%$ & $73\%$ \\
\bottomrule  [1.5pt]
    \end{tabular}
   \caption{Masses $M$ and widths $\Gamma$ of the bottomoniumlike multiplets obtained from the $T$ matrix and $K$ matrix poles. The normalized pole couplings $g^2_{l-1}$ and $g^2_{l+1}$ characterize the relative residue strengths in the $B^{(\ast)}\bar{B}^{(\ast)}$ open-flavor channels with orbital angular momentum $l-1$ and $l+1$, respectively. For narrow isolated resonances, they may
   approximate branching fraction ratios within this truncated channel
   space.}
     \label{Table: b b-bar masses and decay widths}
\end{table}

Except for the $5S$ bottomoniumlike multiplet, the masses extracted from the $T$ matrix and $K$ matrix methods are nearly identical.
Except for the $3P$ charmoniumlike multiplet and the $5S$ and $3D$ bottomoniumlike multiplets, the widths and pole couplings extracted from the $T$ matrix and $K$ matrix methods are compatible and show that the calculated resonances couple more strongly to the tetraquark channel with the higher orbital angular momentum.
The differences between the widths extracted from the $T$ matrix and $K$ matrix in the cases of the $3P$ charmoniumlike multiplet and the $3D$ bottomoniumlike multiplets, as well as the absence of the $5S$ bottomoniumlike pole from the $K$ matrix, are expected.
In fact, in all of those cases, there is a significant interference between a broad resonance and nonresonant threshold contributions.
For the $3P$ charmoniumlike multiplet, a low-momentum $D^{(\ast)}\bar{D}^{(\ast)}$ threshold enhancement contributes to the scattering amplitude near the energy of the broad $3P$ resonance.
For the $5S$ and $3D$ bottomoniumlike multiplets, a low-momentum $B^{(\ast)}\bar{B}^{(\ast)}$ threshold enhancement contributes to the scattering amplitude near the energy of the broad $5S$ and $3D$ resonances.
In these situations, the $K$ matrix poles are expected to differ from the complex $T$ matrix poles, or the correspondence between $K$ matrix and $T$ matrix poles may be entirely broken, as discussed in Section~\ref{subsec:k-matrix}.
Therefore, we obtain an overall coherent picture from the comparison of the $T$ matrix and $K$ matrix approaches. 
Throughout this work, we identify our prediction for the masses and widths with the complex $T$ matrix pole parameters.
Except for the bottomoniumlike $5S$ multiplet, we regard the $K$ matrix pole parameters as a diagnostic for real energies.

\begin{figure*}
\includegraphics[width=10.5cm]{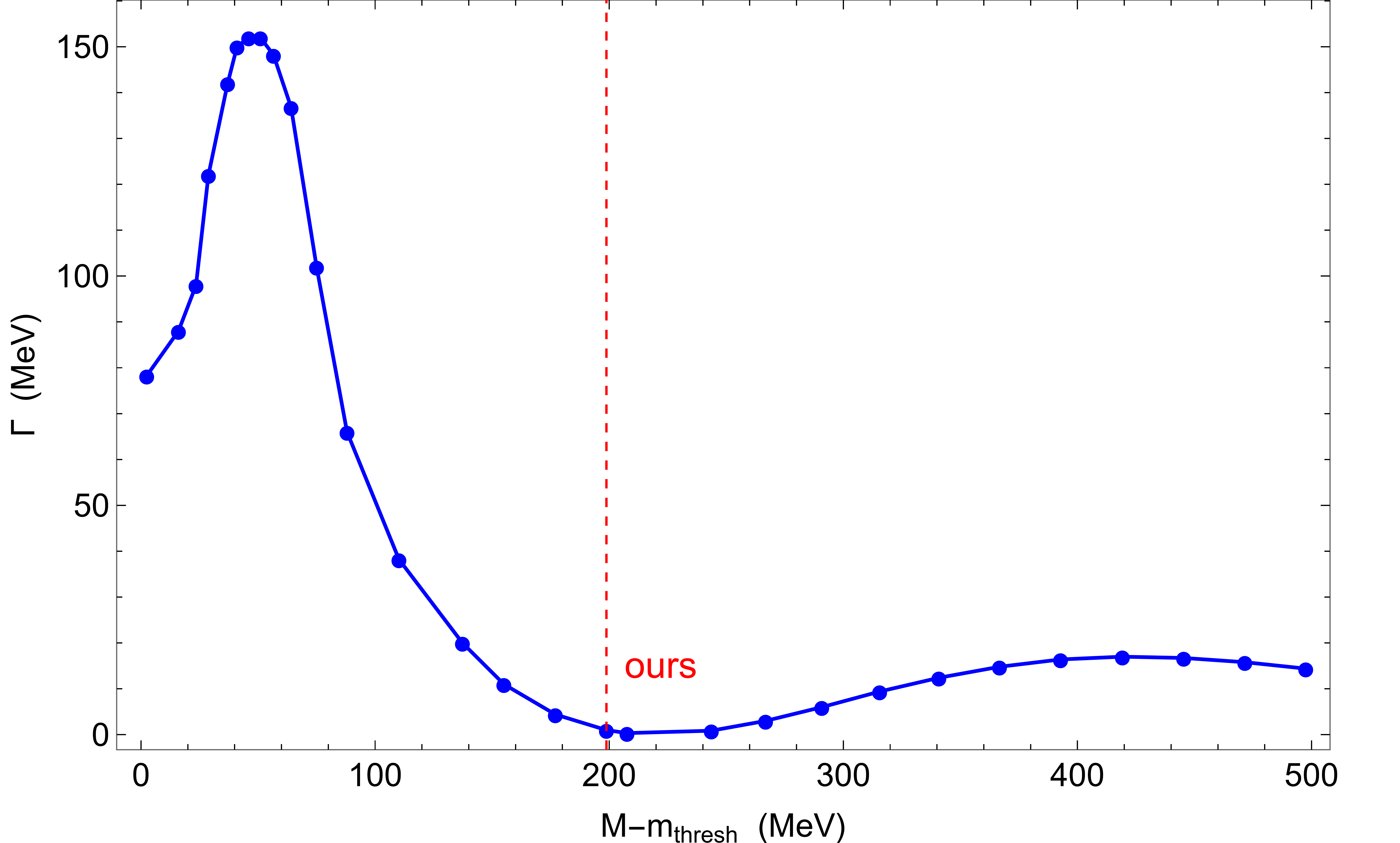}
\caption{
The pole width of the charmoniumlike $3S$ multiplet generated by the included open-charm channels as a function of its mass relative
to the spin--isospin averaged $D^{(\ast)}\bar D^{(\ast)}$ threshold.
The scan is obtained by adding a constant shift to the quarkonium
potential and is used to display the nodal dependence of the
transition amplitude. The red dashed line marks the central pole
position obtained with the potential parameters used in this work.}
\label{fig:3s nodal effects}
\end{figure*}

The charmoniumlike $3S$ pole has a particularly small width into the
included open-charm channels,
$\Gamma_{\rm pole}=1~\mathrm{MeV}$; see
Table~\ref{Table: c c-bar masses and decay widths}. Such a suppression
can arise from a node in the open-flavor transition amplitude of an
excited quarkonium state, an effect already familiar from the Cornell
coupled-channel description
\cite{Eichten:1978tg,Eichten:1979ms}. To illustrate this mechanism, we add a constant shift to the
quarkonium potential and follow the $3S$ pole as its position relative
to the spin--isospin averaged threshold is varied. This scan is used
only as a diagnostic of the energy dependence of the width. The result
is shown in Fig.~\ref{fig:3s nodal effects}. The central pole obtained
with our adopted potential lies approximately $199~\mathrm{MeV}$ above the
threshold, very close to a minimum of the width. The suppression of
the calculated $3S$ width can therefore be traced to a nearby zero of
the transition amplitude. Similar nodal suppressions are likely
responsible for the small widths of the bottomoniumlike $6S$ and
$5P$ poles, see Table~\ref{Table: b b-bar masses and decay widths}.

More generally, the open-flavor widths of the quarkonium-dominated resonances obtained at leading order tend to lie below the widths of their experimental candidates.
This is a direct consequence of the leading-order treatment: the decay amplitudes of excited quarkonia possess nodes as a function of the relative momentum of the decay products, so the predicted width depends sensitively on the position of the pole relative to these nodes.
Since the spin-averaged pole positions are themselves subject to corrections of relative order $v^2$, the widths may change
substantially once spin-dependent splittings displace the poles from their nodal minima, together with the additional decay channels omitted from the present calculation.
The leading-order partial widths are particularly sensitive to omitted
corrections whenever the pole lies near a node of the decay amplitude; they should therefore be regarded as qualitative rather than as systematic lower bounds on the total widths.

For completeness, we finally report the masses of the charmoniumlike $1D$ and $2P$ and the bottomoniumlike $4S$ and $4P$ multiplets, although relative to bound states in our leading-order calculation.
The mass predictions coincide for the $T$ matrix and complex scaling methods, analytically continued to energies below threshold, and are reported in Tables~\ref{tab:ccbelow},~\ref{tab:bbbelow}.
These states are reported here and included in the discussion of Section~\ref{sec:Comparison with experiments} due to their proximity to the open-flavor $S+S$ threshold.
Depending on the $J^{PC}$ quantum numbers, an individual member of these multiplets may remain bound, become a virtual state, or turn into a resonance once the heavy-light spin splitting of the open-flavor threshold is taken into account~\cite{Brambilla:2026zfw,Alasiri:2026vfz}.

\begin{table}[t]
\renewcommand{\arraystretch}{1.2}
\centering
\begin{tabular}{ccc}
\toprule[1.5pt]
Multiplet & $J^{PC}$ & $M$ (MeV) \\
\midrule[1.5pt]
$1D$ &
$2^{-+},(1,2,3)^{--}$ &
$3819$
\\
$2P$ &
$1^{+-},(0,1,2)^{++}$ &
$3946$
\\
\bottomrule[1.5pt]
\end{tabular}
\caption{Masses $M$ of the charmoniumlike bound state multiplets lying below the $D^{(\ast)}\bar{D}^{(\ast)}$ threshold and addressed in Section~\ref{sec:Comparison with experiments} due to their proximity to the threshold. The BOEFT $2P$ multiplet is the shallow tetraquark/open-flavor-dominated multiplet associated with the $\chi_{c1}(3872)$.
The mass predictions coincide for the $T$ matrix and complex scaling methods, analytically continued to energies below threshold.}
\label{tab:ccbelow}
\end{table}

\begin{table}[t]
\renewcommand{\arraystretch}{1.2}
\centering
\begin{tabular}{ccc}
\toprule[1.5pt]
Multiplet & $J^{PC}$ & $M$ (MeV) \\
\midrule[1.5pt]
$4S$ &
$0^{-+},1^{--}$ &
$10595$
\\
$4P$ &
$1^{+-},(0,1,2)^{++}$ &
$10626$
\\
\bottomrule[1.5pt]
\end{tabular}
\caption{
Masses $M$ of the bottomoniumlike bound state multiplets lying below the  $B^{(\ast)}\bar{B}^{(\ast)}$ threshold and addressed in Section~\ref{sec:Comparison with experiments} due to their proximity to the threshold. The BOEFT $4P$ multiplet is the shallow tetraquark/open-flavor-dominated multiplet associated with the $X_b$.
The mass predictions coincide for the $T$ matrix and complex scaling methods, analytically continued to energies below threshold.
}
\label{tab:bbbelow}
\end{table}

\subsection{Internal structure of quarkoniumlike resonances}\label{subsec:probabilities results}

The quarkonium component $Z$ and the tetraquark/open-flavor
BO-channel components $X_i$ obtained with the complex scaling and
$K$ matrix prescriptions are listed in
Tables~\ref{Table: c c-bar table of compositeness} and
\ref{Table: b b-bar table of compositeness} for the charmoniumlike
and bottomoniumlike resonances, respectively. As discussed in
Section~\ref{sec:probabilities}, the same symbols $Z$ and $X_i$ are used
for the two prescriptions to avoid proliferating notation, but their definitions are prescription dependent, and their numerical values
should not be identified with one another.

In the complex scaling prescription, the components are evaluated
from the complex-scaled Gamow state at the resonance pole and are
generally complex
\cite{Sekihara:2015gvw,Garcia-Recio:2015jsa}. 
In the $K$ matrix
prescription, the components are real but may lie outside the interval
$[0,1]$, for example with $Z>1$ and $X_i<0$. Such values are not
unphysical; they reflect the fact that resonance components are not
probabilities.

\begin{table}
    \renewcommand{\arraystretch}{1.2}
    \centering
    \begin{tabular}{crrr} 
        \toprule [1.5pt]
        \multirow{2}{*}{$nl$} & \multirow{2}{*}{Channel} & \multicolumn{2}{c}{$Z$ or $X_i$} \\
        \cmidrule(lr){3-4}
        & & \multicolumn{1}{c}{CSM}  & \multicolumn{1}{c}{$K$ matrix}  \\
        \midrule [1.5pt]
        \multirow{3}{*}{$3P$} & $c\bar{c}$ ($P$-wave)  & $0.89 + 0.26i$ & $1.47$  \\
         & $D^{(\ast)}\bar{D}^{(\ast)} \, \text{($S$-wave)}$ & $0.10 - 0.38i$ & $-0.44$  \\
         & $D^{(\ast)}\bar{D}^{(\ast)} \, \text{($D$-wave)}$ & $0.01 + 0.12i$ & $-0.03$ \\
        \hline
       \multirow{2}{*}{$3S$} & $c\bar{c}$ ($S$-wave)  &  $0.88 + 0.04i$ & $0.89$ \\
       & $D^{(\ast)}\bar{D}^{(\ast)} \, \text{($P$-wave)}$ & $0.12 - 0.04i$ & $0.11$  \\
        \hline
        \multirow{3}{*}{$2D$} & $c\bar{c}$ ($D$-wave)  & $1.03-0.01i$ & $1.03$ \\
         & $D^{(\ast)}\bar{D}^{(\ast)} \, \text{($P$-wave)}$ & $0.03+0.03i$ & $0.00$ \\
         & $D^{(\ast)}\bar{D}^{(\ast)} \, \text{($F$-wave)}$ & $-0.06-0.02i$ & $-0.03$  \\
         \hline
        \multirow{3}{*}{$4P$} & $c\bar{c}$ ($P$-wave)   & $1.00-0.05i$  & $0.99$ \\
         & $D^{(\ast)}\bar{D}^{(\ast)} \, \text{($S$-wave)}$ & $0.01+0.01i$ & $0.00$ \\
         & $D^{(\ast)}\bar{D}^{(\ast)} \, \text{($D$-wave)}$ & $-0.01+0.04i$ & $0.01$ \\
        \bottomrule [1.5pt]
    \end{tabular}
    \caption{Quarkonium component $Z$ and tetraquark/open-flavor
BO-channel components $X_i$ of the charmoniumlike resonances.
The complex scaling method
(CSM) entries are generally complex, whereas the $K$ matrix
entries are real but may lie outside the interval $[0,1]$.
The two columns correspond to different prescriptions and should
not be identified numerically.}
     \label{Table: c c-bar table of compositeness}
\end{table}

\begin{table}
    \renewcommand{\arraystretch}{1.2}
    \centering
    \begin{tabular}{rrrr} 
        \toprule [1.5pt]
        \multirow{2}{*}{$nl$} & \multirow{2}{*}{Channel} & \multicolumn{2}{c}{$Z$ or $X_i$} \\
        \cmidrule(lr){3-4}
        & & \multicolumn{1}{c}{CSM}  & \multicolumn{1}{c}{$K$ matrix}  \\
        \midrule [1.5pt]
        \multirow{2}{*}{$5S$} & $b\bar{b}$ ($S$-wave)  & $-0.04 + 0.37i$ & N/A \\
         & $B^{(\ast)}\bar{B}^{(\ast)} \, \text{($P$-wave)}$ & $1.04 - 0.37i$  & N/A \\
        \hline
        \multirow{3}{*}{$3D$} & $b\bar{b}$ ($D$-wave) & $0.75 + 0.19i$ & $1.46$  \\  
         & $B^{(\ast)}\bar{B}^{(\ast)}\, \text{($P$-wave)}$ & $0.41 - 0.40i$  & $-0.35$ \\
         & $B^{(\ast)}\bar{B}^{(\ast)}\, \text{($F$-wave)}$ & $-0.16 + 0.21i$  & $-0.11$  \\ 
         \hline
        \multirow{3}{*}{$5P$} & $b\bar{b}$ ($P$-wave) & $0.78 - 0.06i$ & $0.78$ \\ 
         & $B^{(\ast)}\bar{B}^{(\ast)} \, \text{($S$-wave)}$ & $0.09 - 0.03i$ & $0.03$ \\
         & $B^{(\ast)}\bar{B}^{(\ast)} \, \text{($D$-wave)}$ & $0.13 + 0.09i$ & $0.19$ \\ 
        \hline
        \multirow{2}{*}{$6S$} & $b\bar{b}$ ($S$-wave)   & $0.86 - 0.09i$ & $0.82$ \\
         & $B^{(\ast)}\bar{B}^{(\ast)} \, \text{($P$-wave)}$ & $0.14 + 0.09i$  & $0.18$ \\
        \hline
        \multirow{3}{*}{$4D$} & $b\bar{b}$ ($D$-wave) & $1.21 - 0.03i$ & $1.15$ \\ 
         & $B^{(\ast)}\bar{B}^{(\ast)} \, \text{($P$-wave)}$ & $-0.07 + 0.08i$ & $-0.05$  \\ 
         & $B^{(\ast)}\bar{B}^{(\ast)}\, \text{($F$-wave)}$ & $-0.14 - 0.05i$ & $-0.10$  \\
         \hline
        \multirow{3}{*}{$6P$} & $b\bar{b}$ ($P$-wave) & $1.16 - 0.06i$  & $1.12$ \\ 
         & $B^{(\ast)}\bar{B}^{(\ast)}\, \text{($S$-wave)}$ & $-0.04 + 0.04i$  & $-0.03$ \\ 
         & $B^{(\ast)}\bar{B}^{(\ast)} \, \text{($D$-wave)}$ & $-0.12 + 0.02i$ & $-0.09$ \\
        \bottomrule [1.5pt]
    \end{tabular}
    \caption{Quarkonium component $Z$ and tetraquark/open-flavor
BO-channel components $X_i$ of the bottomoniumlike resonances.
The complex scaling method
(CSM) entries are generally complex, whereas the $K$ matrix
entries are real but may lie outside the interval $[0,1]$.
The two columns correspond to different prescriptions and should
not be identified numerically.}
     \label{Table: b b-bar table of compositeness}
\end{table}

For a real and normalized representation of the channel content, we
construct the quantities $P_i$ defined in
Eqs.~\eqref{eq:probabilities def Z} and
\eqref{eq:probabilities def X}. They are obtained by taking the
absolute values of the prescription-dependent components and
normalizing their sum to unity. The results are listed in
Tables~\ref{Table: table of c c-bar probabilites} and
\ref{Table: table of b b-bar probabilites} for the charmoniumlike and
bottomoniumlike resonances, respectively.

The $P_i$ provide convenient normalized composition measures for
identifying the dominant channel. For a resonance, however, they are
not probabilities: the absolute-value normalization discards the
complex phases of the complex scaling method components or the signs of the $K$ matrix
components and is itself part of the adopted prescription.

\begin{table}
    \renewcommand{\arraystretch}{1.2}
    \centering
    \begin{tabular}{crrr} 
        \toprule [1.5pt]
        \multirow{2}{*}{$nl$} & \multirow{2}{*}{Channel} & \multicolumn{2}{c}{$P_i$} \\
        \cmidrule(lr){3-4}
        & & \multicolumn{1}{c}{CSM}  & \multicolumn{1}{c}{$K$ matrix}  \\
        \midrule  [1.5pt]
         \multirow{3}{*}{$3P$} & $c\bar{c}$ ($P$-wave)  & $65\%$  &  $76\%$ \\
         & $D^{(\ast)}\bar{D}^{(\ast)} \, \text{($S$-wave)}$  & $27\%$ & $23\%$   \\
         & $D^{(\ast)}\bar{D}^{(\ast)} \, \text{($D$-wave)}$  & $8\%$ & $1\%$   \\
        \hline
        \multirow{2}{*}{$3S$} & $c\bar{c}$ ($S$-wave)  & $87\%$ &  $89\%$  \\
         & $D^{(\ast)}\bar{D}^{(\ast)} \, \text{(P-wave)}$  & $13\%$ & $11\%$  \\
        \hline
        \multirow{3}{*}{$2D$} & $c\bar{c}$ ($D$-wave)  & $90\%$  &  $97\%$  \\
         & $D^{(\ast)}\bar{D}^{(\ast)} \, \text{(P-wave)}$  & $4\%$
        & $0\%$   \\
         & $D^{(\ast)}\bar{D}^{(\ast)} \, \text{(F-wave)}$  & $6\%$ & $3\%$  \\
        \hline
         \multirow{3}{*}{$4P$} & $c\bar{c}$ ($P$-wave)  & $95\%$ &  $99\%$  \\
         & $D^{(\ast)}\bar{D}^{(\ast)} \, \text{($S$-wave)}$  & $2\%$ & $0\%$  \\
         & $D^{(\ast)}\bar{D}^{(\ast)} \text{($D$-wave)}$ & $3\%$ & $1\%$  \\
        \bottomrule [1.5pt]
    \end{tabular}
    \caption{Normalized composition measures $P_i$ of the predicted
charmoniumlike resonances obtained with the complex scaling method
(CSM) and the $K$ matrix prescription. For resonances, the $P_i$ are
prescription-dependent measures of the channel content and not
channel probabilities.}
     \label{Table: table of c c-bar probabilites}
\end{table}

\begin{table}
    \renewcommand{\arraystretch}{1.2}
    \centering
    \begin{tabular}{rrrr} 
        \toprule [1.5pt]
        \multirow{2}{*}{$nl$} & \multirow{2}{*}{Channel} & \multicolumn{2}{c}{$P_i$} \\
        \cmidrule(lr){3-4}
        & & \multicolumn{1}{c}{CSM}  & \multicolumn{1}{c}{$K$ matrix}  \\
        \midrule  [1.5pt]
        \multirow{2}{*}{$5S$} & $b\bar{b}$ ($S$-wave)  & $25\%$ &  N/A  \\
         & $B^{(\ast)}\bar{B}^{(\ast)} \, \text{($P$-wave)}$  & $75\%$ & N/A  \\
        \hline
        \multirow{3}{*}{$3D$} & $b\bar{b}$ ($D$-wave)  & $48\%$ &   $76\%$  \\  
         & $B^{(\ast)}\bar{B}^{(\ast)}\, \text{($P$-wave)}$ & $36\%$  & $18\%$ \\
         & $B^{(\ast)}\bar{B}^{(\ast)}\, \text{($F$-wave)}$  & $\$16\%$ & $6\%$  \\ 
        \hline
        \multirow{3}{*}{$5P$} & $b\bar{b}$ ($P$-wave) & $76\%$ &  $78\%$  \\ 
         & $B^{(\ast)}\bar{B}^{(\ast)} \, \text{($S$-wave)}$  & $9\%$ & $3\%$  \\
         & $B^{(\ast)}\bar{B}^{(\ast)} \, \text{($D$-wave)}$ & $15\%$ & $19\%$ \\ 
        \hline
        \multirow{2}{*}{$6S$} & $b\bar{b}$ ($S$-wave)  & $84\%$ &  $82\%$  \\
         & $B^{(\ast)}\bar{B}^{(\ast)} \, \text{($P$-wave)}$  & $16\%$ & $18\%$  \\
        \hline
        \multirow{3}{*}{$4D$} & $b\bar{b}$ ($D$-wave)  & $82\%$ & $88\%$  \\ 
         & $B^{(\ast)}\bar{B}^{(\ast)} \, \text{($P$-wave)}$  & $8\%$  &  $4\%$   \\ 
         & $B^{(\ast)}\bar{B}^{(\ast)}\, \text{($F$-wave)}$   & $10\%$ & $8\%$  \\
         \hline
        \multirow{3}{*}{$6P$} & $b\bar{b}$ ($P$-wave) & $86\%$ & $91\%$  \\ 
         & $B^{(\ast)}\bar{B}^{(\ast)}\, \text{($S$-wave)}$  & $5\%$ & $2\%$   \\ 
         & $B^{(\ast)}\bar{B}^{(\ast)} \, \text{($D$-wave)}$  & $9\%$ & $7\%$  \\
         \bottomrule [1.5pt]
    \end{tabular}
    \caption{Normalized composition measures $P_i$ of the predicted
bottomoniumlike resonances obtained with the complex scaling method
(CSM) and the $K$ matrix prescription. For resonances, the $P_i$ are
prescription-dependent measures of the channel content and not
channel probabilities.}
     \label{Table: table of b b-bar probabilites}
\end{table}

Both prescriptions give the same dominant-channel classification for
all the resonances listed in
Tables~\ref{Table: table of c c-bar probabilites} and
\ref{Table: table of b b-bar probabilites}, with the exception of the $5S$ bottomoniumlike multiplet for which the $K$-matrix prescription is not applicable since there is no corresponding $K$-matrix pole.
Most of the poles are dominated by their quarkonium component. In the charmoniumlike sector, the dominant quarkonium configurations of the BOEFT $3S$, $2D$, and $4P$ resonance multiplets correspond to the conventional $3S$, $2D$, and $3P$ charmonium assignments, respectively. In the bottomoniumlike sector,
the dominant quarkonium configurations of the BOEFT $5P$, $6S$, $4D$, and
$6P$ resonance multiplets correspond to the conventional $4P$, $5S$, $4D$, and $5P$ bottomonium assignments, respectively.
Thus, away from the included spin--isospin averaged $S+S$ threshold,
the resonance poles retain predominantly the configurations supported
by the confining quarkonium potential, with comparatively small
corrections from the included open-flavor channels.

The largest open-flavor contributions occur for the charmoniumlike
$3P$ and bottomoniumlike $5S$ and $3D$ resonances, whose pole positions are close to the corresponding spin-averaged $S+S$ threshold. The
BOEFT $3P$ charmoniumlike resonance is dominated by the conventional
$2P$ charmonium configuration, but it contains a sizable
$D^{(\ast)}\bar D^{(\ast)}$ component, particularly in the $S$ wave.
The $5S$ bottomoniumlike state is dominated by the $P$-wave $B^{(\ast)}\bar B^{(\ast)}$ configuration, but it contains a sizable $S$-wave bottomonium component.
The largest single component of the BOEFT $3D$
bottomoniumlike resonance is the conventional $3D$ bottomonium
configuration, but its open $B^{(\ast)}\bar B^{(\ast)}$ contribution is substantial.

As commented earlier, the $5S$ bottomoniulike resonance has no corresponding pole in the $K$-matrix, so its composition cannot be calculated using the $K$-matrix prescription.
For the $3D$ bottomoniumlike resonance, the complex scaling method
and $K$ matrix prescriptions give noticeably different numerical composition
measures, with quarkonium contributions of $48\%$ and $76\%$,
respectively. Both prescriptions nevertheless identify the
$b\bar b$ $D$-wave channel as the largest individual component. These  differences reflect the intrinsic distinction between the
T matrix and K matrix pole spectra, which is enhanced by the
nearby nonresonant threshold background, as
discussed in Section~\ref{subsec:spectrum}. It should not be interpreted as a contradiction between the different methods used in this analysis.

The values in Tables~\ref{Table: c c-bar table of compositeness}--%
\ref{Table: table of b b-bar probabilites} are central results for
the potential parametrization adopted in
Section~\ref{subsec:parametrization}. They do not include the model
uncertainty associated with the presently unknown $V_{\Pi_g}$
potential. Since the difference between $V_{\Sigma_g^{+\prime}}$ and
$V_{\Pi_g}$ controls the mixing between the two tetraquark partial
waves, this uncertainty can affect their relative composition, the
normalized pole couplings, and the widths, especially for resonances
with sizable open-flavor components.

These results provide a robust qualitative characterization of the
resonance spectrum: most of the higher poles correspond to predominantly quarkonium states, while the resonances closest to threshold contain the largest open-flavor components. The differences between the complex scaling method and $K$ matrix results quantify part of the prescription dependence inherent in assigning a channel composition to a resonance. Neither the $Z$, $X_i$, nor $P_i$ values should be interpreted as a unique compositeness. Except for the $5S$ bottomoniumlike resonance, they provide complementary, prescription-dependent diagnostics of the dominant channel content.

The $5S$ bottomoniumlike resonance is qualitatively different from the rest of the calculated charmoniumlike and bottomoniumlike states.
Its mass, width, and composition identify it as a $l=0$ multiplet of tetraquark-dominated states with $J^{PC}=0^{-+}$ and $1^{--}$.
We label the $1^{--}$ member of this BOEFT multiplet as $Y_b$.
Its mass lies $50$~MeV above the $B^{(\ast)}\bar{B}^{(\ast)}$ threshold, and its threshold excess is comparable to its width  of $53$~MeV.
It should also be noted that the Schr\"odinger equation for $l=0$ involves a tetraquark channel in orbital $P$-wave.
To characterize its spatial structure, we compute expectation value of the heavy-quark separation squared from the $c$ product of the complex-scaled wavefunction, 
\begin{eqnarray}
        \langle r^2\rangle=\frac{\sum_{i=0}^{N_{\rm open}}\int_0^\infty \mathrm{d}r\,z_\theta'(r)\,z_\theta(r)^4\,\psi_i^\theta(r)^2}{\sum_{i=0}^{N_{\rm open}}\int_0^\infty \mathrm{d}r\,z_\theta'(r)\,z_\theta(r)^2\,\psi_i^\theta(r)^2}.
\end{eqnarray}
For the $Y_b$ state, this gives $\sqrt{\langle r^2\rangle} = (0.8+0.1i)~\mathrm{fm}$.
Although the real part of $\sqrt{\langle r^2\rangle}$ should not be identified with the ordinary root-mean-square heavy-quark separation, it points to a hadronic length scale of order $1/\Lambda_{QCD}$ rather than a shallow bound state with molecular characteristics.       
These features make the $Y_b$ markedly different from the other tetraquarks $X_c$ and $X_b$, which are shallow bound states in the $S$-wave channel.
The composition of this state, consisting of $25\%$ quarkonium and $75\%$ tetraquark contributions, is almost complementary to that of the $4S$ bottomonium-like bound state, with $78\%$ quarkonium and $22\%$ tetraquark probability. The coexistence of the predominantly bottomonium $4S$ bound state and the open-flavor-dominated $5S$ resonance multiplets motivates an investigation of their dynamical relation. The near-complementarity of the displayed composition measures does not, by itself, establish an isolated two-level mixing mechanism.

Finally, in Tables~\ref{Table: table of c c-bar probabilites below},~\ref{Table: table of b b-bar probabilites below} we list the channel probabilities relative to the $1D$, $2P$ charmoniumlike  multiplets and $4S$, $4P$ bottomoniumlike multiplets lying below the open-flavor $S + S$ threshold.
In this case, the channel probabilities are obtained by integrating the squared modulus of the normalized wavefunction components and admit an ordinary probability interpretation.
The probabilities obtained with the $T$ matrix and complex scaling method agree within numerical precision.

\begin{table}
    \renewcommand{\arraystretch}{1.2}
    \centering
    \begin{tabular}{crr} 
        \toprule [1.5pt]
        $nl$ & Channel & $P_i$ \\
        \midrule  [1.5pt]
         \multirow{3}{*}{$1D$} & $c\bar{c}$ ($D$-wave)  & $98\%$  \\
         & $D^{(\ast)}\bar{D}^{(\ast)} \, \text{($P$-wave)}$  & $1\%$\\
         & $D^{(\ast)}\bar{D}^{(\ast)} \, \text{($F$-wave)}$  & $1\%$\\
        \hline
         \multirow{3}{*}{$2P$} & $c\bar{c}$ ($P$-wave)  & $3\%$\\
         & $D^{(\ast)}\bar{D}^{(\ast)} \, \text{($S$-wave)}$  & $96\%$\\
         & $D^{(\ast)}\bar{D}^{(\ast)} \text{($D$-wave)}$ & $1\%$\\
        \bottomrule [1.5pt]
    \end{tabular}
    \caption{Channel probabilities relative to the charmoniumlike bound state multiplets listed in Table~\ref{tab:ccbelow}. These are obtained from normalized bound-state wavefunctions
    and therefore have an ordinary probability interpretation. The normalized channel probabilities obtained with the $T$ matrix and complex scaling methods agree within the numerical precision of the calculation.}
     \label{Table: table of c c-bar probabilites below}
\end{table}

\begin{table}
    \renewcommand{\arraystretch}{1.2}
    \centering
    \begin{tabular}{crrr} 
        \toprule [1.5pt]
         $nl$ & Channel & $P_i$ \\
        \midrule  [1.5pt]
        \multirow{3}{*}{$4S$} & $b\bar{b}$ ($S$-wave)  & $78\%$\\  
         & $B^{(\ast)}\bar{B}^{(\ast)}\, \text{($P$-wave)}$ & $22\%$\\ 
        \hline
        \multirow{3}{*}{$4P$} & $b\bar{b}$ ($P$-wave) & $1\%$  \\ 
         & $B^{(\ast)}\bar{B}^{(\ast)}\, \text{($S$-wave)}$  & $98\%$\\ 
         & $B^{(\ast)}\bar{B}^{(\ast)} \, \text{($D$-wave)}$  & $1\%$\\
         \bottomrule [1.5pt]
    \end{tabular}
    \caption{Channel probabilities relative to the bottomoniumlike bound state multiplets listed in Table~\ref{tab:bbbelow}. These are obtained from normalized bound-state wavefunctions
    and therefore have an ordinary probability interpretation. The      normalized channel probabilities obtained with the $T$ matrix and complex scaling methods agree within the numerical precision of the calculation.}
     \label{Table: table of b b-bar probabilites below}
\end{table}

\subsection{Phase shifts}
\label{sec:shifts}

The phase shifts and the inelasticity parameter are defined through
the $S$-matrix parametrization in Eq.~\eqref{eq:S-matrix} for
$l>0$ and Eq.~\eqref{eq:S-matrix0} for $l=0$. Starting from the
$T$ matrix, the $S$ matrix is obtained using
Eq.~\eqref{eq: S and T relation}. Equivalently, the real-axis
$K$ matrix gives
\begin{equation}
S(E)
=
\bigl[I+iK(E)\bigr]
\bigl[I-iK(E)\bigr]^{-1}.
\label{eq:S-from-K}
\end{equation}
The phase shifts and the inelasticity parameter are then extracted
by comparison with Eqs.~\eqref{eq:S-matrix} and
\eqref{eq:S-matrix0}. Since the $T$ and $K$ matrices are related
algebraically on the physical energy axis, they describe the same
real-axis $S$ matrix. Their numerical agreement, therefore, provides
a consistency check rather than two independent determinations of
the phase shifts.

\begin{figure*}
\centering
\includegraphics[width=\textwidth]{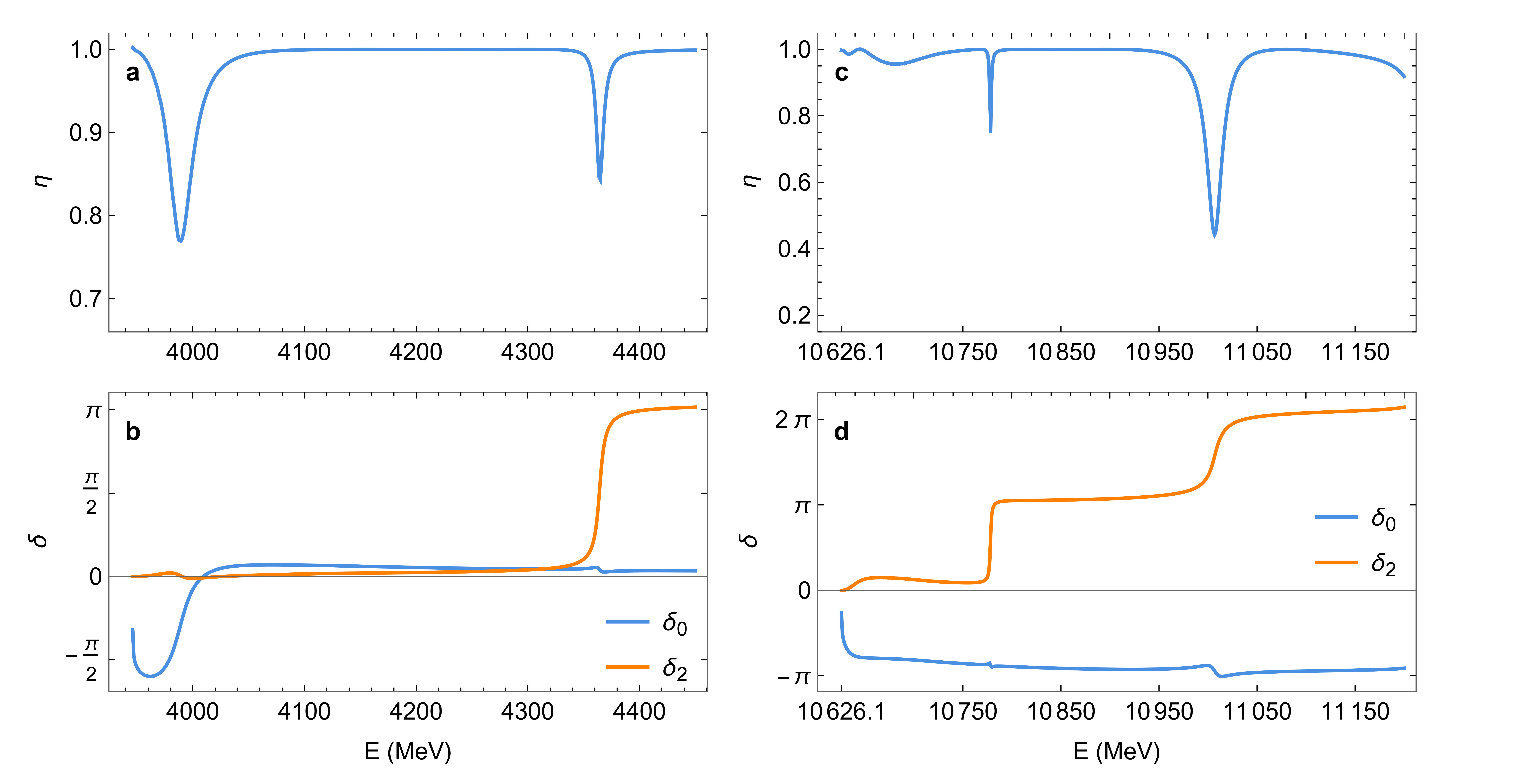}
\caption{\justifying{
Inelasticity parameter $\eta$ (top panels) and
phase shifts $\delta_0$ and $\delta_2$ (bottom panels) in the
$l=1$ coupled-channel sector. The left panels (\text{a-b})
show
$D^{(\ast)}\bar D^{(\ast)}$ scattering and the right panels     (\text{c-d})     show
$B^{(\ast)}\bar B^{(\ast)}$ scattering. The two open channels have
heavy-pair orbital angular momenta $l_{Q\bar Q}=0$ and
$l_{Q\bar Q}=2$. Departures of $\eta$ from unity quantify
transitions between these two included channels. The phase shifts
are displayed using continuous branches and are defined modulo
$\pi$.}}
\label{fig:inleasticity and phase shift for P wave heavy quarkonium}
\end{figure*}

We first focus on the $l=1$ sector, for which the two asymptotically open
tetraquark/open-flavor channels have heavy-pair orbital angular
momenta $l_{Q\bar Q}=0$ and $l_{Q\bar Q}=2$. They will be referred
to as the $S$- and $D$-wave channels, respectively. This sector
contains the BOEFT multiplets corresponding to the shallow $2P$ bound state and the $3P$, $4P$ resonance multiplets in the charmoniumlike sector, and to the shallow $4P$ bound state and the $5P$, and $6P$ resonance multiplets in the bottomoniumlike sector, discussed below.
Fig.~\ref{fig:inleasticity and phase shift for P wave heavy quarkonium}
shows the phase shifts $\delta_0$ and $\delta_2$ and the inelasticity
parameter $\eta$ for the coupled
$D^{(\ast)}\bar D^{(\ast)}$ and
$B^{(\ast)}\bar B^{(\ast)}$ channels. Within the two-channel space included in the calculation, the full $S$ matrix is unitary.
Consequently, departures of $\eta$ from unity measure transitions
between the $S$- and $D$-wave channels; they do not represent loss
of probability into channels omitted from the present $S$ matrix.

Scattering between open-flavor meson-antimeson pairs $M\bar{M}$ in the $S$-wave channel has attracted considerable attention due to its relevance to tetraquark mesons. We plot in Fig.~\ref{fig:inleasticity and phase shift for P wave heavy quarkonium} our values of the $S$-wave phase shift $\delta_0$, the $D$-wave phase shift $\delta_2$, and the inelasticity $\eta$ (corresponding to the coupling between $S$-wave and $D$-wave) for both the $D^{(\ast)}\bar{D}^{(\ast)}$ and $B^{(\ast)}\bar{B}^{(\ast)}$ channels, which are relevant to charmoniumlike and bottomoniumlike mesons, respectively.

As shown in Fig.~\ref{fig:inleasticity and phase shift for P wave heavy quarkonium} (\text{a})-(\text{b}), the $S$-wave and $D$-wave $D^{(\ast)}\bar{D}^{(\ast)}$ phase shifts change rapidly near $3980~\mathrm{MeV}$ and $4360~\mathrm{MeV}$, respectively. These features provide clear signals of the predicted $3P$ and $4P$ charmoniumlike states; see Table~\ref{Table: c c-bar masses and decay widths}.
The corresponding inelasticity $\eta$ dips at those energies, indicating the coupled-channel effects of the $S$- and $D$-wave $D^{(\ast)}\bar{D}^{(\ast)}$ channels through coupling to the $3P$ and $4P$ intermediate states.
The $+\pi$ jump of the $D$-wave $D^{(\ast)}\bar{D}^{(\ast)}$ phase shift suggests that the $4P$ state couples
preferentially to the $D$-wave $D^{(\ast)}\bar{D}^{(\ast)}$ among the included open channels, which is in good agreement with its larger pole coupling to this channel; see Table~\ref{Table: c c-bar masses and decay widths}.
The $S$-wave $D^{(\ast)}\bar{D}^{(\ast)}$ phase shift exhibits a strong variation near the mass of the $3P$ state.
However, the shape of the phase shift does not correspond to the typical $+\pi$ jump of a narrow isolated resonance.
This deviation is attributed to the presence of the nearby bound state $2P$, which is dominated by the $S$-wave tetraquark channel~\cite{Brambilla:2026zfw}.
The presence of the $2P$ bound state introduces a negative ``background'' contribution to the phase shift at large momenta according to Levinson's theorem~\cite{Levinson:407337}
that associates one net unit
of phase accumulation with this bound state.
At low momenta, the existence of such a shallow bound state is associated with a large scattering length $a_0$ in $S$-wave, which in turn produces a negative phase shift around threshold through the relation $\delta_0 \underset{k\to0}{\sim}  -a_0 k$ \cite{Peng:2024yzn,Contessi:2024vae}.
The overall behavior of the $S$-wave $D^{(\ast)}\bar{D}^{(\ast)}$ phase shift remains fully consistent with the larger pole coupling and the normalized composition measure of the $S$-wave $D^{(\ast)}\bar{D}^{(\ast)}$ channels found for the $3P$ state; see Tables~\ref{Table: c c-bar masses and decay widths} and \ref{Table: table of c c-bar probabilites}.

The $S$-wave and $D$-wave $B^{(\ast)}\bar{B}^{(\ast)}$ phase shifts and inelasticity are illustrated in Fig.~\ref{fig:inleasticity and phase shift for P wave heavy quarkonium}(\text{c})--(\text{d}).
The $D$-wave phase shift exhibits the characteristic $+\pi$-jump behavior of a resonance for energies near the masses of the $5P$ and $6P$ states.
The $S$-wave phase shift shows a sharp  drop at low momenta, then it decreases quite smoothly towards $-\pi$ at higher momenta.
This $-\pi$ shift is due to the $4P$ bound state below threshold, which is dominated by the $S$-wave tetraquark component.
The inelasticity dips at energies near the masses of the $5P$ and $6P$ states, indicating the coupled-channel effects of the $S$- and $D$-wave $B^{(\ast)}\bar{B}^{(\ast)}$ channels through coupling to these intermediate states.

\begin{figure*}
\centering
\includegraphics[width=\textwidth]{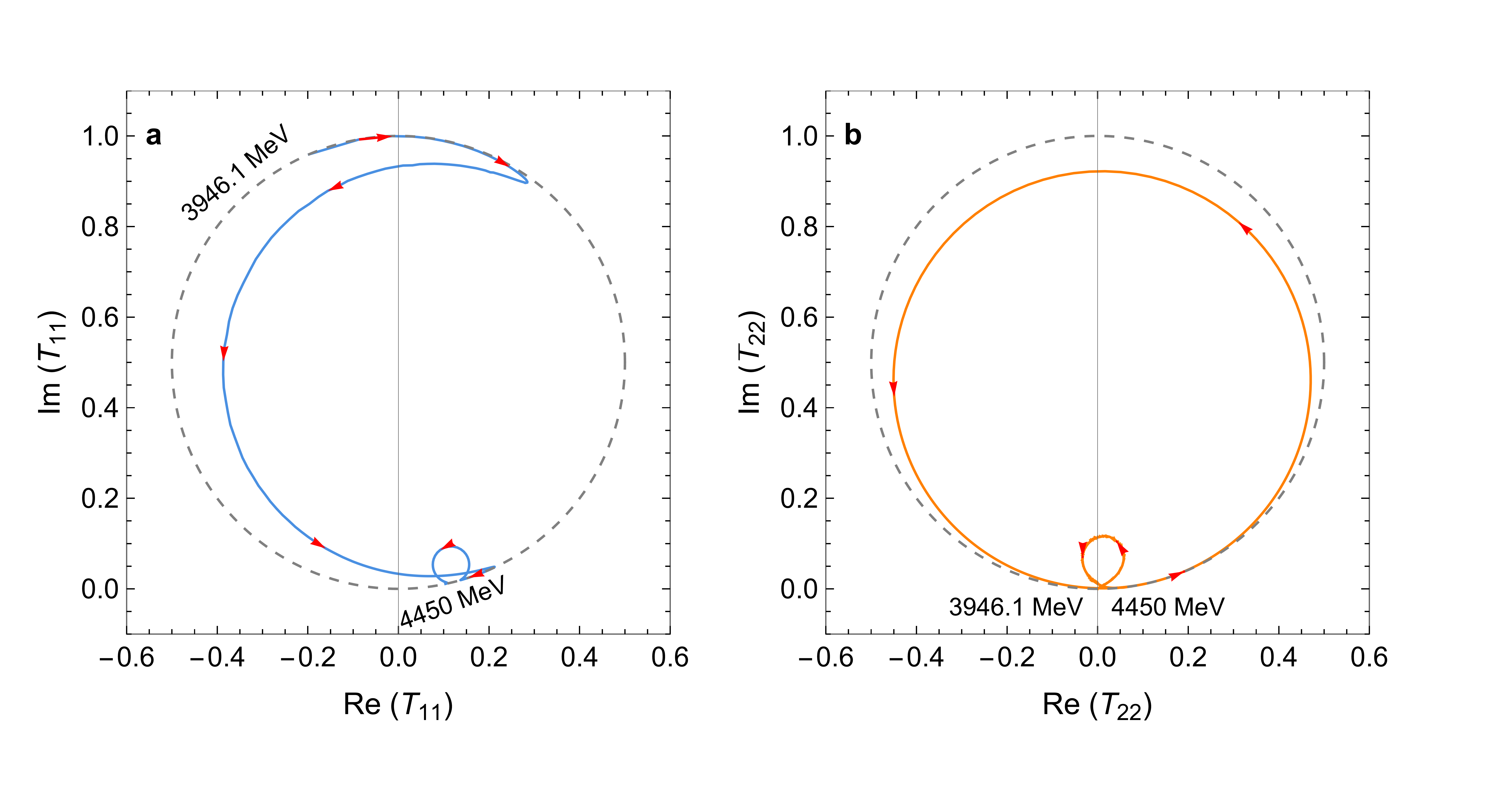}
\caption{\justifying{
Argand trajectories of the diagonal $T$ matrix
elements for the coupled $S$- and $D$-wave
$D^{(\ast)}\bar D^{(\ast)}$ channels. Panel~(a) shows the
$S$-wave amplitude $T_{11}$ and panel~(b) shows the $D$-wave
amplitude $T_{22}$. For the normalization
$S=I+2iT$, the dashed circle is the elastic-unitarity circle
$\left|T_{ii}-i/2\right|=1/2$. A diagonal amplitude may lie inside
this circle because of transitions to the other included partial
wave. The red points mark the endpoints of the energy scan at $E = 3946.5~\mathrm{MeV}$ and $E = 4450~\mathrm{MeV}$, respectively, while the orange points correspond to the real parts of the resonance energies of the $3P$ and $4P$ states.}}
\label{fig:argand diagram}
\end{figure*}

The Argand trajectories for $S$-wave and $D$-wave $D^{(\ast)}\bar{D}^{(\ast)}$ amplitudes are displayed in
Fig.~\ref{fig:argand diagram}. At very low momenta, the $S$-wave
amplitude in panel~(a) moves clockwise from threshold. This motion
is generated by the shallow $2P$ bound-state pole and corresponds
to the negative near-threshold phase shift in the continuous branch
used in Fig.~\ref{fig:inleasticity and phase shift for P wave heavy quarkonium}.
As the energy approaches the broader $3P$ pole, the trajectory
reverses direction and undergoes a counterclockwise motion inside
the elastic-unitarity circle. At higher energies, a further
counterclockwise loop is generated near the $4P$ pole.

The $D$-wave amplitude in panel~(b) instead undergoes
counterclockwise motions near both the $3P$ and $4P$ resonance
regions. The fact that the diagonal trajectories enter the
elastic-unitarity circle reflects the coupling between the $S$-
and $D$-wave channels.

The coexistence of clockwise and counterclockwise motions in the
$S$-wave amplitude, first highlighted in
Ref.~\cite{Bruschini:2022bsh}, provides a direct real-axis
visualization of the interference between the shallow,
tetraquark/open-flavor-dominated $2P$ bound state and the broader,
quarkonium-dominated $3P$ resonance. The two structures are
therefore not generated by separate dynamical descriptions, but
emerge from the same coupled BOEFT equations.

\begin{figure*}
\centering
\includegraphics[width=\textwidth]{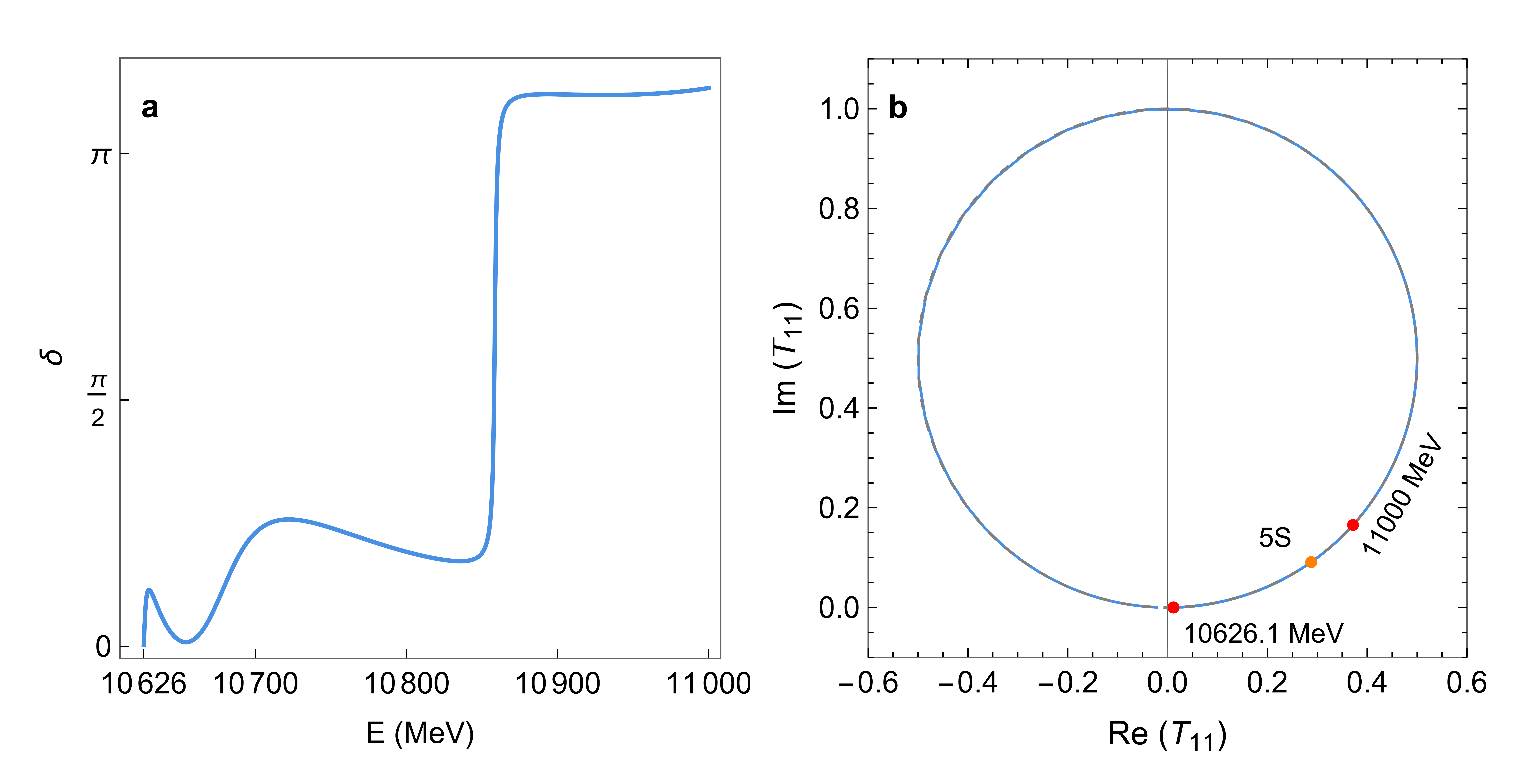}
\caption{\justifying{
Phase shift $\delta_1$ (left panel) and Argand diagram (right panel) in the $l=0$ bottomoniumlike sector, representing $P$-wave $B^{(\ast)}\bar{B}^{(\ast)}$ scattering with heavy-pair orbital angular momenta $l_{Q\bar Q}=1$. The phase shift on panel (a) is displayed using continuous branches and is defined modulo
$\pi$.
The dashed circle on panel (b) is the elastic-unitarity circle $\left|T_{ii}-i/2\right|=1/2$. Since there is only one elastic $P$-wave open channel in this case, the $T$ matrix elements are located on the unitary circle. The red points mark the endpoints of the energy scan at $E = 10626.1~\mathrm{MeV}$ and $E = 11000~\mathrm{MeV}$, respectively, while the orange point corresponds to the real part of the resonance energy of the $5S$ state.
}}
\label{fig:phase shift and Argand diagram for S wave heavy bottomonium}
\end{figure*}

Next, we examine the $l=0$ bottomoniumlike sector, for which the single asymptotically open tetraquark/open-flavor channel has heavy-pair orbital angular momentum $l_{Q\bar{Q}}=1$.
We will refer to it as the $P$-wave channel.
This sector contains the BOEFT multiplets corresponding to the conventional $4S$ bound state relatively near below threshold, as well as the tetraquark-dominated $5S$ resonance and the BOEFT $6S$ resonance, associated with conventional $5S$ bottomonium, above threshold.
Fig.~\ref{fig:phase shift and Argand diagram for S wave heavy bottomonium} shows the phase shift $\delta_1$ and the Argand trajectory for $P$-wave $B^{(\ast)}\bar{B}^{(\ast)}$ scattering.
The $P$-wave phase shift displays a broad excursion in the
$Y_b$ region without crossing $\pi/2$ modulo $\pi$,
consistently with the absence of a real-axis $K$-matrix pole.
The curve describes the full coupled-channel amplitude,
including the effects of the bound and resonant poles and
the nonresonant background. Their separate contributions
have not been isolated here. The higher BOEFT $6S$ pole
produces a rapid phase increase characteristic of a
relatively narrow resonance.

\subsection{Effective range and compositeness}

We next examine the low-energy $S$-wave
$D^{(\ast)}\bar D^{(\ast)}$ amplitude associated with the shallow
$2P$ bound state. For momenta sufficiently small compared with the
inverse range of the interaction, the effective range expansion is
\begin{equation}
k\cot\delta_0(k)
=
-\frac{1}{a_0}
+\frac{r_0}{2}k^2
+\mathcal{O}(k^4),
\label{eq:eff_range}
\end{equation}
where $a_0$ is the scattering length and $r_0$ is the effective
range. Fitting this expression to the calculated phase shift for sufficiently low momenta\footnote{The momentum interval used for this fit is
$k\in[1.4~\mathrm{keV},\,14.0~\mathrm{keV}]$.} gives
\begin{equation}
a_0=15.2~\mathrm{fm},
\label{eq:scattering_length}
\end{equation}
and
\begin{equation}
r_0=1.5~\mathrm{fm}.
\label{eq:effective_range}
\end{equation}

These quantities refer to the leading-order spin-averaged BOEFT
channel and should not be identified directly with the scattering
parameters of the physical neutral
$D^0\bar D^{\ast 0}$ channel.
Here numerical and statistical (i.e., fit) uncertainties are negligible and effects from variations in the fit range or number of subdominant terms included in the effective range expansion are of order $10^{-2}~\mathrm{fm}$.

The fact that the $S$ matrix has a bound-state pole just below threshold, combined with the effective range expansion, relates $a_0$ and $r_0$ to the binding momentum of the bound state
\begin{equation}
    \kappa = \sqrt{2 \mu E_B},
    \label{eq:kappa_Eb}
\end{equation} with $E_B$ for the binding energy.
The elastic $S$-wave scattering amplitude is given by
\begin{equation}
    f_0(k) = \frac{1}{k \cot \delta_0 - i k}.
\end{equation}
We now plug Eq.~\eqref{eq:eff_range} into the above equation, neglecting $\mathcal{O}(k^4)$ terms in the low-momentum limit.
We then introduce the bound state binding momentum $\kappa$, related to the scattering momentum by $k = i \kappa$, and solve for the pole that is closest to the branch point at $k=0$~\cite{vanKolck:2022lqz}
\begin{equation}
    \kappa = \left(1 - \sqrt{1 - 2 r_0 / a_0}\right) / r_0.
    \label{eq:binding_momentum}
\end{equation}
Evaluating the above equation by inserting Eqs.~\eqref{eq:scattering_length} and \eqref{eq:effective_range}, we obtain
\begin{equation}
    \kappa = 13.7~\mathrm{MeV.}
    \label{eq:kappa}
\end{equation}

The weak-binding relation of Ref.~\cite{Weinberg:1965zz} connects the effective range of a shallow bound state to a conventional
field-renormalization parameter $Z$,
\begin{equation}
r_0
=
-\frac{Z}{1-Z}\,\kappa^{-1}
+\mathcal{O}(R_V),
\label{eq:r0_Z}
\end{equation}
where $\mathcal{O}(R_V)$ denotes corrections governed by the finite
range $R_V$ of the interaction.\footnote{In
Ref.~\cite{Weinberg:1965zz}, the force range was assumed to be set
primarily by pion exchange, so that $R_V\sim m_\pi^{-1}$. In BOEFT,
additional length scales may enter, and we therefore retain the
generic notation $R_V$.}

In the idealized weak-binding limit, $Z$ is conventionally
associated with an elementary component and $1-Z$ with a composite
two-body component. 
In the literature, it is often claimed $Z=0$ for a molecular-like state and $Z=1$ for an elementary-particle state.
Accordingly, the quantity $1-Z$ is typically interpreted as the degree of ``compositeness'' of the shallow bound state.

In the present BOEFT calculation,  the
quantity obtained directly from the normalized bound-state
wavefunction has the more specific meaning of a quarkonium-channel
probability. The two notions may be compared in the weak-binding
limit, but they should not be identified before the finite-range
and coupled-channel corrections have been assessed.

Neglecting the correction term in Eq.~\eqref{eq:r0_Z} and solving
the leading relation for $Z$ defines the estimator
\begin{equation}
\widetilde Z
=
\frac{-r_0\kappa}{1-r_0\kappa}.
\label{eq:Z}
\end{equation}
We use a tilde to distinguish this weak-binding estimator from the
quarkonium probability obtained directly from the BOEFT
wavefunction.

The original reference by Weinberg \cite{Weinberg:1965zz} did not advocate for using the leading-order term in Eq.~\eqref{eq:r0_Z} to obtain a quantitative estimate for $Z$ from an experimental value of $r_0$. 
Instead, Ref.~\cite{Weinberg:1965zz} emphasized the qualitative implication of the weak-binding relation: if the effective range is
not large and negative on the scale of the binding length
$\kappa^{-1}$, but remains in the natural range, then the leading
$Z$-dependent term cannot be sizable and $Z$ must be small.
The total effective range need not itself be negative, since the
finite-range correction $\mathcal O(R_V)$ may have either sign.
However, some subsequent studies, e.g., Ref.~\cite{Esposito:2021vhu}, have tried to estimate the compositeness of $\chi_{c1}(3872)$ quantitatively by using the estimator in Eq.~\eqref{eq:Z}, which neglects finite-range corrections.

Applying this to the BOEFT phase shift, using Eqs.~\eqref{eq:effective_range} and \eqref{eq:kappa} in
Eq.~\eqref{eq:Z}, we find
\begin{equation}
\widetilde Z=-0.12.
\label{eq:negative-Z}
\end{equation}
The value outside the interval $[0,1]$ does not signal an
unphysical bound state. Rather, it shows that the leading
weak-binding relation cannot be inverted quantitatively while
neglecting the correction term. Indeed, for a positive effective
range, Eq.~\eqref{eq:Z} cannot produce
$0\leq\widetilde Z\leq1$. The result, therefore, indicates that
finite-range and coupled-channel effects are numerically relevant
for the extraction of a small $Z$.

We can compare these weak-binding estimates with the properties
obtained directly by solving the coupled Schr\"odinger equation
below threshold. The calibrated $2P$ state has binding energy
\begin{equation}
E_B=96~\mathrm{keV},
\end{equation}
which gives $\kappa=13.7~\mathrm{MeV}$ through
Eq.~\eqref{eq:kappa_Eb}, in agreement with the value extracted from
the effective range parameters. This agreement shows that the low-energy scattering amplitude correctly reproduces the position of the shallow bound-state pole.

The root-mean-square heavy-quark separation is
\begin{equation}
\sqrt{\langle r^2\rangle}=10.8~\mathrm{fm}.
\end{equation}
The normalized bound-state wavefunction has a probability of
$97\%$ in the tetraquark/open-flavor BO channels and a quarkonium
probability
\begin{equation}
Z_{Q\bar Q}\equiv Z=0.03.
\end{equation}
The large radius shows that the tetraquark/open-flavor BO channels
are sampled predominantly in their large-distance
meson--antimeson regime. The state, therefore, has molecular
long-distance characteristics, although the same BO channels
interpolate to adjoint-hadron configurations at short distance.

The size of the correction to Eq.~\eqref{eq:r0_Z} can be estimated
within the present model by inserting the directly calculated
quarkonium probability $Z_{Q\bar Q}$ and binding momentum. Defining
\begin{equation}
\Delta r_0
\equiv
r_0
+
\frac{Z_{Q\bar Q}}{1-Z_{Q\bar Q}}\,\kappa^{-1},
\end{equation}
we obtain
\begin{equation}
\Delta r_0=1.96~\mathrm{fm}.
\end{equation}
By comparison, the leading $Z$-dependent weak-binding contribution
is
\begin{equation}
-\frac{Z_{Q\bar Q}}{1-Z_{Q\bar Q}}\,\kappa^{-1}
=
-0.43~\mathrm{fm}.
\end{equation}
The binding length is much larger,
$\kappa^{-1}\simeq14.4~\mathrm{fm}$, but the leading
$Z$-dependent contribution to $r_0$ is suppressed by the small
factor $Z_{Q\bar Q}/(1-Z_{Q\bar Q})$. A finite range correction of
natural hadronic size can therefore dominate the effective range
and make the inversion of Eq.~\eqref{eq:r0_Z}, without accounting for the finite range term, numerically imprecise, even though the state is very shallow.

The effective range expansion thus gives a reliable determination
of the binding momentum, but the estimator $\widetilde Z$ obtained by neglecting finite range corrections should not be interpreted as a
quantitative, model-independent determination of the quarkonium
probability or of the compositeness. Its robust implication is
qualitative: the absence of a large negative effective range is
consistent with a shallow state dominated by its open-flavor
component, while the precise value of that component requires the
full coupled-channel wavefunction.

These results show that the non-universal corrections to Eq.~\eqref{eq:r0_Z} can generally be of the same order as the universal contribution, even for a shallow bound state.
Therefore, a value $\widetilde{Z}$ obtained from $a_0$ and $r_0$ using Eqs.~\eqref{eq:binding_momentum} and \eqref{eq:Z} should not be regarded as a quantitative prediction for the elementarity.
This conclusion is in line with the qualitative approach by Weinberg \cite{Weinberg:1965zz}, but it is in contrast with the claims of some subsequent studies, e.g., Ref.~\cite{Esposito:2021vhu}, that one can obtain quantitative and model-independent values of $Z$ from experimental values of $a_0$ and $r_0$.

The phase shifts, effective range parameters, and partial-wave structure quoted in this subsection are central results for the
potential parametrization adopted in
Section~\ref{subsec:parametrization}. 

Taken together, the phase shifts and Argand trajectories provide a
real-axis representation of the bound-state and resonance poles
found in the complex-energy analysis. In particular, they display
directly the interference between the shallow
tetraquark/open-flavor-dominated $2P$ state and the broader
quarkonium-dominated $3P$ resonance generated by the same coupled
BOEFT dynamics.\\

\subsection{Sensitivity to the adjoint meson mass}
\label{sec:adjoint meson fine tuning}

The adjoint meson mass
$\bar\Lambda_{1^{--}}^{\rm HL}$ in the heavy-light scheme, or equivalently $\Lambda_{1^{--}}^{\rm C}=\bar\Lambda_{1^{--}}^{\rm HL}-V_0$ in the Cornell scheme, is the only parameter in the present calculation that is calibrated to experimental spectroscopy. The remaining potential parameters are determined from lattice QCD input, the known short- and long-distance BOEFT constraints, continuity conditions, and the modeling assumptions
specified in Section~\ref{subsec:parametrization}. 

The value of the adjoint meson mass adopted in the present work is
fixed by requiring the spin-averaged charmoniumlike $2P$ multiplet associated with
the $\chi_{c1}(3872)$ to reproduce its experimentally established
very shallow character. In the mass and threshold convention used here, this condition places the multiplet $96~\mathrm{keV}$ below
the spin--isospin averaged
$D^{(\ast)}\bar D^{(\ast)}$ threshold and gives
\begin{equation}
\bar\Lambda_{1^{--}}^{\rm HL}=-0.113~\mathrm{GeV}.
\end{equation}
This is therefore the appropriate central value for the present
spin-averaged calculation.

To illustrate the sensitivity of the shallow states, we also repeat
the calculation using
\begin{equation}
\bar\Lambda_{1^{--}}^{\rm HL}=-0.134~\mathrm{GeV},
\end{equation}
the value employed in the previous below-threshold study
\cite{Brambilla:2026zfw}\footnote{In the below-threshold study of
Ref.~\cite{Brambilla:2026zfw}, the adjoint meson mass was calibrated with respect to a spin-averaged heavy-light threshold parametrized also with a $1/m_Q$ correction at variance 
with the fully static parameterization adopted here (see
Section~\ref{subsec:parametrization} and the accompanying footnote).}.

When inserted into the equations and
threshold convention of the present work, this value gives a binding
energy of $764~\mathrm{keV}$ for the same spin-averaged charmoniumlike $2P$ 
multiplet. It should not be regarded as an alternative central
calibration in the present analysis, since it does not reproduce the
very small binding energy imposed here. Its role is instead to show
how the properties of a near-critical state change when its binding
energy is increased within the sub-MeV regime.

The comparison between the two calculations probes the response of
the shallow bound-state and resonance tetraquark-dominated multiplets when the $2P$ multiplet binding energy is increased from approximately $0.1~\mathrm{MeV}$ to $0.8~\mathrm{MeV}$. This
should not be interpreted as assigning an uncertainty interval to
the physical binding energy. The central prediction of the present
work is the $96~\mathrm{keV}$ result obtained with
$\bar\Lambda_{1^{--}}^{\rm HL}=-0.113~\mathrm{GeV}$. The second calculation is
included to expose the near-critical sensitivity to the
adjoint meson mass.

We find that the quarkonium-dominated part of the spectrum is essentially unaffected by this variation.
The masses of the quarkonium-dominated multiplets shift by amounts much
smaller than the relativistic and spin-dependent uncertainties assigned to them in Section~\ref{sec:Comparison with experiments}, and their dominant composition is unchanged. The organization of the spectrum into HQSS multiplets and the assignment of the higher states are therefore robust against variations of the adjoint meson mass.

The shallow near-threshold states as well as the tetraquark-dominated resonance show a much stronger dependence on the adjoint meson mass. As summarized in Table~\ref{Table: calibration new versus old}, lowering
$\bar\Lambda_{1^{--}}^{\rm HL}$ from $-0.113$ to $-0.134~\mathrm{GeV}$ increases the binding energy of the charmoniumlike $2P$ multiplet from
$96$ to $764~\mathrm{keV}$ and reduces its root-mean-square
heavy-quark separation from $10.8$ to $4.2~\mathrm{fm}$. Its
quarkonium probability increases from $3\%$ to $7\%$, while the state remains strongly dominated by the tetraquark/open-flavor BO
channels.

\begin{table}
    \renewcommand{\arraystretch}{1.2}
    \setlength{\tabcolsep}{5pt}
    \centering
   \begin{tabular}{ccccccccccc}
        \toprule [1.5pt]
        \multirow{2}{*}{$\bar\Lambda_{1^{--}}^{\rm HL}$ (GeV)} & \multicolumn{5}{c}{$\chi_{c1}(3872)$} & \multicolumn{5}{c}{$X_b$}  \\
         \cmidrule(lr){2-6} \cmidrule(lr){7-11} 
       & $E_B$ (keV) & $\sqrt{\langle r^2\rangle}$ (fm) & $Z$ & $X_1$ & $X_2$ & $E_B$ (keV) & $\sqrt{\langle r^2\rangle}$ (fm) & $Z$ & $X_1$ &  $X_2$ \\
     \midrule [1.5pt]
        $-0.113$ & $\hphantom{0}96$ & $10.8$ & $3\%$ & $96\%$ & $1\%$ & $\hphantom{0}233$ & $4.9$ & $1\%$ & $98\%$ & $1\%$ \\
        $-0.134$ & $764$ & $\hphantom{0}4.2$ & $7\%$ & $92\%$ & $1\%$ & $1669$ & $2.3$ & $2\%$ & $96\%$ & $2\%$\\
    \bottomrule  [1.5pt]
    \end{tabular}
   \caption{Binding energy $E_B$, root-mean-square heavy-quark
separation $\sqrt{\langle r^2\rangle}$, quarkonium probability
$Z=P_0$, and tetraquark probabilities $X_i=P_i$ ($i=1,2$ denote the $S$- and $D$-wave tetraquark channels, respectively) of the shallow charmoniumlike $2P$ multiplet associated with the
$\chi_{c1}(3872)$ and of its bottomoniumlike $4P$ counterpart,
denoted by $X_b$. Notice that for states below threshold normalized composition measures $P_i$ admit a probabilistic interpretation. The value
$\bar\Lambda_{1^{--}}^{\rm HL}=-0.113~\mathrm{GeV}$ is the central value adopted
in this work, fixed so that the spin-averaged charmoniumlike
multiplet lies $96~\mathrm{keV}$ below the spin--isospin averaged
threshold. The results for
$\bar\Lambda_{1^{--}}^{\rm HL}=-0.134~\mathrm{GeV}$ are shown only as a sensitivity
test based on the value used in Ref.~\cite{Brambilla:2026zfw}; in the
present equations, it produces a binding energy of
$764~\mathrm{keV}$.}
     \label{Table: calibration new versus old}
\end{table}

The bottomoniumlike $4P$ multiplet displays the same qualitative
response: its binding energy increases from $233~\mathrm{keV}$ to
$1.669~\mathrm{MeV}$, its root-mean-square separation decreases from
$4.9$ to $2.3~\mathrm{fm}$, and its bottomonium probability changes
from $1\%$ to $2\%$. These results demonstrate the sensitivity of
the spatial properties and pole positions of shallow states without
altering their dominant open-flavor character in the two calculations
considered.
This behavior has a simple physical origin: for a shallow near-threshold state the binding momentum, and hence the radius, depends strongly on the depth of the
supporting potential, which is controlled by $\bar{\Lambda}_{1^{--}}^{\rm HL}$. 

\begin{table}
    \renewcommand{\arraystretch}{1.2}
    \setlength{\tabcolsep}{5pt}
    \centering    
   \begin{tabular}{cccccc}
        \toprule [1.5pt]
        \multirow{2}{*}{$\bar\Lambda_{1^{--}}^{\rm HL}$ (GeV)} & \multicolumn{5}{c}{$Y_b$} \\
         \cmidrule(lr){2-6}
       & $M - m_{B^{(\ast)}\bar B^{(\ast)}}$ (MeV) & $\Gamma$ (MeV) & $\sqrt{\langle r^2\rangle}$ (fm) & $P_0$ & $P_1$ \\
     \midrule [1.5pt]
        $-0.113$ & $50$ & $53$ & $0.8+0.1 i$ & $25\%$ & $75\%$ \\
        $-0.134$ & $42$ & $44$ & $0.8 + 0.1 i$ & $27\%$ & $73\%$ \\
    \bottomrule  [1.5pt]
    \end{tabular}
   \caption{
   Energy relative to the $B^{(\ast)}\bar B^{(\ast)}$ threshold $M - m_{B^{(\ast)}\bar B^{(\ast)}}$, width into the included $B^{(\ast)}\bar B^{(\ast)}$ channels $\Gamma$, complex root-mean-square heavy-quark
separation $\sqrt{\langle r^2\rangle}$, normalized quarkonium channel composition measure
$P_0$, and normalized $P$-wave tetraquark channel composition measure $P_1$ of the broad bottomoniumlike $5S$ resonance $Y_b$ (and its HQSS partner).
The value $\bar\Lambda_{1^{--}}^{\rm HL}=-0.113~\mathrm{GeV}$ is the central value adopted in this work. The results for
$\bar\Lambda_{1^{--}}^{\rm HL}=-0.134~\mathrm{GeV}$ are shown only as a sensitivity
test based on the value used in Ref.~\cite{Brambilla:2026zfw}.
}
    \label{Table: Yb calibration new versus old}
\end{table}

As for the bottomoniumlike $5S$ multiplet, its dependence on the adjoint meson mass is summarized in Table~\ref{Table: Yb calibration new versus old}. Its mass decreases by $8$ MeV, and similarly its open-flavor width decreases by $9$ MeV, showing larger absolute pole shifts than the two shallow bound-state multiplets, but a much smaller relative change of its threshold separation.
At both tested parameter values, it remains an above-threshold, open-flavor-dominated resonance. Its normalized composition and the complex root-mean square heavy-quark separation change little at the quoted precision.

The comparison demonstrates the near-critical nature of the shallow multiplets and it investigates the dependence of the $Y_b$ resonance on the adjoint-meson mass. The value
$\bar\Lambda_{1^{--}}^{\rm HL}=-0.113~\mathrm{GeV}$ is selected in the present calculation because it reproduces the experimentally motivated very small binding of the multiplet associated with the
$\chi_{c1}(3872)$. The calculation with
$\bar\Lambda_{1^{--}}^{\rm HL}=-0.134~\mathrm{GeV}$ is not an equally preferred
calibration, but a diagnostic showing that a modest change of the adjoint meson mass can increase the binding of a shallow multiplet by almost an order of magnitude and substantially reduce the spatial size of a near-threshold state, or change by around $10~\mathrm{MeV}$ the mass and decay width of a resonant one.

The central physical conclusion is therefore twofold. First, with
the calibration appropriate to the present spin-averaged framework,
the $\chi_{c1}(3872)$ multiplet is an exceptionally shallow and
spatially extended state. Second, the precise binding energies and radii of such near-critical states and the pole location of tetraquark-dominated resonances are highly sensitive to the
adjoint meson mass, whereas the organization and dominant channel assignment of the higher predominantly quarkonium spectrum are considerably more stable. A lattice QCD determination of the lowest $1^{--}$ adjoint-meson energy would test the present calibration directly and sharpen the predictions for the shallow bottomoniumlike counterpart $X_b$.

\section{Hybrid levels}
\label{sec:hybrids}

In addition to conventional quarkonium and tetraquark/open-flavor
configurations, quarkonium hybrids may contribute to the quarkoniumlike spectrum in the energy region considered here. The
coupled equations solved in Sections~\ref{sec:BOEFT}--\ref{sec:results} do not include hybrid
channels. We therefore calculate the corresponding BOEFT
hybrid multiplets and place them alongside the quarkonium--open-flavor
spectrum as spectroscopic levels
\cite{Brambilla:2022hhi,Oncala:2017hop}.

This treatment should not be understood as implying that hybrids are
dynamically decoupled from all the other configurations at leading
order. Although hybrid--quarkonium mixing is suppressed by
$1/m_Q$, hybrid static energies can mix at leading order with
tetraquark BO channels carrying the same BO quantum numbers. These
mixings are not included here because the required nonperturbative
mixing potentials are not presently known.

Quarkonium hybrids, schematically denoted by $Q\bar Q g$, are
color-singlet states in which the heavy $Q\bar Q$ pair is coupled to
an excited gluonic configuration. In the short-distance limit, the heavy quark-antiquark pair acts as a pointlike color-octet source, and the light field
configuration approaches a gluelump. The lowest hybrid BO
potentials, $V_{\Sigma_u^-}$ and $V_{\Pi_u}$, belong at short
distance to the multiplet associated with the lowest
$k^{PC}=1^{+-}$ gluelump.

The hybrid static potentials are better constrained than the
tetraquark static potentials because of extensive lattice QCD results
are available, particularly in pure SU(3) gauge theory;
see Ref.\cite{Alasiri:2024nue} and references therein.
 At short distances,
$V_{\Sigma_u^-}$ and $V_{\Pi_u}$ become degenerate and approach a
repulsive color-octet potential offset by the $1^{+-}$ gluelump
mass. In the pure-gauge description, they rise linearly
at large distances and are reasonably represented by the appropriate
excitations of the effective-string spectrum
\cite{Juge:2002br}. The resulting hybrid multiplets are obtained
from the coupled $\Sigma_u^-$--$\Pi_u$ Schr\"odinger equations of
BOEFT \cite{Berwein:2015vca}. The equations and potential
parametrizations \cite{Alasiri:2024nue}  used here are summarized in Appendix~\ref{app:Hybrids}.

An important feature of the hybrid static potentials is that they lie above the spin-isospin-averaged $S+S$ threshold. 
Consequently, there is an energy gap between the open-flavor threshold and the lowest-lying hybrid states. 

The hybrid potential
$V_{\Sigma_u^-}$ shares the same BO quantum numbers as an isoscalar
tetraquark BO channel that approaches a repulsive color-octet
potential offset by the $0^{-+}$ adjoint meson mass at short
distance and the $S+S$ static-light meson--antimeson threshold at
large distance \cite{Berwein:2024ztx}. The two channels, therefore
mix already at leading order in BOEFT
\cite{Bruschini:2023tmm,TarrusCastella:2024zps}.

This mixing provides the BOEFT mechanism through which hybrid states
couple to pairs of $S$-wave heavy-light mesons. Once it is included,
the discrete levels obtained from the confining hybrid
potentials are expected in general to become resonance poles with
open-flavor widths
\cite{Bruschini:2023tmm,TarrusCastella:2024zps}.
The mixing between the hybrid and tetraquark potentials with quantum numbers $\Sigma_u^-$ is responsible for the decays of  hybrids into a pair of $S$-wave heavy mesons, contrary to the conventional wisdom from decay models \cite{Tanimoto:1982eh, Page:1996rj, Kou:2005gt, McNeile:2006bz, Woss:2020ayi, Farina:2020slb}.  
This was first observed in Ref.~\cite{Bruschini:2023tmm} and subsequently analyzed within BOEFT in Ref.~\cite{TarrusCastella:2024zps}.  

The $V_{\Sigma_u^-}$ and $V_{\Pi_u}$ hybrid potentials also share
their BO quantum numbers with tetraquark channels approaching
$S+P$ heavy-meson thresholds. Their leading-order mixing produces
string-breaking avoided crossings and may be particularly important
for hybrid levels close to the upper boundary of the energy window
considered here \cite{Berwein:2024ztx}. No lattice QCD determination
of the required hybrid--tetraquark mixing potentials is presently
available. A quantitative calculation of the corresponding resonance
poles, therefore, lies beyond the scope of this work.

A separate source of mixing occurs between quarkonium and hybrid
states with common $J^{PC}$ quantum numbers. For the low-lying
hybrids associated with $V_{\Sigma_u^-}$ and $V_{\Pi_u}$, this
mixing is generated by an ${\cal O}(1/m_Q)$ spin-dependent operator
coupled to the chromomagnetic field. It can produce sizable mass
shifts and violations of the leading-order HQSS multiplet structure
\cite{Oncala:2017hop}. Since it is subleading in the heavy-quark
expansion, hybrid--quarkonium mixing is not included in the present
leading-order calculation.

Consequently, the masses listed below are eigenvalues of the
 hybrid BOEFT Hamiltonian. They should be regarded as
spectroscopic levels and not as predictions for physical hybrid
resonance pole masses. In particular, the present calculation does
not determine their open-flavor pole widths, pole couplings, or
hybrid--tetraquark--quarkonium composition.

The charmonium and bottomonium hybrid multiplets lying
below the spin--isospin averaged $S+P$ boundary are shown in
Tables~\ref{tab:ccbarhybspectrum} and
\ref{tab:bbbarhybspectrum}, respectively. 
The shapes of the hybrid static potentials are taken from the
pure-SU(3) parametrization of Ref.~\cite{Alasiri:2024nue}.
Their common additive normalization is fixed in
Appendix~\ref{app:Hybrids} by matching the ABM quarkonium potential
to the quarkonium potential used in the present work. This gives
the flavor-independent shift
$\delta_g=-1.23\pm0.10~\mathrm{GeV}$.

\begin{table}[t]
\renewcommand{\arraystretch}{1.2}
\centering
\begin{tabular}{ccc}
\toprule[1.5pt]
Multiplet & $J^{PC}$ & $M_{\rm ref}$ (MeV) \\
\midrule[1.5pt]
$H_1$ &
$1^{--},(0,1,2)^{-+}$ &
$4180$
\\
$H_2$ &
$1^{++},(0,1,2)^{+-}$ &
$4300$
\\
\bottomrule[1.5pt]
\end{tabular}
\caption{
Central masses $M_{\rm ref}$ and $J^{PC}$ quantum numbers of the
 charmonium-hybrid multiplets whose central
values lie below the spin--isospin averaged $S+P$ threshold boundary.
The masses are calculated accounting for the common matching shift
$\delta_g=-1.232~{\rm GeV}$ defined in
Appendix~\ref{app:Hybrids}. The masses have a common matching uncertainty of $\pm 0.100~{\rm GeV}$ 
as discussed in Eq.~\eqref{eq:hybrid-matching-uncertainty}; it is not an independent uncertainty for each multiplet. 
The uncertainty bands of some levels extend across the nominal $S+P$ threshold boundary. The quoted uncertainty does not include
mass shifts or widths induced by hybrid--tetraquark or
hybrid--quarkonium mixings. The next relevant multiplet, $H_4$, has a
central mass of approximately $4371~{\rm MeV}$ and lies above the
nominal $S+P$ threshold boundary, so it is not included in our analysis. The charmonium-hybrid multiplets $H_3$, $H_5$ lie at still higher masses.}
\label{tab:ccbarhybspectrum}
\end{table}

\begin{table}[t]
\renewcommand{\arraystretch}{1.2}
\centering
\begin{tabular}{ccc}
\toprule[1.5pt]
Multiplet & $J^{PC}$ & $M_{\rm ref}$ (MeV) \\
\midrule[1.5pt]
$H_1$ &
$1^{--},(0,1,2)^{-+}$ &
$10749$
\\
$H'_1$ &
$1^{--},(0,1,2)^{-+}$ &
$10940$
\\
$H_2$ &
$1^{++},(0,1,2)^{+-}$ &
$10813$
\\
$H'_2$ &
$1^{++},(0,1,2)^{+-}$ &
$11027$
\\
$H_3$ &
$0^{++},1^{+-}$ &
$11013$
\\
$H_4$ &
$2^{++},(1,2,3)^{+-}$ &
$10862$
\\
$H_5$ &
$2^{--},(1,2,3)^{-+}$ &
$10916$
\\
\bottomrule[1.5pt]
\end{tabular}
\caption{
Central masses $M_{\rm ref}$ and $J^{PC}$ quantum numbers of the
 bottomonium-hybrid multiplets whose central
values lie below the spin--isospin averaged $S+P$ boundary.
A prime denotes the first radial excitation of the corresponding
hybrid BOEFT multiplet. The masses are calculated accounting for the common matching shift $\delta_g=-1.232~{\rm GeV}$ defined in
Appendix~\ref{app:Hybrids}. The masses have a common matching uncertainty of $\pm 0.100~{\rm GeV}$ 
as discussed in Eq.~\eqref{eq:hybrid-matching-uncertainty}; it is not an independent error for each multiplet. The uncertainty bands of some levels extend across the nominal $S+P$ threshold boundary. The quoted uncertainty does not include mass shifts or widths generated by the omitted hybrid--tetraquark and hybrid--quarkonium mixings.
The next relevant hybrid multiplets all lie above the $S+P$ threshold boundary, so are not included in our analysis.
}
\label{tab:bbbarhybspectrum}
\end{table}

The proximity of one of these levels to a quarkonium or
open-flavor multiplet should therefore be interpreted as indicating
a region in which omitted mixing effects may be important, rather
than as establishing a pure-hybrid assignment. The hybrid levels
will be used in this limited sense in the comparison with the
experimental spectrum.

\section{Comparison with the experimental spectrum}\label{sec:Comparison with experiments}
In this section, we compare the spin-averaged quarkonium--tetraquark/open-flavor
BOEFT multiplets obtained in Section~\ref{sec:results}, together with the hybrid levels of Section~\ref{sec:hybrids}, with the available charmoniumlike and bottomoniumlike experimental spectra. The purpose of this comparison is primarily to investigate whether the observed candidates can be organized into HQSS multiplets and to identify the regions in which dynamics omitted from the present calculation may be important. The associations discussed below should therefore be understood as possible multiplet assignments within the accuracy and channel content of the present leading-order calculation, rather than as unique state-by-state identifications.

The terms ``quarkonium dominated'' and ``tetraquark/open-flavor-dominated'' refer to the dominant channel content of
the calculated BOEFT states, as characterized for resonances in
Tables~\ref{Table: table of c c-bar probabilites} and~\ref{Table: table of b b-bar probabilites} and for bound states in Tables~\ref{Table: table of c c-bar probabilites below} and
~\ref{Table: table of b b-bar probabilites below}. The tetraquark BO channels interpolate between short-distance adjoint-hadron configurations and large-distance heavy-light meson--antimeson configurations. The hybrid levels are listed in Tables~\ref{tab:ccbarhybspectrum} and \ref{tab:bbbarhybspectrum}. Their proximity to an experimental state may indicate that hybrid dynamics and omitted mixing effects are relevant, but it does not by itself establish a predominantly hybrid assignment.

For clarity, the radial labels of the BOEFT multiplets in
Tables~\ref{Table: c c-bar masses and decay widths} and
\ref{Table: b b-bar masses and decay widths} do not always coincide with the
conventional quarkonium notation. In the charmoniumlike sector, the
predominantly quarkonium BOEFT multiplets $3P$ and $4P$ correspond,
respectively, to the conventional $2P$ and $3P$ quarkonium multiplets, whereas the BOEFT $2P$ multiplet is the shallow tetraquark/open-flavor-dominated multiplet associated with the $\chi_{c1}(3872)$. Similarly, in the bottomoniumlike sector, the predominantly quarkonium BOEFT multiplets $6S$, $5P$, and $6P$ correspond, respectively, to the conventional $5S$, $4P$, and $5P$ bottomonium multiplets, whereas the BOEFT $4P$ multiplet is the shallow tetraquark/open-flavor-dominated multiplet containing the predicted $X_b$.

\begin{figure*}[p]
\centering
\includegraphics[width=1.0\textwidth]{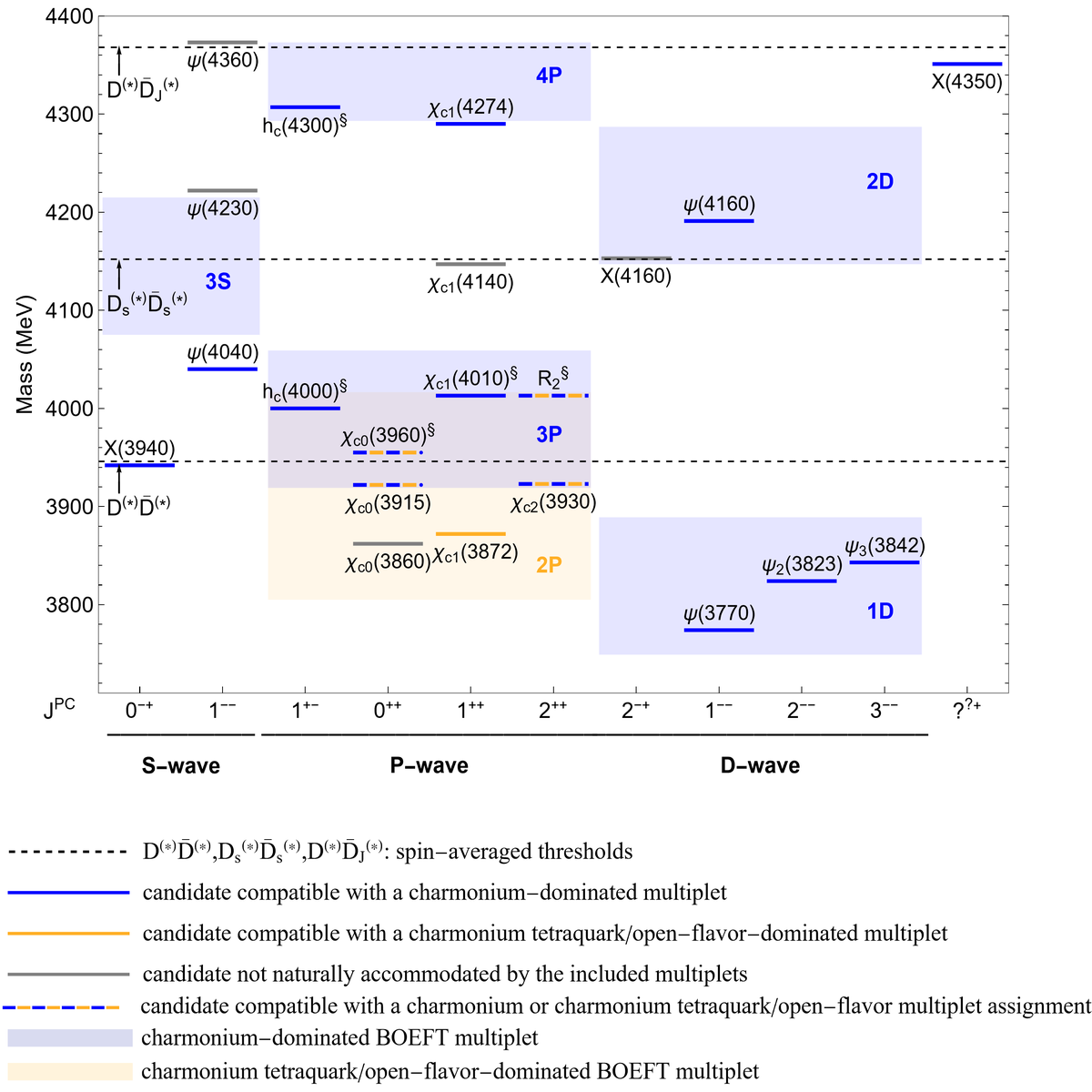}
\caption{\justifying{Comparison between the charmoniumlike multiplets predicted in BOEFT (shaded boxes) by solving  Eqs.~\eqref{eq:diabatic},~\eqref{eq:diabaticl0} and the experimental spectrum (continuous lines) above the $D \bar{D}$ threshold and
up to the $D^{(\ast)} \bar{D}_J^{(\ast)}$ spin-isospin averaged thresholds.
The shaded bands are order-of-magnitude estimates of corrections omitted from the leading-order calculation and are not statistical confidence intervals.
For charmonium-dominated multiplets, they estimate relativistic and
spin-dependent effects ($135~{\rm MeV}$), while for charmonium tetraquark/open-flavor-dominated multiplets, they estimate the effects of physical heavy-light meson spin splittings ($210~{\rm MeV}$).
Candidate associations indicate possible multiplet assignments within the present accuracy and channel content.
Experimental states not listed in the PDG are marked by a ``§'' sign.
}}  
\label{fig:cc sector with experiments}
\end{figure*}

\begin{figure*}[p]
\centering
\includegraphics[width=0.96\textwidth]{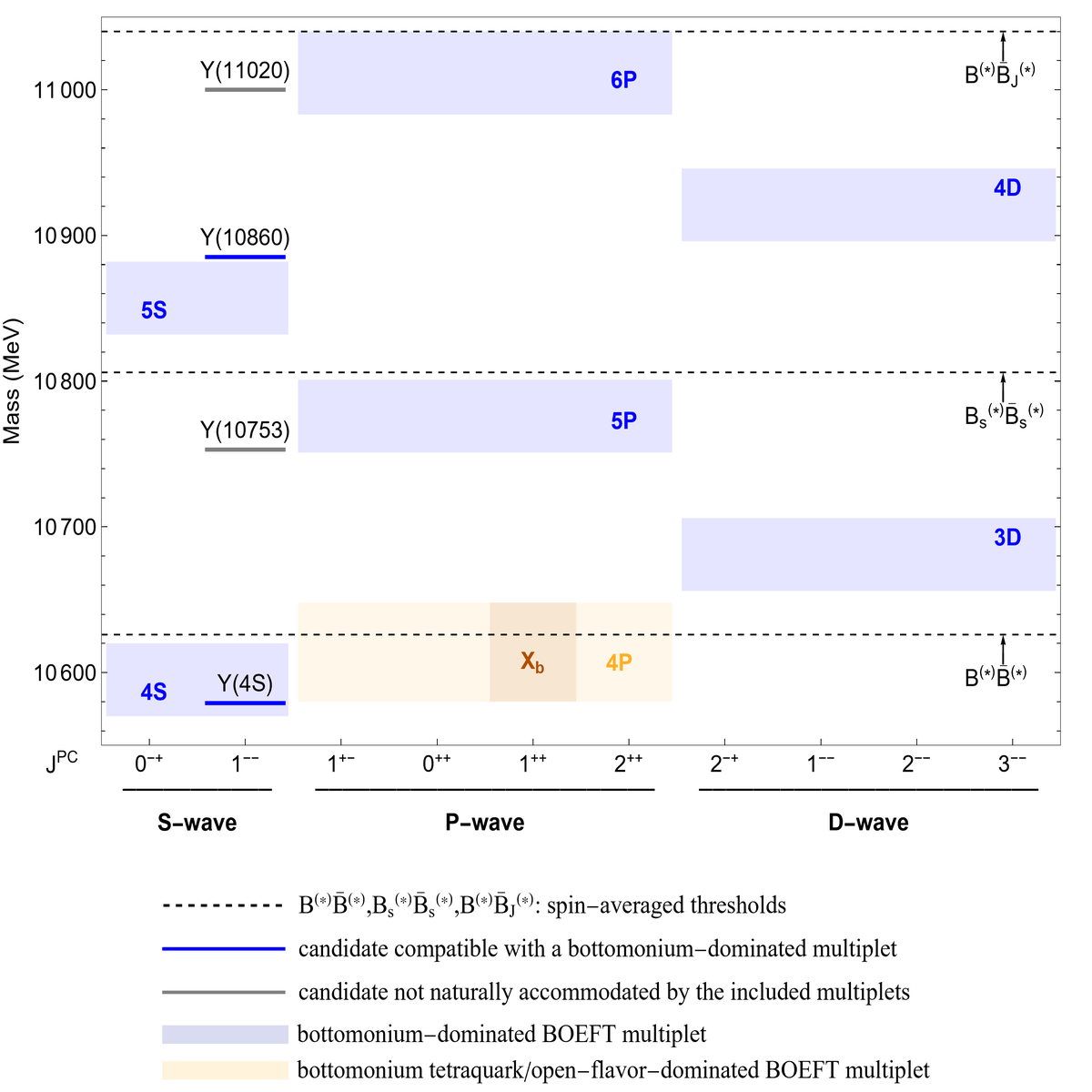}
\caption{\justifying{Comparison between the bottomoniumlike multiplets predicted in BOEFT (shaded boxes) by solving  Eqs.~\eqref{eq:diabatic},~\eqref{eq:diabaticl0} and the experimental spectrum (continuous lines) above the $B \bar{B}$ threshold and
up to the $B^{(\ast)} \bar{B}_J^{(\ast)}$ spin-isospin averaged thresholds.
The shaded bands are order-of-magnitude estimates of corrections omitted from the leading-order calculation and are not statistical confidence intervals.
For bottomonium-dominated multiplets, they estimate relativistic and
spin-dependent effects ($45~{\rm MeV}$), while for bottomonium tetraquark/open-flavor-dominated multiplets, they estimate the effects of physical heavy-light meson spin splittings ($70~{\rm MeV}$).
Candidate associations indicate possible multiplet assignments within the
present accuracy and channel content.
}}
\label{fig:bb sector with experiments}
\end{figure*}

\begin{figure*}[p]
\centering
\includegraphics[width=1.0\textwidth]{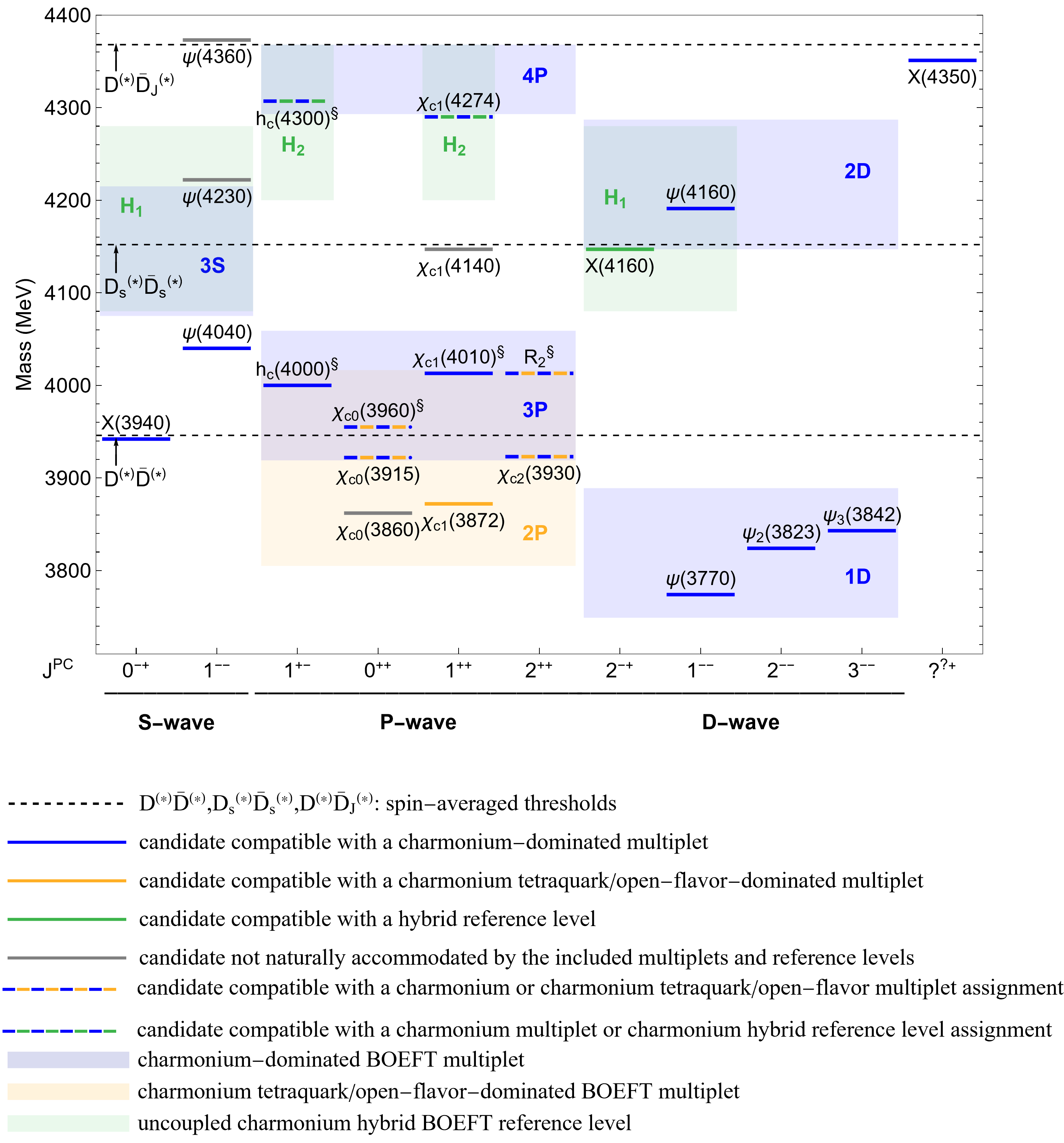}
\caption{\justifying{
Comparison between the charmoniumlike multiplets predicted in BOEFT (shaded boxes) by solving Eqs.~\eqref{eq:diabatic},~\eqref{eq:diabaticl0},~\eqref{eq:diffeq-hyb}, including also the charmonium hybrid levels predictions, and the experimental spectrum (continuous lines) above the $D \bar{D}$ threshold and up to the $D^{(\ast)} \bar{D}_J^{(\ast)}$ spin-isospin averaged thresholds.
For the charmonium hybrid levels, the displayed band represents
the common $\pm 100~\mathrm{MeV}$ uncertainty associated with
matching the pure-gauge hybrid static energies to the additive heavy-light scheme used in the present work. This uncertainty
is fully correlated among the hybrid multiplets. It does not
include the potentially important mass shifts or open-flavor
widths generated by the omitted leading-order
hybrid--tetraquark mixing or by the $1/m_Q$-suppressed
hybrid--quarkonium mixing.
The remaining shaded bands are order-of-magnitude
estimates of omitted relativistic and spin-dependent effects ($135~{\rm MeV}$) for charmonium-dominated multiplets, or heavy-light
spin-splitting effects ($210~{\rm MeV}$) for charmonium tetraquark/open-flavor-dominated multiplets, and are not statistical confidence intervals.
Candidate associations indicate possible multiplet assignments within the
present accuracy and channel content.
Experimental states not listed in the PDG are marked by a ``§'' sign.
}}
\label{fig:cc sector with hybrids with experiments}
\end{figure*}

\begin{figure*}[p]
\centering
\includegraphics[width=0.96\textwidth]{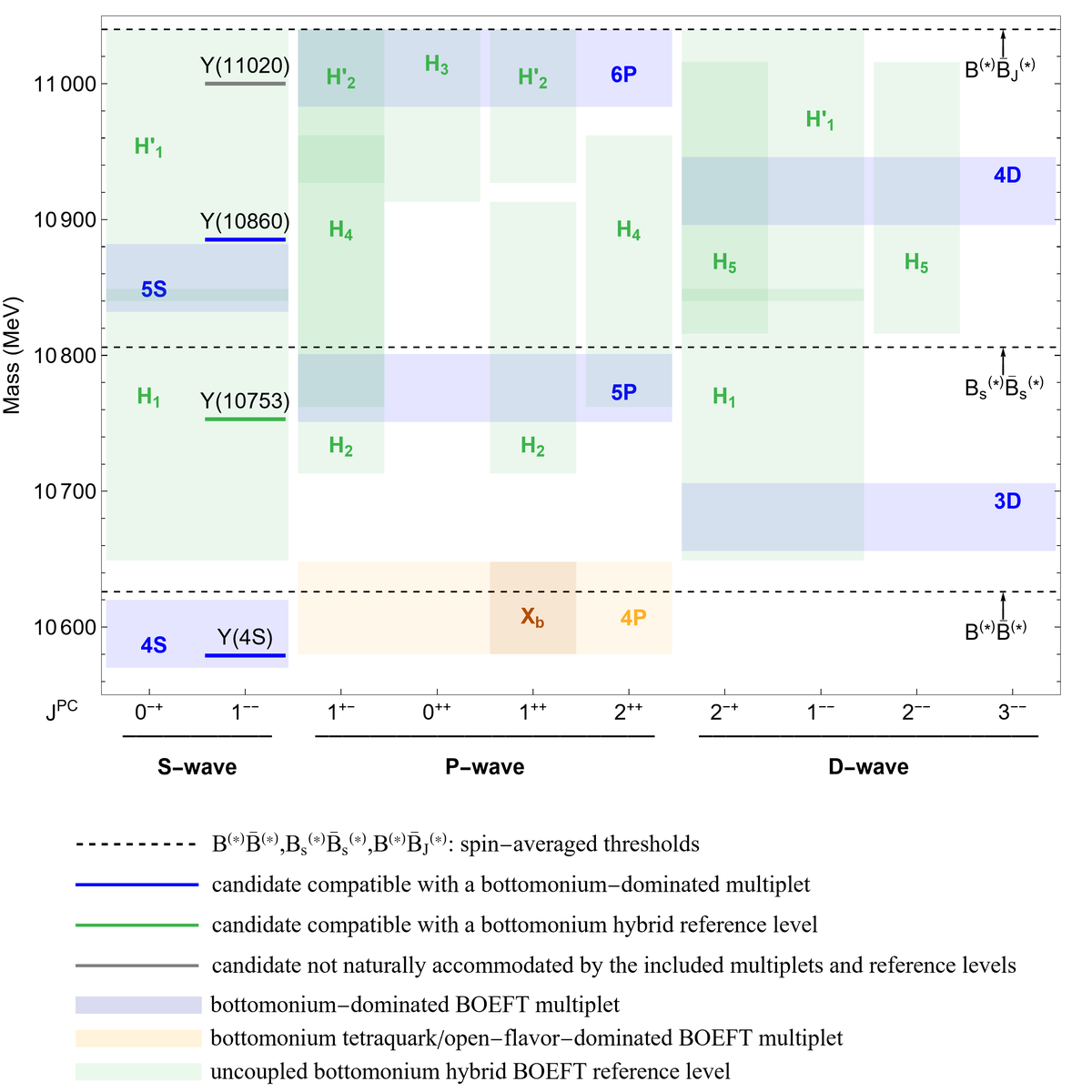}
\caption{\justifying{
Comparison between the bottomoniumlike multiplets predicted in BOEFT (shaded boxes) by solving Eqs.~\eqref{eq:diabatic},~\eqref{eq:diabaticl0},~\eqref{eq:diffeq-hyb}, including also the bottomonium hybrid levels predictions, and the experimental spectrum (continuous lines) above the $B \bar{B}$ threshold and up to the $B^{(\ast)} \bar{B}_J^{(\ast)}$ spin-isospin averaged thresholds.
For the bottomonium hybrid levels, the displayed band represents
the common $\pm 100~\mathrm{MeV}$ uncertainty associated with
matching the pure-gauge hybrid static energies to the additive
heavy-light scheme used in the present work. This uncertainty
is fully correlated among the hybrid multiplets. It does not
include the potentially important mass shifts or open-flavor
widths generated by the omitted leading-order
hybrid--tetraquark mixing or by the $1/m_Q$-suppressed hybrid--quarkonium mixing.
The remaining shaded bands are order-of-magnitude estimates of omitted  relativistic and spin-dependent effects ($45~{\rm MeV}$) for bottomonium-dominated multiplets, or heavy-light
spin-splitting effects ($70~{\rm MeV}$) for bottomonium tetraquark/open-flavor-dominated multiplets, and are not statistical confidence intervals.
Candidate associations indicate possible multiplet assignments within the present accuracy and channel content.
}}
\label{fig:bb sector with hybrids with experiments}
\end{figure*}

The coupled-channel calculation applies in the energy region between the
lowest spin--isospin averaged nonstrange $S+S$ threshold,
$D^{(\ast)}\bar D^{(\ast)}$ or $B^{(\ast)}\bar B^{(\ast)}$, and the first
spin-averaged $S+P$ threshold,
$D^{(\ast)}\bar D_J^{(\ast)}$ or $B^{(\ast)}\bar B_J^{(\ast)}$. The comparison with experiments is shown in
Figs.~\ref{fig:cc sector with experiments} and
\ref{fig:bb sector with experiments} without the hybrid levels, and
in Figs.~\ref{fig:cc sector with hybrids with experiments} and
\ref{fig:bb sector with hybrids with experiments} with those levels included.
Unless otherwise stated, the threshold lines shown in these figures are spin averaged and, for the nonstrange thresholds, also isospin averaged.

For completeness, the figures also display the charmoniumlike $1D$ and $2P$ multiplets and the bottomoniumlike $4S$ and $4P$ multiplets, although these are composed of bound states, due to their proximity to the spin--isospin averaged $S+S$ threshold in the leading-order calculation. We report the masses relative to these multiplets in Tables~\ref{tab:ccbelow},~\ref{tab:bbbelow}.
Physical heavy-light spin splittings and quarkonium spin-dependent corrections resolve each
spin-averaged HQSS multiplet into members having different positions relative to the physical thresholds. Depending on its quantum numbers and proximity to a threshold, an individual member may remain bound, become a virtual state, or
appear as a resonance with an open-flavor width. The comparison should
therefore be understood at the multiplet level and does not imply that every member of a spin-averaged bound multiplet necessarily becomes a physical resonance.

The shaded bands in
Figs.~\ref{fig:cc sector with experiments}--\ref{fig:bb sector with hybrids with experiments}
represent order-of-magnitude estimates of effects omitted from the
leading-order calculation; they are not statistical confidence intervals. For predominantly quarkonium multiplets, we estimate relativistic and spin-dependent corrections of relative order $v^2$, corresponding to energy corrections of order $m_Qv^4$, approximately $135~{\rm MeV}$ in the charmoniumlike sector and $45~{\rm MeV}$ in the bottomoniumlike sector. These estimates do not
include potentially state-dependent effects from nearby omitted thresholds or from quarkonium--hybrid mixing.
For $S$-wave tetraquark/open-flavor-dominated multiplets, the bands estimate the uncertainty from the spin-splitting structure of the tetraquark multiplet by computing threshold spin-splitting in perturbation theory~\cite{Brambilla:2024imu,Brambilla:2026zfw}. This amounts to uncertainties of $210~{\rm MeV}$ and $70~{\rm MeV}$, respectively, in the charmoniumlike and bottomoniumlike sectors.
For a very shallow state, however, these splittings may move individual HQSS partners through nearby
thresholds and may change the analytic nature of a pole from bound to virtual or resonant~\cite{Alasiri:2026vfz}. The displayed bands should therefore not be interpreted as ordinary uncertainties on a fixed type of pole.
Both the $0^{-+}$ and $1^{--}$ members of the bottomoniumlike $5S$ tetraquark/open-flavor-dominated multiplet receive no corrections from threshold spin-splitting in perturbation theory, being this the only $P$-wave tetraquark/open-flavor-dominated resonance of our model~\cite{Brambilla:2026zfw}. 
 This result does not imply the absence of spin effects: the order $1/m_b$ interaction couples the leading configurations to additional BO channels and generates pole shifts at second order, with possible enhancement when the relevant energy separations are small. Direct higher-order spin interactions and spin-independent inverse-mass corrections are also not considered. 
 Therefore, we estimate the error band from the relativistic corrections of order $v^2$, giving order $m_b v^4$ energy corrections, as for the predominantly bottomonium multiplets. 
The size of this band should be regarded as illustrative, rather than as a calculated spin spread or a quantitative total uncertainty.
None of these bands includes the model dependence associated with the
presently unknown $V_{\Pi_g}$ potential and the
short-distance part of the $V_{\Sigma_g^{+\prime}}$ potential. This additional uncertainty may be relevant for states with sizable open-flavor components.

For the hybrid levels, the displayed band is associated with the fully correlated uncertainty of $\pm 100~\mathrm{MeV}$ in matching the pure-gauge hybrid static energies to the heavy-light scheme of our calculations (see appendix~\ref{app:Hybrids}).
It is common to both the charmoniumlike and bottomoniumlike sectors, and it does not include the potentially important mass shifts and open-flavor widths generated by the omitted leading-order hybrid--tetraquark mixing, nor the effects of the $1/m_Q$-suppressed hybrid--quarkonium mixing.

Throughout this section, the calculated quantity $\Gamma$ denotes only the pole width generated by the spin-averaged, nonstrange $S+S$ open-flavor channels included in the coupled equations. It is not a prediction for the physical total width, which may also receive contributions from physical spin-splitted thresholds, hidden-strange and $S+P$ channels, hadronic transitions, electromagnetic decays, and hybrid-induced channels. Comparisons with measured
total widths are therefore necessarily qualitative unless these limitations are stated explicitly. Unless otherwise specified, the experimental $J^{PC}$ quantum numbers, masses, and widths are taken from the Particle Data Group~\cite{ParticleDataGroup:2024cfk}.

The heading of each box gives the complete HQSS multiplet.
Only the experimental associations adopted for the discussion are
listed underneath; an omitted entry does not mean that the
corresponding multiplet member is absent from the calculation.
Where possible candidates exist but the present analysis does not
select among their interpretations, we use the phrase
``No unique experimental assignment.''

\subsection{Charmoniumlike sector}\label{sec:Comparison for charm}

The charmoniumlike spectrum between the spin--isospin averaged $S+S$ and
$S+P$ thresholds contains candidates with several different $J^{PC}$ quantum numbers. We organize the discussion according to the calculated BOEFT multiplets and compare their masses, dominant channel content, and 
quantum numbers with the available experimental information. We first discuss the predominantly quarkonium multiplets, followed by the shallow
tetraquark/open-flavor-dominated $2P$ multiplet and the hybrid levels. We then consider the remaining experimental candidates that are not naturally accommodated by the channels included in the present calculation. Unless otherwise stated, the corresponding BOEFT results are summarized in Tables~\ref{Table: c c-bar masses and decay widths}, \ref{tab:ccbelow}, \ref{Table: table of c c-bar probabilites}, \ref{Table: table of c c-bar probabilites below}, and
\ref{tab:ccbarhybspectrum}.

\subsubsection {Quarkonium-dominated BOEFT multiplets}
\label{subsubsec:charmonium}
\begin{center}
\noindent\fbox{%
    \parbox{.95\textwidth}{%
    \bigskip
\centering
\textbf{BOEFT multiplet $\bm{1D}$ (quarkonium multiplet $\bm{1D}$), $\bm{J^{PC}=2^{-+},1^{--},2^{--},3^{--}}$}
\[
\setlength\arraycolsep{20pt}
\def\arraystretch{1.4}
\begin{array}{lll}
\arraycolsep=1.4pt
1^{--}\to\psi(3770): & M=3773.7\pm0.7\text{ MeV}, & \Gamma=27.2\pm1.0\text{ MeV;}\\
2^{--}\to\psi_2(3823): & M=3823.51\pm0.34\text{ MeV,} & \Gamma < 2.9\text{ MeV;}\\
3^{--}\to\psi_3(3842): & M=3842.71\pm0.16\pm0.12\text{ MeV,} & \Gamma = 2.8\pm0.6\text{ MeV.}
\end{array}
\]
}
}
\end{center}

The BOEFT $1D$ multiplet lies below the spin--isospin averaged
$D^{(\ast)}\bar D^{(\ast)}$ threshold and has a calculated spin-averaged mass
of $3819~{\rm MeV}$. The well-established $\psi(3770)$, $\psi_2(3823)$, and
$\psi_3(3842)$ provide natural candidates for the
$1\,{}^3D_1$, $1\,{}^3D_2$, and $1\,{}^3D_3$ charmonium states,
respectively~\cite{Brambilla:2026zfw}. This association is supported by their
quantum numbers, their relative masses, and the established conventional
charmonium interpretation of this below-threshold multiplet.

At leading order, all members of the BOEFT multiplet are degenerate and lie below the common spin-isospin averaged open-charm threshold, so that the width generated by the open-flavor channels included in the present calculation is zero.
Physical heavy-light and quarkonium spin splittings resolve the multiplet and place its individual members differently relative to the physical thresholds (some states may go above the $D\bar{D}$ threshold, but all remain below the $D\bar{D}^{*}$ threshold).
The $\psi(3770)$ can decay into $D\bar D$ in a relative $P$ wave and consequently has a sizable open-charm width. The decay of the $\psi_2(3823)$ into
$D\bar D$ is forbidden by angular-momentum and parity conservation, and so its dominant decays are expected to proceed through quarkonium channels, consistent with the PDG observations. The $\psi_3(3842)$ can decay into $D\bar D$ only in an $F$ wave and is therefore
centrifugally suppressed. This pattern is qualitatively consistent with the
observed broad $\psi(3770)$ and narrow $\psi_2(3823)$ and $\psi_3(3842)$.
The same BOEFT multiplet contains an additional spin-singlet $1\,{}^1D_2$ state with $J^{PC}=2^{-+}$, for which no established experimental candidate is presently available. Its mass is expected to lie in the vicinity
of the other members of the physical $1D$ multiplet, once spin-dependent corrections are included.

\bigskip
\begin{center}
\noindent\fbox{%
    \parbox{.95\textwidth}{%
    \bigskip
\centering
\textbf{BOEFT multiplet $\bm{3P}$ (quarkonium multiplet $\bm{2P}$), $\bm{J^{PC}=1^{+-},0^{++},1^{++},2^{++}}$}
\[
\setlength\arraycolsep{20pt}
\def\arraystretch{1.4}
\begin{array}{lll}
\arraycolsep=1.4pt
1^{+-}\to h_c(4000): & M= 4000^{+17}_{-14}\,^{+29}_{-22}\text{ MeV,} & \Gamma = 184^{+71}_{-45}\,^{+96}_{-61}\text{ MeV;} \\
0^{++}\to\chi_{c0}(3960): & M= 3956\pm 11\text{ MeV,} & \Gamma = 43 \pm 15\text{ MeV.} \\
1^{++}\to \chi_{c1}(4010): & M= 4012.5^{+3.6}_{-3.9}\,^{+4.1}_{-3.7}\text{ MeV,} & \Gamma = 62.7^{+7.0}_{-6.4}\,^{+6.4}_{-6.6}\text{ MeV;} \\
2^{++}\to R_2(4014):& M=4014.3 \pm 4.3\text{ MeV,} & \Gamma = 4\pm12.5 \text{ MeV.} \\
\end{array}
\]
}
}
\end{center}

The BOEFT $3P$ resonance is predominantly quarkonium and corresponds, in
conventional quark-model notation, to the $2P$ charmonium multiplet. Its central pole mass is $3988~{\rm MeV}$, while its pole width into the included spin-averaged, nonstrange $S+S$ open-charm channels is $28~{\rm MeV}$. The latter quantity is not a prediction for the physical total widths of the individual HQSS partners, whose positions relative to the physical $D\bar D$, $D\bar D^\ast$, and $D^\ast\bar D^\ast$ thresholds are resolved only after spin-dependent effects are included.

The recently observed $h_c(4000)$ and $\chi_{c1}(4010)$ states by LHCb in the decay $B^+\to D^{*\pm}D^{\mp}K^+$ \cite{LHCb:2024vfz} are possible candidates for the $2^1P_1$ and $2^3P_1$ charmonium states, respectively. 
The corresponding masses and widths, taken from Ref.~\cite{LHCb:2024vfz}, are consistent with our predictions within the sizable experimental and theoretical uncertainties.

The state $\chi_{c0}(3960)$ was observed by LHCb in the decay process
$B^+\rightarrow D_s^+D_s^-K^+$ \cite{LHCb:2022aki}. Its mass, quantum
numbers, and decay width are compatible with the conventional
$2^3P_0$ charmonium assignment within the present leading-order accuracy. However, after including spin
splittings, this state lies close to the $D_s\bar{D}_s$ threshold.
Moreover, its observation in the $D_s^+D_s^-$ invariant-mass spectrum
suggests that coupled-channel effects associated with the
$D_s\bar{D}_s$ channel, which is not included in our calculation, may further influence its properties.  

The $R_2(4014)$ state was observed by Belle in the process
$\gamma\gamma\rightarrow\gamma\psi(2S)$ with a global significance of
only $2.8\sigma$~\cite{Belle:2021nuv}. The state was analyzed under
$J^{PC}=0^{++}$ and $2^{++}$ hypotheses. Assuming
$J^{PC}=2^{++}$, this state is a possible candidate for the conventional $2^3P_2$ charmonium state, with its measured mass compatible with our predictions within the current uncertainties. After including spin splittings, the state lies below
the $D^{*}\bar{D}^{*}$ threshold but above the $D\bar{D}$ and
$D\bar{D}^{*}$ thresholds. Consequently, it can decay into $D\bar{D}$ or $D\bar{D}^{*}$ channels, which require a relative $D$-wave due to angular momentum and parity conservation, leading to a suppressed decay rate. This provides a natural explanation for the small observed width of the state. 

The scalar and tensor sectors remain intrinsically ambiguous in the present
leading-order comparison. One possible organization associates
$\chi_{c0}(3915)$ and $\chi_{c2}(3930)$ with the shallow
tetraquark/open-flavor-dominated $2P$ multiplet addressed below and
$\chi_{c0}(3960)$ and $R_2(4014)$ with the predominantly quarkonium $3P$
multiplet. The alternative organization interchanges these two pairs. Since the
corresponding theoretical bands overlap after the estimated spin-dependent
corrections are included, and since hidden-strange and physical open-charm
thresholds are omitted, masses alone do not establish a unique choice between
these assignments.
Nevertheless, within our leading-order calculation, the $3P$ multiplet assignment for both  $\chi_{c0}\left(3960\right)$ and $R_2\left(4014\right)$ may be somewhat preferred based on the mass.

\bigskip
\begin{center}
\noindent\fbox{%
    \parbox{.95\textwidth}{%
    \bigskip
\centering
\textbf{BOEFT multiplet $\bm{3S}$ (quarkonium multiplet $\bm{3S}$), $\bm{J^{PC}=0^{-+},1^{--}}$}
\[
\setlength\arraycolsep{20pt}
\def\arraystretch{1.4}
\begin{array}{lll}
\arraycolsep=1.4pt
0^{-+}\to X(3940): & M= 3942\pm9\text{ MeV,} & \Gamma = 43^{+28}_{-18}\text{ MeV;} \\
1^{--} \to \psi(4040): & M= 4040\pm4\text{ MeV,} & \Gamma = 84\pm12\text{ MeV.} \\
\end{array}
\]
}
}
\end{center}

Assuming that the $X(3940)$ and the recently reported $\eta_c(3945)$ signal~\cite{LHCb:2024vfz} refer to the same state, we adopt the favored assignment
$J^{PC}=0^{-+}$. The $X(3940)$ and the $\psi(4040)$ then provide possible
spin-singlet and spin-triplet candidates for the conventional $3S$ charmonium
multiplet. This quantum-number organization is natural, although the comparison with the spin-averaged BOEFT pole mass is not quantitatively accurate: the experimental masses lie approximately $200~{\rm MeV}$ and $100~{\rm MeV}$, respectively, below the calculated central mass of $4145~{\rm MeV}$. The discrepancy is especially large for the spin-singlet candidate and indicates that the assignment cannot be regarded as a precise leading-order mass prediction.

The calculated $3S$ pole width is only about $1~{\rm MeV}$, whereas the measured
total widths of the proposed candidates are substantially larger. This
difference should not be interpreted as a direct failure of an open-charm
partial-width prediction. As shown in Fig.~\ref{fig:3s nodal effects}, the central
spin-averaged pole lies close to a node of the transition amplitude, so even a
moderate displacement of the pole can produce a large change in the width into
the included open-charm channel. In addition, the experimental total widths contain hadronic-transition, electromagnetic, and other strong-decay contributions absent from the present calculation. The leading-order result therefore predicts a nodally suppressed open-charm pole width at the calculated mass, rather than a sharply determined physical total width for either experimental state.

The $3S$ pole at $4145~{\rm MeV}$ lies 
$35~{\rm MeV}$ below the $H_1$ hybrid level at $4180~{\rm MeV}$. This proximity does not establish a hybrid
assignment for the $X(3940)$ or the $\psi(4040)$, but it indicates
that the omitted hybrid--quarkonium and hybrid--tetraquark
mixings may produce additional mass shifts and widths in this
region.

\bigskip
\begin{center}
\noindent\fbox{%
    \parbox{.95\textwidth}{%
    \bigskip
\centering
\textbf{BOEFT multiplet $\bm{2D}$ (quarkonium multiplet $\bm{2D}$), $\bm{J^{PC}=2^{-+},1^{--},2^{--},3^{--}}$}
\[
\setlength\arraycolsep{20pt}
\def\arraystretch{1.4}
\begin{array}{lll}
\arraycolsep=1.4pt
1^{--}\to\psi(4160): & M= 4191\pm5\text{ MeV,} & \Gamma = 69\pm 10\text{ MeV;} \\
\end{array}
\]
}
}
\end{center}

The $\psi(4160)$ is a natural candidate for the spin-triplet
$2\,{}^3D_1$ charmonium state within the accuracy of the present
spin-averaged calculation. Its measured mass lies approximately $30$~MeV below the calculated central pole mass and is well inside the estimated
quarkonium uncertainty band.

The experimental total width is larger than the pole width generated by the included spin-averaged, nonstrange $S+S$ channels. This difference cannot be assigned uniquely to spin-dependent corrections. Physical threshold splittings, additional open-flavor channels, hadronic transitions, and the dependence of the decay amplitude on the shifted pole position may all contribute. The width comparison should therefore be regarded as qualitative.

The $X(4160)$ lies in the same mass region and, for the favored
$J^{PC}=2^{-+}$ assignment, may be compared both with the
spin-singlet $2\,{}^1D_2$ member of the predominantly quarkonium
$2D$ multiplet and with the $2^{-+}$ member of the $H_1$ hybrid multiplet. Its central mass lies approximately
$27~{\rm MeV}$ below the $H_1$ level and
$64~{\rm MeV}$ below the $2D$ pole. The mass comparison, therefore,
slightly favors proximity to the hybrid level, while both assignments remain compatible with the corresponding theoretical bands. Its semi-inclusive quarkonium decay pattern has also been argued to be compatible with significant hybrid dynamics~\cite{Brambilla:2022hhi}.

The $H_1$ and $2D$ central levels themselves are separated by only
$37~{\rm MeV}$. This identifies the $4.15$--$4.22~{\rm GeV}$
region as one in which hybrid--quarkonium and hybrid--open-flavor mixings may be particularly important. The physical
$\psi(4160)$ and $X(4160)$ may therefore contain different
mixtures of quarkonium, hybrid, and open-flavor configurations.

\bigskip

\begin{center}
\noindent\fbox{%
    \parbox{.95\textwidth}{%
    \bigskip
\centering
\textbf{BOEFT multiplet $\bm{4P}$ (quarkonium multiplet $\bm{3P}$), $\bm{J^{PC}=1^{+-},0^{++},1^{++},2^{++}}$}
\[
\setlength\arraycolsep{20pt}
\def\arraystretch{1.4}
\begin{array}{lll}
\arraycolsep=1.4pt
1^{+-}\to h_c(4300): & M= 4307.3^{+6.4}_{-6.6}\,^{+3.3}_{-4.1}\text{ MeV,} & \Gamma = 58^{+28}_{-16}\,^{+28}_{-25}\text{ MeV;} \\
1^{++}\to \chi_{c1}(4274): & M= 4290\pm 6\text{ MeV,} & \Gamma = 54\pm 8\text{ MeV; }\\
\left(0^{++} / 2^{++}\right)\to X(4350): & M = 4351\pm 5\text{ MeV,} & \Gamma = 13^{+18}_{-10}\text{ MeV; } 
\end{array}
\]
}
}
\end{center}

The BOEFT $4P$ pole is predominantly quarkonium and corresponds, in
conventional quark-model notation, to the $3P$ charmonium multiplet. Its central pole mass is $4364~{\rm MeV}$ and its pole width into the included spin-averaged, nonstrange $S+S$ channels is $8~{\rm MeV}$. Since this multiplet lies close to the first $S+P$ threshold and to several hybrid levels, its comparison with individual experimental candidates is necessarily less direct than for the lower multiplets.

The $\chi_{c1}(4274)$ is a possible candidate for the
$3\,{}^3P_1$ charmonium state, although its measured mass lies below the
calculated central value. Its decay into $J/\psi\,\phi$ also suggests that hidden-strange dynamics may contribute significantly. The omitted
$D_s^{(\ast)}\bar D_s^{(\ast)}$ channels may therefore affect its mass,
composition, and physical total width.

The $h_c(4300)$ state was recently observed by LHCb in the decay $B^+\to D^{*\pm}D^{\mp}K^+$ \cite{LHCb:2024vfz}. It is a plausible
candidate for the spin-singlet $3^1P_1$ charmonium state. Its experimental mass lies approximately $50~{\rm MeV}$ below our central
prediction but remains compatible within the estimated uncertainties. 
The measured total width is larger than the pole width generated by
the included spin-averaged nonstrange $S+S$ channels. In view of the
nearby omitted thresholds, spin-dependent effects, and additional
decay channels, this comparison should be regarded as qualitative.

The $X(4350)$ state was observed by Belle in the process $\gamma\gamma\rightarrow\phi J/\psi$ with a statistical significance of
$3.9\,\sigma$ \cite{Belle:2009rkh}. Its quantum numbers have not yet been established, with both $J^{PC}=0^{++}$ and $2^{++}$ remaining
viable assignments. Based on its measured mass and width, the state is a possible candidate for either the $3^3P_0$ or the $3^3P_2$
charmonium state. An interpretation as the $H_4[2^{++}]$ hybrid is disfavored by the predicted semi-inclusive decay widths to
quarkonium \cite{Brambilla:2022hhi}.
Neither comparison establishes a unique assignment in the absence of the
relevant mixing dynamics.

The discrepancies in the mass and the widths of these states suggest that important effects are still missing from the present
leading-order calculation. In particular, the central mass of the $4P$ multiplet is approximately $64~{\rm MeV}$ above the $H_2$ hybrid
multiplet and only about $7~{\rm MeV}$ below the $H_4$ hybrid multiplet (see Tables~\ref{Table: c c-bar masses and decay widths},~\ref{tab:ccbarhybspectrum} and the discussion of the $H_4$ level below). Consequently, hybrid-quarkonium mixing, especially with the nearby $H_4$ multiplet, may significantly
modify the physical states in this region. In addition, the proximity of the $4P$ multiplet to the $S+P$ charmed-meson threshold may lead to additional coupled-channel effects and modify the resonance
properties.

\subsubsection {Tetraquark/open-flavor-dominated BOEFT multiplets}
\label{subsubsec:Tetraquark charmonium}

\begin{center}
\noindent\fbox{%
    \parbox{.95\textwidth}{%
    \bigskip
\centering
\textbf{BOEFT multiplet $\bm{2P}$ ($\bm{\chi_{c1}(3872)}$ and its HQSS partners), $\bm{J^{PC}=1^{+-},0^{++},1^{++},2^{++}}$}
\[
\setlength\arraycolsep{20pt}
\def\arraystretch{1.4}
\begin{array}{lll}
\arraycolsep=1.4pt
0^{++}\to\chi_{c0}(3915): & M= 3922.1\pm1.8\text{ MeV,} & \Gamma = 20\pm 4\text{ MeV;} \\
1^{++}\to\chi_{c1}(3872): & M=3871.64 \pm 0.06\text{ MeV,} & \Gamma = 1.19\pm0.21 \text{ MeV;} \\
2^{++}\to \chi_{c2}(3930): & M= 3922.5\pm1.0\text{ MeV,} & \Gamma = 35.2\pm2.2\text{ MeV.} \\
\end{array}
\]
}
}
\end{center}

The BOEFT $2P$ multiplet is qualitatively different from the predominantly quarkonium $3P$ multiplet discussed above. With the central calibration adopted in this work, $\bar\Lambda_{1^{--}}^{\rm HL}=-0.113~\mathrm{GeV}$, it forms a shallow bound
state $96~\mathrm{keV}$ below the common spin--isospin averaged
$D^{(\ast)}\bar D^{(\ast)}$ threshold. This value of the adjoint meson mass is fixed so that the spin-averaged multiplet associated with the $\chi_{c1}(3872)$ reproduces its experimentally established very shallow character in the spin-isospin averaged threshold convention of the present calculation.

The normalized bound-state wavefunction has a quarkonium probability of
approximately $3\%$ and a combined probability of approximately $97\%$ in the tetraquark/open-flavor BO channels. Its root-mean-square heavy-quark separation, $\sqrt{\langle r^2\rangle}=10.8~\mathrm{fm}$, is much larger than a natural hadronic length scale. The state, therefore, samples predominantly the large-distance meson--antimeson region of the tetraquark BO channels and has molecular long-distance characteristics. It should not, however, be described as a purely meson--antimeson configuration or as a permanently compact $c\bar c q\bar q$ configuration: the same BO channels interpolate continuously to short-distance adjoint-hadron configurations and also mix with the quarkonium channel through string breaking.

The association with the physical $\chi_{c1}(3872)$ is made at the multiplet level. The leading-order calculation contains a common spin-averaged open-charm threshold, whereas the physical $\chi_{c1}(3872)$ lies extremely close to the neutral $D^0\bar D^{\ast 0}$ threshold. The calibration, therefore, uses the observed shallow nature of this state but should not be interpreted as a direct equality between the calculated spin-averaged mass and the measured physical mass.

Because the spin-averaged $2P$ multiplet is bound below the only open-flavor threshold included in the present coupled equations, its calculated pole width into those channels is zero. This result is not a prediction that the physical HQSS partners have vanishing total widths. Physical heavy-light spin splittings resolve the common threshold into several distinct thresholds and may move different members of the multiplet to different Riemann sheets. An individual partner may remain bound, become a virtual state, or appear as a resonance.
Furthermore, its physical total width may receive contributions from
spin-split open-charm channels, hidden-strange channels, hadronic transitions, electromagnetic decays, and other processes absent from the present calculation.

The $\chi_{c1}(3872)$ is the established $1^{++}$ experimental state motivating this multiplet. Within BOEFT, its long-distance molecular behavior and small quarkonium probability emerge dynamically from the complete quarkonium--tetraquark/open-flavor coupled equations.

On the basis of mass and quantum numbers, we identify $\chi_{c0}\left(3915\right)$ as a possible candidate for the spin-triplet $0^{++}$ member of the $2P$ multiplet, assuming that $\chi_{c0}(3960)$ is identified with the conventional $2^3P_0$ charmonium state.  Its measured mass is compatible with this assignment within the estimated uncertainties. The discrepancy in the width is due to the absence of spin splittings in our coupled-channel framework. Once spin splittings are included, $\chi_{c0}(3915)$ moves above the $D\bar{D}$ threshold and can decay into this channel through an $S$-wave, accounting for its observed nonzero width. Moreover, since the state lies close to the $D_s\bar{D}_s$ threshold after spin splittings are included, additional coupled-channel effects from this channel, which are not considered in our calculation, may further influence its properties.  

The $\chi_{c2}(3930)$ is a candidate for the spin-triplet $2^{++}$ member of the  $2P$ multiplet, with its measured mass  compatible with our predictions within the current uncertainties.  After including spin splittings, this state lies between the $D\bar{D}^\ast$ and $D^\ast\bar{D}^\ast$ thresholds. 
Consequently, it can decay into the $D\bar{D}$ and $D\bar{D}^\ast$ channels, which require a relative $D$-wave due to angular-momentum and parity conservation, leading to nonzero widths.

An alternative interpretation, illustrated in
Fig.~\ref{fig:cc sector with experiments}, is to identify
$\chi_{c0}(3915)$ and $\chi_{c2}(3930)$ with the conventional
$2^3P_0$ and $2^3P_2$ charmonium states, respectively. In this scenario,
$\chi_{c0}(3960)$ and $R_2(4014)$ are instead interpreted as the possible candidates for $0^{++}$ and $2^{++}$ members of the $\chi_{c1}(3872)$ multiplet, respectively. Under this assignment, the measured masses and widths are compatible with our predictions for the $2^3P_0$ and $2^3P_2$ charmonium states within the current uncertainties. We therefore regard it as an ambiguous $2P/3P$ candidate rather than a clean BOEFT assignment.

The $2P$ BOEFT multiplet also contains a spin-singlet state with
$J^{PC}=1^{+-}$, for which no established experimental candidate is presently available.

\subsubsection {Hybrid BOEFT multiplets}
\label{subsubsec:Hybrid charmonium}

The hybrid multiplets discussed in this subsection are the eigenstates of the hybrid BOEFT Hamiltonian of Section~\ref{sec:hybrids}.
Our mass predictions for charmonium hybrid multiplets lying below the spin-isospin averaged $S + P$ threshold are summarized in Table~\ref{tab:ccbarhybspectrum}.
They are used as spectroscopic levels and are not predictions for physical hybrid resonance poles.
In particular, the present calculation does not determine their open-flavor pole widths, pole couplings, or hybrid--tetraquark--quarkonium compositions.
Proximity between an experimental candidate and a hybrid level may indicate that hybrid dynamics is relevant, but it does not establish a predominantly hybrid assignment.

\begin{center}
\noindent\fbox{%
    \parbox{.95\textwidth}{%
    \bigskip
\centering
\textbf{Hybrid multiplet} $\bm{H_1(4180)}$, $\bm{J^{PC}=1^{--},0^{-+},1^{-+},2^{-+}}$
\[
\setlength\arraycolsep{20pt}
\def\arraystretch{1.4}
\begin{array}{lll}
\arraycolsep=1.4pt
2^{-+}\to X(4160): & M= 4153^{+23}_{-21}\text{ MeV,} & \Gamma = 136^{+60}_{-35}\text{ MeV.} \\
\end{array}
\]
}
}
\end{center}

For the $X(4160)$, the quantum numbers have not been firmly established,
although $J^{PC}=2^{-+}$ is presently favored. Its mass and preferred quantum numbers place it close to the spin-triplet $2^{-+}$ member of the $H_1$  multiplet. In addition, the semi-inclusive decay pattern into quarkonium calculated in Ref.~\cite{Brambilla:2022hhi} is compatible with significant hybrid dynamics. The $X(4160)$ may therefore be associated plausibly with the $H_1$ region.

However, the state also lies within the theoretical
band of the predominantly quarkonium $2D$ multiplet, whose spin-singlet
$2\,{}^1D_2$ member has the same $J^{PC}=2^{-+}$ quantum numbers.
In this case, the agreement with our predicted $2D$ multiplet mass is only marginal within the estimated uncertainties, with the experimental mass lying approximately $70~{\rm MeV}$ below our prediction. Furthermore, the observed width is up to an order of magnitude larger than our calculated value.  Given the proximity of the $2D$ charmonium and $H_1$ hybrid multiplets, hybrid-quarkonium mixing may play an important role in this region, and the physical $X(4160)$ state may contain both hybrid and quarkonium components rather than corresponding to a pure configuration.

\bigskip

\begin{center}
\noindent\fbox{%
\parbox{.95\textwidth}{%
\bigskip
\centering
\textbf{Hybrid multiplet
$\bm{H_2(4300)}$,
$\bm{J^{PC}=1^{++},0^{+-},1^{+-},2^{+-}}$}
\[
\setlength\arraycolsep{20pt}
\def\arraystretch{1.4}
\begin{array}{lll}
1^{+-}\to h_c(4300): & M= 4307.3^{+6.4}_{-6.6}\,^{+3.3}_{-4.1}\text{ MeV,} & \Gamma = 58^{+28}_{-16}\,^{+28}_{-25}\text{ MeV;} \\
1^{++}\to \chi_{c1}(4274): & M= 4290\pm 6\text{ MeV,} & \Gamma = 54\pm 8\text{ MeV; }
\end{array}
\]
}}
\end{center}
The $h_c(4300)$ and $\chi_{c1}(4274)$ lie approximately
$7~{\rm MeV}$ and $10~{\rm MeV}$, respectively, from the central
$H_2$ hybrid level. Both states have quantum numbers
contained in the $H_2$ HQSS multiplet. At the same time, they are
plausible candidates for members of the predominantly quarkonium
$4P$ multiplet. Their masses, therefore, do not select uniquely
between quarkonium and hybrid interpretations.

The central $4P$ pole lies $64~{\rm MeV}$ above
$H_2$, and the corresponding uncertainty bands overlap.
Hybrid--quarkonium mixing may consequently be important for the
physical $h_c(4300)$ and $\chi_{c1}(4274)$ states. Hidden-strange
and $S+P$ open-charm channels, which are omitted from the present
calculation, may provide additional mass shifts and widths,
particularly for the $\chi_{c1}(4274)$.

The semi-inclusive quarkonium decay calculation of
Ref.~\cite{Brambilla:2022hhi}, performed for unmixed hybrid states, gives
\[
 \Gamma_{\rm semi-inc.}\left(H_2[1^{++}]\right)
 =55^{+54}_{-31}~\mathrm{MeV},
 \qquad
 \Gamma_{\rm semi-inc.}\left(H_2[(0,1,2)^{+-}]\right)
 =18^{+18}_{-10}~\mathrm{MeV}.
\]
These quantities provide lower bounds on the corresponding total widths within that separate decay analysis. They are not widths obtained from the hybrid spectrum calculated here and may be modified by the mixing effects omitted in the present treatment.

\begin{center}
\noindent\fbox{%
    \parbox{.95\textwidth}{%
    \centering
    \bigskip
\textbf{Hybrid multiplet $\bm{H_4(4371)}$}, $\bm{J^{PC}=2^{++},1^{+-},2^{+-},3^{+-}}$\\[1em]
No unique experimental assignment.\\
\vspace{1em}
}
}
\end{center}

The next $H_4$ hybrid level has a central mass of
approximately $4371~{\rm MeV}$, about $7~{\rm MeV}$ above the
predominantly quarkonium $4P$ pole mass. Its central value lies above
the nominal spin-averaged $S+P$ boundary, and it is therefore not
included in Table~\ref{tab:ccbarhybspectrum}. Nevertheless, its
matching band overlaps the upper part of the energy region and
the mass of the $X(4350)$. This proximity indicates that hybrid
dynamics may be relevant, but a quantitative interpretation
requires the omitted $S+P$ tetraquark BO channels and their
mixing with quarkonium and hybrid configurations.

The semi-inclusive decay analysis of Ref.~\cite{Brambilla:2022hhi} gives
\[
 \Gamma_{\rm semi-inc.}\left(H_4[2^{++}]\right)
 =65^{+57}_{-33}~\mathrm{MeV},
 \qquad
 \Gamma_{\rm semi-inc.}\left(H_4[(1,2,3)^{+-}]\right)
 =22^{+19}_{-11}~\mathrm{MeV}.
\]
As for the $H_2$ multiplet, these are lower bounds obtained for unmixed hybrid states in a separate decay calculation, rather than physical pole widths predicted by the hybrid Hamiltonian used here. The proximity of the $H_4$ and $4P$ levels should therefore be understood as evidence that omitted mixing may be important, not as establishing a pure-hybrid assignment for any particular experimental candidate.

\subsubsection {Experimental candidates not naturally accommodated by the included channels}
\label{subsubsec:ccbarOther}

\begin{center}
\noindent\fbox{%
    \parbox{.95\textwidth}{%
\[
\setlength\arraycolsep{20pt}
\def\arraystretch{1.4}
\begin{array}{llll}
\arraycolsep=1.4pt
\bm{\chi_{c0}(3860)}: & J^{PC}=0^{++}, & M=3862^{+50}_{-35}\text{ MeV,} & \Gamma = 200^{+180}_{-110}\text{ MeV.}
\end{array}
\]
}
}
\end{center}

The $\chi_{c0}(3860)$ was reported by Belle in
$e^+e^-\to J/\psi D\bar D$, whereas no corresponding signal was found in the LHCb analysis of $B^+\to D^+D^-K^+$. Its mass and large experimental
uncertainty overlap with the theoretical bands of both the shallow $2P$ and predominantly quarkonium $3P$ multiplets.

If the $\chi_{c0}(3860)$, $\chi_{c0}(3915)$, and $\chi_{c0}(3960)$ are all confirmed as distinct scalar states, however, the present calculation contains only two $0^{++}$ multiplets in this mass region and cannot accommodate all three simultaneously. Additional scalar poles could arise from physical open-charm and hidden-strange channels, spin-dependent interactions, or other BO potentials not included here. We therefore do not propose a definite multiplet assignment for the $\chi_{c0}(3860)$.

\bigskip
\begin{center}
\noindent\fbox{%
    \parbox{.95\textwidth}{%
\[
\setlength\arraycolsep{20pt}
\def\arraystretch{1.4}
\begin{array}{llll}
\arraycolsep=1.4pt
\bm{\chi_{c1}(4140)}: & J^{PC}=1^{++}, & M= 4146.5\pm3.0\text{ MeV,} & \Gamma = 19^{+7}_{-5}\text{ MeV.}
\end{array}
\]
}
}
\end{center}

The $\chi_{c1}(4140)$ is not naturally associated with the nonstrange quarkonium--tetraquark/open-flavor multiplets calculated here. Its proximity to hidden-strange thresholds and its observation in the $J/\psi\,\phi$ channel indicate that BO channels approaching $D_s^{(\ast)}\bar D_s^{(\ast)}$ thresholds may be important.

Such a channel may produce a state with hidden-strange tetraquark character at shorter distances and meson--antimeson long-distance characteristics near the corresponding threshold. The present calculation does not include this channel and therefore cannot determine the pole position, composition, or width of the $\chi_{c1}(4140)$. A pure association with the $H_2[1^{++}]$ level is also disfavored by the semi-inclusive quarkonium-decay analysis of Ref.~\cite{Brambilla:2022hhi}, although hybrid
admixtures cannot be excluded once the omitted mixings are included.

\bigskip
\begin{center}
\noindent\fbox{%
    \parbox{.95\textwidth}{%
\[
\setlength\arraycolsep{20pt}
\def\arraystretch{1.4}
\begin{array}{llll}
\arraycolsep=1.4pt
\bm{\psi(4230)}: & J^{PC}=1^{--}, & M= 4221.7\pm2.5\text{ MeV,} & \Gamma = 51\pm 8\text{ MeV.}
\end{array}
\]
}
}
\end{center}

The $\psi(4230)$ is not naturally accommodated as an additional state by the channels included in the present calculation. The vector members of the calculated $3S$ and $2D$ multiplets have natural experimental candidates in the $\psi(4040)$ and $\psi(4160)$.
The $\psi(4230)$ also lies approximately $42~{\rm MeV}$ above
the central $H_1$ hybrid level and hence well inside
its common matching band. The mass proximity alone, therefore,
does not exclude important hybrid dynamics. The semi-inclusive
quarkonium-decay analysis of Ref.~\cite{Brambilla:2022hhi},
however, disfavors a simple unmixed $H_1$ interpretation.
Together with the proximity of the first $S+P$ open-charm
thresholds, this points toward a state involving hybrid and
$S+P$ open-flavor dynamics rather than a pure hybrid.

In particular, several molecular and
coupled-channel analyses associate the $\psi(4230)$ with dynamics involving the $D\bar D_1$ threshold~\cite{Wang:2013cya, Ji:2022blw, Peng:2022nrj, vonDetten:2024eie, Dong:2026fdi}. The corresponding
tetraquark BO potentials approaching $S+P$ meson pairs are absent from the present channel basis.

The $\psi(4230)$, therefore, points specifically toward the need to include $S+P$ tetraquark/open-flavor BO channels, potentially together with their mixing with nearby quarkonium and hybrid configurations. Its absence from the present spectrum should not be interpreted as evidence against the BOEFT framework.

\bigskip
\begin{center}
\noindent\fbox{%
    \parbox{.95\textwidth}{%
\[
\setlength\arraycolsep{20pt}
\def\arraystretch{1.4}
\begin{array}{llll}
\arraycolsep=1.4pt
\bm{\psi(4360)}: & J^{PC}=1^{--}, & M= 4373\pm7\text{ MeV,} & \Gamma = 124\pm 13\text{ MeV.}
\end{array}
\]
}
}
\end{center}

The $\psi(4360)$ lies at, or slightly above, the upper $S+P$ boundary of the energy region described by the present coupled equations. It therefore cannot be expected to be described reliably using only the lowest nonstrange $S+S$ tetraquark/open-flavor channels.

Its mass, quantum numbers, and semi-inclusive quarkonium-decay pattern are compatible with important hybrid dynamics~\cite{Brambilla:2022hhi}. Moreover, hybrid static energies may mix at leading order with tetraquark BO potentials approaching the $S+P$ thresholds. The physical $\psi(4360)$ may consequently contain hybrid, quarkonium, and $S+P$ open-flavor components. A quantitative assignment requires the additional BO channels and mixing potentials that are omitted from the present calculation.

\subsection{Bottomoniumlike sector}\label{sec:Comparison for bottom}

The experimental bottomoniumlike spectrum in the energy region considered
here is considerably sparser than the corresponding charmoniumlike spectrum.
The established candidates in this interval all have vector quantum numbers $J^{PC}=1^{--}$. We therefore compare them primarily with the vector members of the calculated $S$- and $D$-wave BOEFT multiplets and with the $1^{--}$ members of the hybrid multiplets.

The absence of experimental candidates for many of the predicted non-vector HQSS partners is not in tension with the calculation, since these sectors have not yet been explored with comparable experimental coverage. In particular,
the predicted $5P$ and $6P$ bottomoniumlike multiplets do not contain a
$1^{--}$ member and therefore have no direct counterpart among the presently established vector states. We also discuss the shallow
tetraquark/open-flavor-dominated $4P$ multiplet containing the predicted
$X_b$, and finally the $\Upsilon(11020)$, whose interpretation remains
ambiguous.

Unless otherwise noted, our predictions for the BOEFT multiplet are summarized in Tables~\ref{Table: b b-bar masses and decay widths},  \ref{tab:bbbelow}, \ref{Table: table of b b-bar probabilites}, \ref{Table: table of b b-bar probabilites below}, and \ref{tab:bbbarhybspectrum}.

\subsubsection {Quarkonium-dominated BOEFT multiplets}
\label{subsubsec:bottomonium}

\begin{center}
\noindent\fbox{%
    \parbox{.95\textwidth}{%
    \bigskip
\centering
\textbf{BOEFT multiplet $\bm{4S}$ (quarkonium multiplet $\bm{4S}$), $\bm{J^{PC}= 0^{-+}, 1^{--}}$}
\[
\setlength\arraycolsep{20pt}
\def\arraystretch{1.4}
\begin{array}{lll}
\arraycolsep=1.4pt
1^{--}\to\Upsilon(4S): & M=10579.4\pm1.2\text{ MeV} & \Gamma=20.5\pm2.5\text{ MeV.}\\
\end{array}
\]
}
}
\end{center}

The BOEFT $4S$ multiplet has a calculated spin-averaged mass of
$10595~\mathrm{MeV}$ and lies below the common spin--isospin averaged
$B^{(\ast)}\bar B^{(\ast)}$ threshold. The well-established
$\Upsilon(4S)$ is the natural physical candidate for its spin-triplet
$4\,{}^3S_1$ member.

The leading-order calculation gives zero width into the included open-bottom channel because the common spin-averaged multiplet lies below the common spin-averaged threshold. The physical $\Upsilon(4S)$, by contrast, lies above the $B\bar B$ threshold and can decay into this channel in a relative $P$ wave. Its observed nonzero width therefore reflects the physical threshold structure and spin-dependent mass corrections that are absent from the leading-order spin-averaged calculation. The calculated zero width should not
be interpreted as a prediction for its physical total width.

The same BOEFT multiplet contains a spin-singlet $4\,{}^1S_0$ state with
$J^{PC}=0^{-+}$ and mass in the vicinity of the $4S$ multiplet, 
for which no established experimental candidate is presently
available.

\bigskip

\begin{center}
\noindent\fbox{%
    \parbox{.95\textwidth}{%
    \bigskip
\centering
\textbf{BOEFT multiplet $\bm{3D}$ (quarkonium multiplet $\bm{3D}$), $\bm{J^{PC}=2^{-+},1^{--},2^{--},3^{--}}$}
\\[1em]No established candidates.\\
\vspace{1em}
}
}
\end{center}

No established experimental candidate is presently available for the vector
member of the BOEFT $3D$ multiplet. In both resonance-composition
prescriptions in Table~\ref{Table: table of b b-bar probabilites},
 the $b\bar b$ $D$-wave channel is the largest individual
component, although the nearby open-bottom threshold produces a substantial $B^{(\ast)}\bar B^{(\ast)}$ contribution. The  actual numerical composition may be prescription dependent for this pole and should not be interpreted as a unique probability.

The central $3D$ pole lies approximately $70~\mathrm{MeV}$
below the $H_1$ hybrid level.
This separation is larger than in some
of the more strongly overlapping regions discussed below, but omitted mixing may still affect individual members with common $J^{PC}$. The absence of observed candidates for the vector and non-vector partners is not problematic, given the limited experimental information in this sector.

\begin{center}
\noindent\fbox{%
    \parbox{.95\textwidth}{%
    \bigskip
\centering
\textbf{BOEFT multiplet $\bm{6S}$ (quarkonium multiplet $\bm{5S}$), $\bm{J^{PC}= 0^{-+}, 1^{--}}$}
\[
\setlength\arraycolsep{20pt}
\def\arraystretch{1.4}
\begin{array}{lll}
\arraycolsep=1.4pt
1^{--}\to\Upsilon(10860): & M=10855.2^{+2.6}_{-1.6}\text{ MeV} & \Gamma=37\pm4\text{ MeV.}\\
\end{array}
\]
}
}
\end{center}

The $\Upsilon(10860)$ is a candidate for the $1^{--}$ member of the BOEFT $6S$ multiplet, whose dominant conventional quarkonium configuration is $5\,{}^3S_1$. Its spin-singlet partner has dominant conventional configuration $5\,{}^1S_0$.

The calculated pole width into the included spin-averaged, nonstrange $S+S$ open-bottom channel is approximately $2~\mathrm{MeV}$, substantially smaller than the measured total width. As discussed in Section~\ref{subsec:spectrum}, the transition amplitude of this excited state lies close to a nodal suppression, so the included open-bottom width can change strongly under comparatively small
shifts of the pole position. In addition, the experimental total width contains contributions from physical spin-split thresholds, hidden-strange channels, hadronic transitions, and other processes absent from the calculation. The difference should therefore not be attributed to spin splittings alone.

The $6S$ pole lies approximately $109~\mathrm{MeV}$ above and $82~{\rm MeV}$ below the $H_1$ and $H_1^\prime$ hybrids levels, respectively, and so hybrid-quarkonium mixing may be important.  
The proximity of the $B_s^{(\ast)}\bar B_s^{(\ast)}$ thresholds provides an additional possible source of mass shifts and widths.

\begin{center}
\noindent\fbox{%
    \parbox{.95\textwidth}{%
    \bigskip
\centering
\textbf{BOEFT multiplet $\bm{4D}$ (quarkonium multiplet $\bm{4D}$), $\bm{J^{PC}=2^{-+},1^{--},2^{--},3^{--}}$}
\\[1em]No unique experimental assignment.\\
\vspace{1em}
}
}
\end{center}

No unique experimental candidate is assigned to the vector member of the BOEFT $4D$ multiplet. The calculated pole is predominantly quarkonium in both
composition prescriptions, with smaller open-bottom components,
see Table~\ref{Table: table of b b-bar probabilites}.

The central $4D$ pole is only about $21~{\rm MeV}$ below the
$H'_1$ hybrid level and is nearly degenerate with the
$H_5$ level, which lies approximately $3~{\rm MeV}$ below it
(see Table~\ref{tab:bbbarhybspectrum}).
Different members of the $4D$ BOEFT multiplet may therefore be
affected by different hybrid multiplets. In particular, the
$1^{--}$ member can mix with the $1^{--}$ member of $H'_1$,
whereas the $2^{-+}$ and $2^{--}$ members have quantum numbers
in common with members of $H_5$.

The proximity of the quoted central masses, therefore, does not imply
a common shift of the entire $4D$ multiplet.

The $\Upsilon(11020)$ is a possible vector candidate in this mass region, but its interpretation is also compatible with the $H'_1$ level and with dynamics associated with the nearby $S+P$ thresholds. We therefore discuss it separately below rather than assigning it uniquely to the $4D$ multiplet.

\begin{center}
\noindent\fbox{%
    \parbox{.95\textwidth}{%
    \bigskip
\centering
\textbf{Non-vector quarkoniumlike candidates}\\[1em]
No established candidates.\\
\vspace{1em}
}
}
\end{center}

The remaining $5P$ and $6P$ bottomoniumlike multiplets contain no $1^{--}$ member (see Fig.~\ref{fig:bb sector with experiments}). There are no experimentally observed states associated with these multiplets. Based on the composition in Table~\ref{Table: table of b b-bar probabilites}, the states in these multiplets correspond to the conventional $4P$ and $5P$ bottomonium states.

The vicinity of several hybrid levels (i.e., $H_2$, $H_2^\prime$, $H_3$) and the proximity of the $B_s^{(\ast)}\bar B_s^{(\ast)}$ and $S+P$ thresholds identify the non-vector sectors as particularly sensitive targets for future calculations of hybrid--quarkonium mixing and coupled-channel dynamics, including the effects of currently omitted thresholds.

\subsubsection {Tetraquark/open-flavor-dominated BOEFT multiplets}
\label{subsubsec:Tetraquark bottomonium}

\begin{center}
\noindent\fbox{%
    \parbox{.95\textwidth}{%
    \bigskip
\centering
\textbf{BOEFT multiplet $\bm{4P}$ ($\bm{X_b}$ and its HQSS partners), $\bm{J^{PC}=1^{+-},0^{++},1^{++},2^{++}}$} \\[1em]
No established candidates.\\
\vspace{1em}
}
}
\end{center}

With the central adjoint-meson calibration adopted in this work,
$\bar\Lambda_{1^{--}}^{\rm HL}=-0.113~\mathrm{GeV}$, the bottomoniumlike BOEFT $4P$
multiplet forms a shallow bound state $233~\mathrm{keV}$ below the common
spin--isospin averaged $B^{(\ast)}\bar B^{(\ast)}$ threshold. The
$J^{PC}=1^{++}$ member of this multiplet is denoted by $X_b$.

The normalized bound-state wavefunction has a bottomonium probability of
approximately $1\%$ and a combined probability of approximately $99\%$ in the tetraquark/open-flavor BO channels. Its root-mean-square heavy-quark separation is $4.9~\mathrm{fm}$, demonstrating that it samples predominantly the large-distance meson--antimeson regime of these BO channels. The $X_b$ therefore has molecular long-distance characteristics, while remaining a state of the complete tetraquark/open-flavor BO channel rather than a separately introduced meson--meson configuration.

The values $E_B=233~\mathrm{keV}$ and
$\sqrt{\langle r^2\rangle}=4.9~\mathrm{fm}$ are the central predictions of the present calculation. The result obtained with
$\bar\Lambda_{1^{--}}^{\rm HL}=-0.134~\mathrm{GeV}$, for which the binding increases to $1.669~\mathrm{MeV}$ and the radius decreases to $2.3~\mathrm{fm}$, is only a sensitivity test and is not an alternative central calibration.

Physical heavy-light spin splittings will resolve the spin-averaged $4P$
multiplet into several thresholds and may move its individual HQSS partners onto different Riemann sheets. Depending on the quantum numbers, a partner may remain bound, become virtual, or appear as a resonance. The spin-averaged binding energy alone, therefore, does not determine the physical pole nature of every member.

No experimental candidate has yet been established for the $X_b$ or for its
$J^{PC}=1^{+-},0^{++},1^{++},2^{++}$ HQSS partners. Given the sparse
experimental bottomoniumlike spectrum, this absence is not problematic. The prediction nevertheless provides a particularly sensitive target for future experiments and for a lattice QCD determination of the lowest
$1^{--}$ adjoint meson mass.

The production cross-section of the $X_b$ has been calculated in 
\cite{Brambilla:2026ujo}.

\begin{center}
\noindent\fbox{%
    \parbox{.95\textwidth}{%
    \bigskip
\centering
\textbf{BOEFT multiplet $\bm{5S}$ ($Y_b$ and its HQSS partner), 
$\bm{J^{PC}= 0^{-+}, 1^{--}}$}
\[
\setlength\arraycolsep{20pt}
\def\arraystretch{1.4}
\begin{array}{lll}
\arraycolsep=1.4pt
1^{--}\to\Upsilon(10753): & M=10756.6\pm2.8\text{MeV} & \Gamma=29\pm9\text{ MeV.}\\
\end{array}
\]
}
}
\end{center}
The BOEFT $5S$ resonance is predominantly tetraquark with quarkonium normalized composition measure of approximately $25\%$, and combined
tetraquark/open-flavor normalized composition measure of approximately
$75\%$. Its central pole mass is $10676~{\rm MeV}$ and its pole width into the included spin-averaged, nonstrange $S+S$ channels is $53~{\rm MeV}$. We label the $1^{--}$ member of the multiplet as $Y_b$. Since this multiplet lies close to the $H_1$ hybrid level as well as the BOEFT $3D$ multiplet, a unique experimental candidate assignment with this multiplet is necessarily less direct compared to other multiplets.

The $\Upsilon(10753)$ state, which 
was seen by Belle in $e^+e^-\rightarrow \Upsilon(nS)\pi^+\pi^-$ 
\cite{Belle:2019cbt}, can be tentatively associated with the
spin-triplet $1^{--}$ member of the $5S$ multiplet. 
Its measured mass is approximately $81~{\rm MeV}$ above the calculated spin-averaged pole. The calculated and experimental widths are of the same order, but quantitative compatibility has not been established. Physical thresholds and the additional mixing and decay mechanisms omitted here must be assessed.
The measured mass of the $\Upsilon(10753)$ also lies close to the hybrid $H_1$ level as well as in  the broader mass region of the BOEFT $3D$ and $5S$ multiplets. Its proximity to these levels indicates that mixing with hybrid and quarkonium may be relevant.

In Ref.~\cite{TarrusCastella:2021pld}, under an unmixed $H_1$ interpretation, the $\Upsilon(10753)\to\Upsilon(1S)\pi^+\pi^-$ transition was predicted to show a pronounced structure near the $f_0$ mass. The absence of such a pronounced
structure in the Belle~II data weakens this simple unmixed interpretation, while leaving the tetraquark interpretation viable~\cite{BelleII:2024mjm}.  It does not exclude hybrid admixture, since the physical state may contain quarkonium and open-flavor components omitted in that prediction.

The $\Upsilon(10753)$ also lies close to hidden-strange open-bottom
thresholds. The $B_s\bar B_s$ channel couples to a $1^{--}$ state in a
relative $P$ wave, which suppresses its near-threshold decay rate, but
hidden-strange channels may still generate mass shifts and contribute to the physical pole structure. A quantitative interpretation, therefore, requires hybrid--quarkonium mixing, hybrid--tetraquark mixing, and the relevant hidden-strange BO channels.

Additionally, Belle II recently observed a rapid enhancement in the $e^+e^-\rightarrow B^\ast \bar{B}^\ast$ cross-section around $4$~MeV above the $B^\ast \bar{B}^\ast$ threshold, corresponding to a mass of around $10653$ MeV \cite{Belle-II:2024niz, Belle-II:2025ubm}. This unconfirmed state is expected to have a very small width, as its decay into the $B^{\left(*\right)}\bar{B}^{\left(*\right)}$ threshold proceeds only through a relative $P$-wave.  A connection between this enhancement and the vector member of our $5S$ multiplet is a possibility for a physical-threshold amplitude analysis, rather than an identification from the spin-averaged pole mass alone. In this scenario, $\Upsilon(10753)$ may instead be interpreted as the possible candidate for the spin-triplet $1^{--}$ member of the $H_1\left(10749\right)$ hybrid multiplet.

\subsubsection {Hybrid BOEFT multiplets}
\label{subsubsec:Hybrid bottomonium}

\begin{center}
\noindent\fbox{%
    \parbox{.95\textwidth}{%
    \bigskip
\centering
\textbf{Hybrid multiplet $\bm{H_1(10749)}$, $\bm{J^{PC}=1^{--},0^{-+},1^{-+},2^{-+}}$}
\\[1em]No unique experimental assignment.\\
\vspace{1em}
}
}
\end{center}

There is no unique assignment for the $1^{--}$ state of the $H_1$ multiplet. Both the $\Upsilon(10753)$ and the $\Upsilon(10860)$ lie nearby, but may also be assigned to the $5S$ and $6S$ BOEFT multiplets. The vicinity of this level to the predominantly quarkonium $3D$ and $6S$ multiplets, as well as to the predominantly open-flavor $5S$ multiplet, hints at the importance of considering hybrid-quarkonium and hybrid-tetraquark mixing mechanisms to more accurately determine the mass of this level.
Moreover, the vicinity of the $B_s^{(\ast)}\bar B_s^{(\ast)}$ threshold provides an additional channel that may further modify the pole location associated with the level.

\begin{center}
\noindent\fbox{%
    \parbox{.95\textwidth}{%
    \bigskip
\centering
\textbf{Hybrid multiplet $\bm{H_1^\prime(10940)}$,  $\bm{J^{PC}=1^{--},0^{-+},1^{-+},2^{-+}}$}
\\[1em] No unique experimental assignment.\\
}
}
\end{center}

No experimental state is assigned uniquely to the state of the $H'_1$ multiplet.
The $\Upsilon(11020)$ lies nearby and is a possible candidate, but it is also compatible with the vector member of the predominantly quarkonium $4D$ multiplet and with dynamics associated with the first $S+P$ open-bottom thresholds.

The central $H'_1$ level lies only about
$21~{\rm MeV}$ above the predominantly quarkonium $4D$ pole.
The $\Upsilon(11020)$ lies about $60~{\rm MeV}$ above
the $H'_1$ central value and about $81~{\rm MeV}$ above the $4D$
pole. Its mass is therefore compatible with both theoretical
regions once the corresponding uncertainties and omitted mixing
effects are considered. The close $4D$--$H'_1$ spacing makes a
mixed quarkonium--hybrid interpretation particularly plausible.

The separate semi-inclusive decay analysis of
Ref.~\cite{Brambilla:2022hhi} gives
\begin{eqnarray}
    \Gamma_{\rm semi-inc.}\left(H'_1[1^{--}]\right)
 =23^{+11}_{-7}~\mathrm{MeV},
 \qquad
 \Gamma_{\rm semi-inc.}\left(H'_1[(0,1,2)^{-+}]\right)
 =48^{+31}_{-19}~\mathrm{MeV}.
\end{eqnarray}

These are lower bounds for unmixed hybrid states in that analysis, rather than physical pole widths predicted by the present hybrid spectrum.

\begin{center}
\noindent\fbox{%
    \parbox{.95\textwidth}{%
    \bigskip
\centering
\textbf{Non-vector hybrid states}
\[
\setlength\arraycolsep{20pt}
\def\arraystretch{1.4}
\begin{array}{lll}
\arraycolsep=1.4pt
\text{No established candidates.} \\
\end{array}
\]
}
}
\end{center}

The remaining $H_2$, $H_3$, $H_4$, and $H_5$ bottomonium hybrid
multiplets contain no $1^{--}$ member (see Table~\ref{tab:bbbarhybspectrum}). Since the experimentally established
bottomoniumlike candidates in this interval are presently all vectors, no
direct experimental assignments can be made for these multiplets. Their
overlap with the predicted non-vector quarkonium multiplets nevertheless marks regions in which omitted mixing may become important once the corresponding experimental sectors are explored.

We also obtain several non-vector hybrid levels close to quarkonium multiplets. The $H_2$ level lies approximately $35~{\rm MeV}$ above the predominantly quarkonium $5P$ pole, while $H'_2$ and $H_3$ lie approximately $20~{\rm MeV}$ and $6~{\rm MeV}$ above the $6P$ pole, respectively. Most notably, $H_5$ is nearly degenerate with the $4D$ multiplet. These near-degeneracies identify the corresponding non-vector sectors as particularly sensitive targets for future calculations of
hybrid--quarkonium mixing.

\subsubsection {Experimental candidate not uniquely accommodated by the included channels}
\label{subsubsec:bbbarOther}

\begin{center}
\noindent\fbox{%
    \parbox{.95\textwidth}{%
\[
\setlength\arraycolsep{20pt}
\def\arraystretch{1.4}
\begin{array}{llll}
\arraycolsep=1.4pt
\bm{\Upsilon(11020)}: & J^{PC}=1^{--}, & M= 11000\pm 4\text{ MeV,} & \Gamma = 24^{+8}_{-6} \text{ MeV.}
\end{array}
\]
}
}
\end{center}

The $\Upsilon(11020)$ is not assigned uniquely within the present
spin-averaged calculation. Its mass lies near the $H'_1$ hybrid 
level, above the predominantly quarkonium $4D$ pole, and close to the first spin-averaged $B^{(\ast)}\bar B_J^{(\ast)}$ thresholds; see
Fig.~\ref{fig:bb sector with hybrids with experiments}.

A conventional $4\,{}^3D_1$ interpretation is possible once the expected
spin-dependent and relativistic mass corrections are included. A state with substantial hybrid content is also plausible because of the nearby $H'_1$ level, although the semi-inclusive quarkonium-decay analysis of Ref.~\cite{Brambilla:2022hhi} disfavors a simple pure-hybrid interpretation.
Finally, tetraquark BO potentials approaching the $S+P$ thresholds may
generate an additional pole or mix strongly with the nearby quarkonium and
hybrid configurations.

The $\Upsilon(11020)$ should therefore be regarded as an ambiguous candidate for a mixed quarkonium--hybrid--open-flavor state. Its presence in this region points to the need for the omitted $S+P$ threshold channels and hybrid mixing potentials rather than selecting a unique assignment within the present truncated calculation. Its measured total width should not be compared directly with any one of the calculated nonstrange $S+S$ pole widths.

\section{Comparison with lattice QCD and related BO studies
}
\label{sec:Comparison with literature}

In this section, we compare the qualitative organization of the
BOEFT spectrum with previous lattice QCD calculations and with
related coupled-channel studies based on Born--Oppenheimer
potentials. These comparisons are not one-to-one. The calculations
differ in their light-quark masses, operator bases, finite-volume
treatment, included thresholds, heavy-quark spin dependence, and
definitions of resonance composition. We therefore focus on common
spectroscopic patterns and on the origin of significant differences,
rather than on precise level-by-level agreement.

\subsection{Lattice QCD}\label{subsec:Comparison with lattice}

The spectroscopy calculations of
Refs.~\cite{HadronSpectrum:2012gic,Cheung:2016bym,Ryan:2020iog}
used large bases of predominantly single-hadron quarkonium and
hybrid interpolating operators, without explicit open-flavor
meson--antimeson operators. They were designed primarily to resolve
quarkonium and hybrid excitations rather than to determine
near-threshold coupled-channel scattering amplitudes.

The charmonium calculations of
Refs.~\cite{HadronSpectrum:2012gic,Cheung:2016bym} find the
lowest hybrid multiplet approximately $1.2$--$1.3~\mathrm{GeV}$
above the lowest charmonium levels. Their overall organization of
the quarkonium and hybrid spectra is qualitatively compatible with
the corresponding quarkonium-dominated multiplets and hybrid levels obtained in BOEFT. They also place the vector
$3S$ charmonium excitation above the experimental $\psi(4040)$,
similarly to the central value of our spin-averaged $3S$ pole.

Reference~\cite{Ryan:2020iog} finds the lowest bottomonium hybrid
multiplet approximately $1.5~\mathrm{GeV}$ above the bottomonium ground
state and an overall pattern broadly compatible with our
bottomonium-hybrid levels. No clear levels corresponding
to some of our higher $S$-wave BOEFT multiplets are resolved in that
calculation.

None of these single-hadron spectroscopy calculations resolves a
level that can be directly associated with the shallow,
tetraquark/open-flavor-dominated BOEFT multiplets. This absence does
not by itself exclude such states. It may reflect their weak overlap
with the adopted operator bases and the absence of an explicit
finite-volume meson--antimeson scattering analysis. Conversely, an
enlarged operator basis cannot create additional physical levels;
it permits finite-volume eigenstates with previously weak overlaps
to be resolved more reliably.

The $J^{PC}=1^{++}$ charmoniumlike sector was studied with bases
containing both $c\bar c$ and open-charm meson--antimeson
interpolators in
Refs.~\cite{Prelovsek:2013cra,Padmanath:2015pkj}. A finite-volume level associated with a shallow bound-state pole was found
approximately $10~\mathrm{MeV}$ below the lattice
$D\bar D^\ast$ threshold and interpreted as a candidate for the
$\chi_{c1}(3872)$. The enlarged operator-basis analysis of
Ref.~\cite{Padmanath:2015pkj} again found this candidate when both
$c\bar c$ and $D\bar D^\ast$ operators were included. The existence
of a near-threshold state with important open-charm dynamics is
qualitatively consistent with the BOEFT $2P$ multiplet.

The low-momentum lattice amplitude gives
$a_0=-1.7\pm0.4~\mathrm{fm}$ and
$r_0=0.5\pm0.1~\mathrm{fm}$ in the convention of
Ref.~\cite{Prelovsek:2013cra}.\footnote{%
Ref.~\cite{Prelovsek:2013cra} adopts the opposite convention for the sign of the scattering length. Its result should therefore be
compared with $a_0=1.7\pm0.4~\mathrm{fm}$ in the convention used in
Eq.~\eqref{eq:scattering_length}.}
These values should not be compared directly with
Eqs.~\eqref{eq:scattering_length} and
\eqref{eq:effective_range} as if the two calculations described the
same threshold problem. The lattice state is bound by several MeV, the calculation uses an unphysical light-quark mass and a physical
$D\bar D^\ast$ channel, whereas the present leading-order BOEFT
calculation considers a spin-isospin averaged threshold and a binding energy of $96~\mathrm{keV}$. The substantially larger scattering length in
BOEFT primarily reflects its much smaller binding momentum, while
differences in the effective range may also receive finite-range,
spin-dependent, and quark-mass corrections.

\begin{figure}
\centering 
\includegraphics*[width=13.5cm,clip=true]{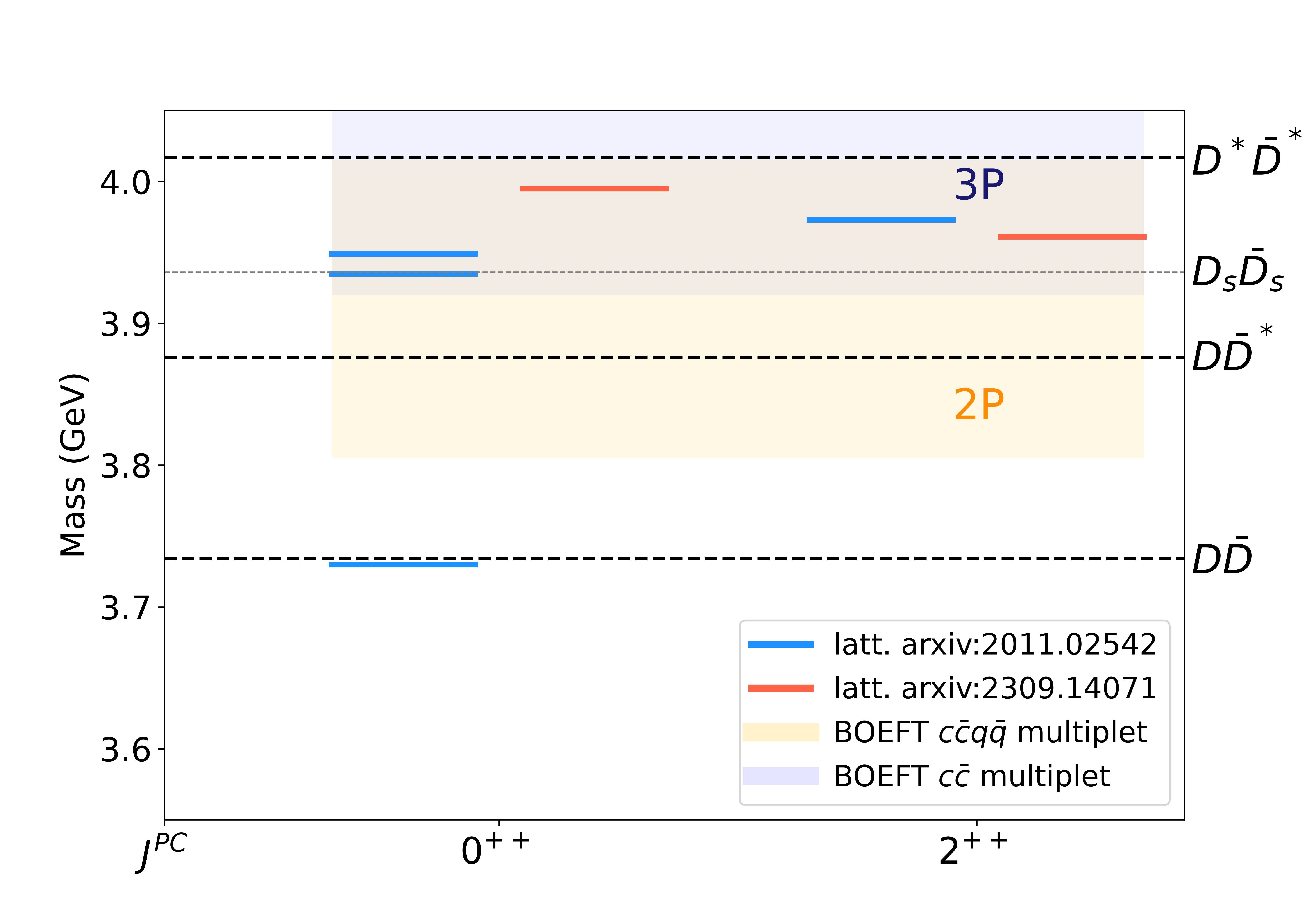}
\caption{Comparison of the lattice QCD results of
Refs.~\cite{Prelovsek:2020eiw,Wilson:2023anv} with the corresponding
spin-averaged BOEFT $2P$ and $3P$ multiplets in the
$J^{PC}=0^{++}$ and $2^{++}$ sectors in the energy region between 3.6 GeV and 4.1 GeV.
The comparison is qualitative: the lattice calculations employ physical open-flavor threshold channels, unphysical
light-quark masses, and different coupled-channel bases of interpolating operators, whereas the BOEFT bands represent leading-order HQSS predictions.}
\label{fig:lattice comparison}
\end{figure}

In Ref.~\cite{Prelovsek:2020eiw}, the charmoniumlike spectra with $J^{PC}=0^{++}$ and $2^{++}$ have been computed using lattice QCD with a basis of interpolating operators that included quarkonium and open-flavor meson-antimeson pairs; see Fig.~\ref{fig:lattice comparison}.
Reference~\cite{Prelovsek:2020eiw} calculates three $J^{PC}=0^{++}$ charmoniumlike states in the energy region between $3.6$ GeV and $4.1$ GeV, compared to only two predicted states in our leading-order BOEFT calculation.
The first, dubbed  $\chi_{c0}^{D\bar D}$, is located a few MeV below the $D\bar D$ threshold.
The second, dubbed $\chi_{c0}^{D_s\bar D_s}$, is a narrow resonance just below the $D_s\bar D_s$ threshold.
The third, $\chi_{c0}^\prime$, is a broad resonance above the $D_s\bar D_s$ threshold, which has a strong coupling to the $D \bar{D}$ threshold.
These states should be compared to our predicted $2P$ and $3P$ states in the same energy range.
The difference in the number of predicted states could be attributed to the lack of charm-strange meson-antimeson channels in our calculation, which would be necessary to describe a possible hidden-charm--hidden-strange tetraquark BOEFT multiplet.
Reference~\cite{Prelovsek:2020eiw} also calculates a single $J^{PC}=2^{++}$ charmoniumlike state in the energy region between $3.6$ GeV and $4.1$ GeV, compared to the two predicted states in our leading-order BOEFT calculation.
This state can be associated with either one of our $2P$ or $3P$ BOEFT multiplets.

A different picture emerges from
Ref.~\cite{Wilson:2023anv}, which performed a coupled-channel
analysis of the charmonium $0^{++}$ and $2^{++}$ sectors and found
one prominent resonance in each channel. The finite-volume levels
associated with these poles have substantial overlap with
$c\bar c$ interpolators, and the resulting resonances were
interpreted as predominantly conventional charmonium. Their masses
are qualitatively compatible with the quarkonium-dominated BOEFT
$3P$ multiplet. No additional pole clearly associated with the
threshold-dominated BOEFT $2P$ multiplet was resolved.

More recently, Ref.~\cite{Wilson:2026bhu} studied near-threshold
$D\bar D$ scattering as a function of the light-quark mass, and found
weak interactions, with no additional bound-state or resonance
singularity between the deeply bound $\chi_{c0}(1P)$ level and the
$D_s\bar D_s$ threshold in the setup considered there. This result
is in tension with the additional scalar poles reported in
Ref.~\cite{Prelovsek:2020eiw} and illustrates that the lattice description of this region is not yet uniform.

Taken together, the lattice studies support the conventional
quarkonium and hybrid organization over much of the spectrum. The number and nature of additional $0^{++}$ and $2^{++}$ poles remain
calculation dependent.

\subsection{Related coupled--channel Born--Oppenheimer studies }\label{subsec:Comparison with BO}

References~\cite{Bicudo:2019ymo,Bicudo:2020qhp,Bicudo:2022ihz}
studied the bottomoniumlike spectrum using coupled radial
Schr\"odinger equations containing a confining quarkonium potential,
spin-averaged meson-pair channels, and a mixing potential extracted
from lattice static energies. Their coupled-channel structure is
closely related to Eqs.~\eqref{eq:diabatic},~\eqref{eq:diabaticl0}, but the potential inputs
differ substantially from those used here.

In those studies, the meson-pair diabatic potentials are represented
by constant threshold energies. They therefore do not implement the
short-distance repulsive color-octet behavior of the tetraquark BO
potentials and effectively take the $\Sigma_g^{+\prime}$ and
$\Pi_g$ channels to be degenerate. Their quarkonium--meson-pair
mixing potential is also stronger and concentrated at shorter distances than the mixing adopted in the present calculation, which
is constrained by the more recent string-breaking data used in
Section~\ref{subsec:parametrization}.

The following comparison focuses on the above-threshold spectrum.
The corresponding below-threshold comparison is discussed in
Ref.~\cite{Brambilla:2026zfw}.

References~\cite{Bicudo:2019ymo,Bicudo:2020qhp} find an additional
dynamically generated $S$-wave resonance near $10774~\mathrm{MeV}$ that may correspond to the $5S$ pole in the present BOEFT spectrum, although the mass and width of that resonance are markedly larger than those of the $5S$ state in the present calculation. They also obtain a higher $S$-wave bottomonium resonance with a mass comparable to our
$6S$ pole, but with an open-flavor width of about $22~\mathrm{MeV}$,
compared with approximately $2~\mathrm{MeV}$ in the present
calculation.

These differences may be attributed to the stronger mixing potential used in Ref.~\cite{Bicudo:2019ymo,Bicudo:2020qhp} with respect to our analysis, constrained from the lattice data of Ref.~\cite{Bali:2005fu}.
\footnote{Moreover, Ref.~\cite{Bicudo:2019ymo} noted that the two-meson creation operator used in Ref.~\cite{Bali:2005fu} probes not only the $\Sigma_g^+$ quantum numbers relevant to the quarkonium-tetraquark mixing but also the $\Pi_g$ and $\Sigma_u^-$ quantum numbers. This study resampled the correlation functions of Ref.~\cite{Bali:2005fu} and found a mixing potential that is approximately twice the value of the string-breaking rate calculated in Ref.~\cite{Bali:2005fu}. The same kind of two-meson creation operator was used also in Refs.~\cite{Bulava:2019iut,Bulava:2024jpj}, but it is unclear at this time if the resampling proposed in Ref.~\cite{Bicudo:2019ymo} applies to these references as well.}
Other factors for these differences may be related to the constant-threshold approximation, differences in the heavy-quark masses and potential normalization, and the inclusion of the hidden-strange threshold channels in their calculations. 

Reference~\cite{Bicudo:2022ihz} extends these calculations to the $P$-, $D$-, and $F$-wave bottomonium spectra, accounting for the coupling to spin-isospin averaged open-flavor thresholds.
Our predictions for the spectrum are qualitatively compatible with those of Ref.~\cite{Bicudo:2022ihz}.
Note that the inclusion in Ref.~\cite{Bicudo:2022ihz} of the hidden-bottom, hidden-strange meson-antimeson channel, which is absent from our calculation, has a very significant impact on the calculated spectrum, shifting in general by a few tens of MeV both the pole masses and widths of the resonant states.

Our finding that quarkoniumlike states far above threshold are dominated by their conventional quarkonium component is in contrast with Refs.~\cite{Bicudo:2020qhp,Bicudo:2022ihz}, which report dominant tetraquark probabilities for most of the same states.
Part of this difference originates from the stronger and different mixing potential adopted in those works.
A more fundamental source, however, is the definition of the composition itself. In Refs.~\cite{Bicudo:2020qhp,Bicudo:2022ihz}, the quarkonium and tetraquark probabilities are obtained by integrating the modulus squared of the radial wavefunction, evaluated at the real part of the complex pole energy, out to a fixed radius $R_\text{max}=2.4$~fm.
For a state above threshold, this prescription is cutoff-dependent and is not an analytic continuation of the bound-state probability, so its value reflects in part the arbitrary choice of $R_\text{max}$.
By contrast, the complex scaling and $K$ matrix definitions employed here reduce to the standard bound-state probabilities in the appropriate limit and require no spatial cutoff. The residual scheme dependence is confined to the well-understood ambiguity in the resonances normalization discussed in Section~\ref{sec:probabilities}.
We therefore regard the present quarkonium-dominance of the far-from-threshold states as a more robust statement about their internal structure.

Finally, it is worth commenting on the results of a recent study that calculated charmoniumlike resonances above threshold in the BO approximation with nonperturbative spin-dependent effects \cite{Alasiri:2026vfz}.
This study proposed a model for the tetraquark spin-dependent potential inside the Schr\"odinger equation, and considered different potential parameters that in all cases produced a $1^{++}$ tetraquark bound state just below the $D^\ast\bar{D}$ threshold.
In all of the models considered in Ref.~\cite{Alasiri:2026vfz}, the $0^{++}$ state in the tetraquark multiplet was a virtual state.
In the models where the coupling between quarkonium and tetraquarks was not included, the $1^{+-}$ state was either a virtual state or a resonance, and the $2^{++}$ state was a resonance.
In the models where the coupling between quarkonium and tetraquarks was included, the $1^{+-}$ state was a virtual state, and the $2^{++}$ state was either a virtual state or a resonance.
These results, which cannot be directly compared to the leading-order BOEFT calculation performed here, motivate us to further investigate spin-dependent effects for above threshold quarkoniumlike resonances in future studies\footnote{The quarkoniumlike spectrum below the spin-isospin-averaged $M^{(\ast)}\bar{M}^{(\ast)}$ threshold, accounting for the $1/m_Q$ tetraquark spin-dependent potential inside the Schr\"odinger equation, has been computed in~\cite{Brambilla:2026zfw} within the BOEFT framework in the charmoniumlike and bottomoniumlike sectors.}.

\section{Conclusions}

\label{sec:conclusions}

In this work, we have investigated the isoscalar hidden-charm and
hidden-bottom spectra in the energy region between the lowest
spin--isospin averaged nonstrange $S+S$ and $S+P$ open-flavor
thresholds within Born--Oppenheimer effective field theory
(BOEFT)~\cite{Brambilla:2017uyf,Berwein:2024ztx}. At leading order in
the heavy-quark expansion, the quarkonium static potential mixes,
through the string-breaking interaction constrained by lattice QCD~\cite{Bali:2005fu,Bulava:2019iut,Bulava:2024jpj}, with the lowest
isoscalar tetraquark/open-flavor BO potentials carrying the same BO
quantum numbers. These potentials are constrained by QCD symmetries,
their short- and long-distance behavior, and the available lattice
input. The only parameter calibrated to experimental spectroscopy is
the lowest $1^{--}$ adjoint meson mass. In the mass and threshold
convention of the present leading-order calculation, its central value
is $\bar\Lambda_{1^{--}}^{\rm HL}=-0.113~\mathrm{GeV}$, fixed so that the
spin-averaged $2P$ multiplet associated with the $\chi_{c1}(3872)$
lies $96~\mathrm{keV}$ below the common spin--isospin averaged
$D^{(\ast)}\bar D^{(\ast)}$ threshold.

We have determined the bound states and resonance poles using
complementary $T$ matrix, $K$ matrix, and complex scaling methods. The agreement between the analytically continued $T$ matrix poles and the complex-scaled eigenvalues provides a nontrivial numerical validation
of the pole extraction, while the $K$ matrix gives a complementary
real-axis characterization. The calculated phase shifts,
inelasticities, and Argand trajectories provide a further
representation of the same pole spectrum. In particular, the
charmoniumlike $S$-wave amplitude displays the interference directly
between the shallow open-flavor-dominated $2P$ bound state and the
broader quarkonium-dominated $3P$ resonance, showing that both arise
from the same coupled dynamics.

The central result is a global heavy-quark-spin-symmetry (HQSS) multiplet organization of the spectrum rather than a collection of independent state-by-state predictions. Between the two spin-averaged thresholds, we obtain the BOEFT charmoniumlike resonance multiplets $3P$, $3S$, $2D$, and $4P$, and the BOEFT bottomoniumlike resonance multiplets $5S$, $3D$, $5P$, $6S$, $4D$, and $6P$. The two prescriptions used to characterize their channel content show that most of these poles are associated with predominantly quarkonium states, with the largest open-flavor components occurring for the resonances closest to threshold. Thus, a recognizable sequence of comparatively small radius, quarkonium-dominated multiplets persists above the first open-flavor threshold, while threshold dynamics produces its largest distortions only for selected nearby states.

Strikingly, the same coupled equations also generate exceptional
shallow and spatially extended states. For the central calibration,
the spin-averaged charmoniumlike $2P$ multiplet associated with the
$\chi_{c1}(3872)$ has a binding energy of $96~\mathrm{keV}$, a
root-mean-square heavy-quark separation of $10.8~\mathrm{fm}$, and a
quarkonium probability of approximately $3\%$. The quarkonium component of the state is essential to reproduce the updated value of the $\chi_{c1}(3872)$ radiative decays ratio $\mathcal{R}_{\gamma \psi} = \Gamma_{\chi_{c1}(3872) \rightarrow \gamma \psi(2S)}/\Gamma_{\chi_{c1}(3872) \rightarrow \gamma J/\psi}$, reported by LHCb in~\cite{LHCb:2024tpv}, under the assumption that the radiative decays proceed through this component~\cite{Brambilla:2024imu}.
Subsequent studies, which instead identified the $\chi_{c1}(3872)$ with the conventional $\chi_{c1}(2P)$ quarkonium state, combined an effective field theory approach to treat the charmonium radiative decays with heavy-quark-spin-symmetry to provide a prediction for $\mathcal{R}_{\gamma \psi}$ compatible with the LHCb result~\cite{Colangelo:2025uhs}. 

Its bottomoniumlike
$4P$ counterpart containing the predicted $X_b$ has a binding energy
of $233~\mathrm{keV}$, a root-mean-square separation of
$4.9~\mathrm{fm}$, and a bottomonium probability of approximately
$1\%$. Their wavefunctions sample predominantly the large-distance
meson--antimeson regime of the tetraquark/open-flavor BO channels and
therefore have molecular long-distance characteristics. These states
are nevertheless solutions of complete BO channels that interpolate
from short-distance color-octet $Q\bar Q$ sources dressed by light
degrees of freedom to heavy-light meson--antimeson thresholds at large
distance~\cite{Berwein:2024ztx,Braaten:2024tbm,Brambilla:2026zfw}.
Compact-tetraquark and molecular characteristics consequently describe
different spatial regimes of the same BO channel rather than two
permanently separate microscopic configurations.

To expose the near-critical sensitivity of these shallow states, we
have also repeated the calculation with
$\bar\Lambda_{1^{--}}^{\rm HL}=-0.134~\mathrm{GeV}$, the value used in the previous below-threshold study~\cite{Brambilla:2026zfw}. In the equations and threshold convention of the present work, this is only a sensitivity test and not an alternative central calibration or an endpoint of an uncertainty interval. It increases the binding of the charmoniumlike $2P$ multiplet from $96$ to $764~\mathrm{keV}$, reduces its radius from $10.8$ to $4.2~\mathrm{fm}$, and increases its quarkonium
probability from $3\%$ to $7\%$. For the $X_b$ multiplet, the binding
increases from $233~\mathrm{keV}$ to $1.669~\mathrm{MeV}$, the radius
decreases from $4.9$ to $2.3~\mathrm{fm}$, and the bottomonium
probability changes from $1\%$ to $2\%$. The comparison, therefore,
shows how the properties of a shallow bound state change when the binding is increased from approximately $0.1$ to
$0.8~\mathrm{MeV}$. The dominant open-flavor character of both
shallow multiplets is stable, while the ordering, multiplet
organization, and dominant channel content of the higher predominantly quarkonium spectrum are even less sensitive. The strong response of the shallow poles is thus a consequence of their proximity to the critical condition for binding, not an instability of the complete BOEFT spectrum.
Despite the high sensitivity of the properties of these shallow bound states from the specific value of the adjoint meson mass, the size of these states remains considerably larger in comparison to their quarkonium-dominated equivalents, whose typical radii do not exceed $1-2~\mathrm{fm}$.

In addition to the shallow bound multiplets, the coupled equations predict an additional $l = 0$ bottomoniumlike resonance with $M = 10676$~MeV and $\Gamma=53$~MeV, with a substantial tetraquark/open-flavor content according to the complex-scaled channel measures.
The complex pole-size diagnostic indicates a natural hadronic scale, while the \(P\)-wave phase shift provides its real-axis characterization; the resonance remains above threshold and open-flavor dominated at both tested adjoint-meson parameter values.
It has no identified real-axis $K$-matrix-pole counterpart.
Its $1^{--}$ member, denoted by $Y_b$, has a degenerate $0^{-+}$ partner at leading order.
It extends the spectrum beyond the sequence of predominantly quarkonium poles.
Its experimental identification requires the physical spin-resolved thresholds and the additional interactions omitted from the present calculation.

For bound states, the components of the normalized wavefunction have
the ordinary probability interpretation. For resonances, by contrast,
there is no unique probabilistic compositeness. The components obtained
from complex scaling are generally complex, whereas those obtained
from the $K$ matrix prescription are real but may lie outside the
interval $[0,1]$. 
The quantities $P_i$ introduced in this work are
therefore normalized composition measures, not resonance
probabilities. The two prescriptions nevertheless agree on the
dominant-channel classification, when both applicable: most higher resonances are
quarkonium dominated, while the largest open-flavor contributions
occur closest to threshold. Their numerical differences quantify part
of the prescription dependence intrinsic to assigning a composition
to a resonance.

The normalized pole couplings provide distinct information: they
characterize the relative residue strengths with which a pole couples
to the asymptotically open channels. For narrow and isolated
resonances, they may approximate branching fraction ratios only within
the truncated set of included channels and should not be identified
with physical branching fractions. Likewise, the widths reported here
are only the pole widths generated by the included spin-averaged,
nonstrange $S+S$ channels. They are not predictions for physical total
widths, which may also receive contributions from physical heavy-light
spin splittings, hidden-strange and $S+P$ open-flavor channels,
hadronic transitions, electromagnetic decays, and hybrid-induced
channels. In addition to the corrections associated with the omitted physical
channels and spin-dependent interactions, the quoted central results retain a model dependence from the presently unknown $V_{\Pi_g}$ potential and from the modeled short-distance part of
$V_{\Sigma_g^{+\prime}}$. This dependence is expected to be small for strongly quarkonium-dominated poles, but it may affect the masses,
widths, normalized pole couplings, and relative partial-wave composition of states with sizable open-flavor components.

The effective range analysis provides an additional characterization
of the shallow charmoniumlike state. The calculated scattering length
is $a_0=15.2~\mathrm{fm}$, and the effective range is
$r_0=1.5~\mathrm{fm}$.
The binding momentum extracted from these scattering parameters agrees with that obtained directly from
the coupled Schr\"odinger equation. Inverting the leading weak-binding relation while neglecting finite-range corrections gives
$\widetilde Z=-0.12$, whereas the direct BOEFT wavefunction gives
$Z_{Q\bar Q}=0.03$. The negative estimator does not indicate an
unphysical state. Rather, it shows that the leading weak-binding
relation~\cite{Weinberg:1965zz,vanKolck:2022lqz} cannot be inverted
quantitatively, when natural finite-range and coupled-channel corrections are omitted. In the present case, the nominal leading
contribution to the effective range is suppressed by the small factor
$Z_{Q\bar Q}/(1-Z_{Q\bar Q})$, allowing natural-range corrections to
dominate even though the state is extremely shallow. The
effective range expansion, therefore, determines the binding momentum
reliably, and the Weinberg relation supports the qualitative
conclusion of a predominantly open-flavor state, but it does not give
a precise model-independent extraction of the small quarkonium
component.

The comparison with the observed hidden-charm and hidden-bottom
spectra should be understood at the level of possible HQSS-multiplet
assignments rather than as a set of unique state-by-state
identifications. Together with the hybrid levels, the coupled quarkonium--open-flavor multiplets provide plausible
associations for essentially all established isoscalar candidates in the energy window considered, or identify the additional dynamics needed for their description. The hybrid levels are not complete
predictions for physical hybrid pole positions, open-flavor widths,
pole couplings, or hybrid--tetraquark--quarkonium compositions: the
leading hybrid--tetraquark mixings and the $1/m_Q$-suppressed
hybrid--quarkonium mixing are not included
here~\cite{Berwein:2015vca,Oncala:2017hop,TarrusCastella:2024zps,
Bruschini:2023tmm}. Their proximity to quarkonium or open-flavor
multiplets instead mark regions where the omitted mixing may be
important.

Earlier coupled BO analyses addressed substantial portions of the
isoscalar bottomonium spectrum
\cite{Bicudo:2019ymo,Bicudo:2020qhp,Bicudo:2022ihz}. To our
knowledge, the present work is the first QCD-constrained analysis to
attempt, within a single dynamical and HQSS-multiplet framework, a
simultaneous organization of essentially all established isoscalar
hidden-charm and hidden-bottom candidates across the full energy
interval between the spin--isospin averaged nonstrange $S+S$ and
$S+P$ open-flavor thresholds. This statement concerns the global
multiplet organization, not unique assignments for every individual
candidate: where the present channel basis does not naturally
accommodate a state, the comparison identifies the additional BO
sectors and mixing potentials likely required.

The remaining candidates provide specific information about this
missing dynamics. States correlated with heavy-strange thresholds,
such as the $\chi_{c1}(4140)$, point toward hidden-strange
tetraquark/open-flavor BO channels. States in the region of the first
$S+P$ thresholds, notably the $\psi(4230)$ and $\psi(4360)$, indicate
the need for tetraquark BO potentials approaching $S+P$ heavy-meson
pairs, together with their mixing with nearby quarkonium and hybrid
configurations. Molecular interpretations proposed for the
$\chi_{c1}(4140)$
\cite{Liu:2009ei,Ding:2009vd,Karliner:2016ith} and for the
$\psi(4230)$
\cite{Wang:2013cya,Ji:2022blw,Peng:2022nrj,
vonDetten:2024eie,Dong:2026fdi} are naturally compatible with this
BOEFT perspective: once the corresponding BO channels are included,
their static energies approach the relevant meson--antimeson
thresholds at large distances and may dynamically generate extended
states.

The comparison with existing lattice QCD studies is broadly
consistent with the conventional quarkonium and hybrid organization
over much of the spectrum
\cite{HadronSpectrum:2012gic,Cheung:2016bym,Ryan:2020iog} and supports
the existence of a near-threshold $1^{++}$ charmoniumlike state
\cite{Prelovsek:2013cra,Padmanath:2015pkj}. However, the number and nature of additional scalar and tensor poles are not yet uniform among
coupled-channel lattice analyses
\cite{Prelovsek:2020eiw,Wilson:2023anv,Wilson:2026bhu}. The absence of a level in a calculation cannot be attributed uniquely to one omitted
operator type: finite-volume states with weak overlap may not be
resolved by a restricted basis, while explicit meson-pair operators
and controlled finite-volume scattering analyses remain important for
threshold-dominated systems.

\begin{figure*}
\centering
\includegraphics[width=0.95\textwidth]{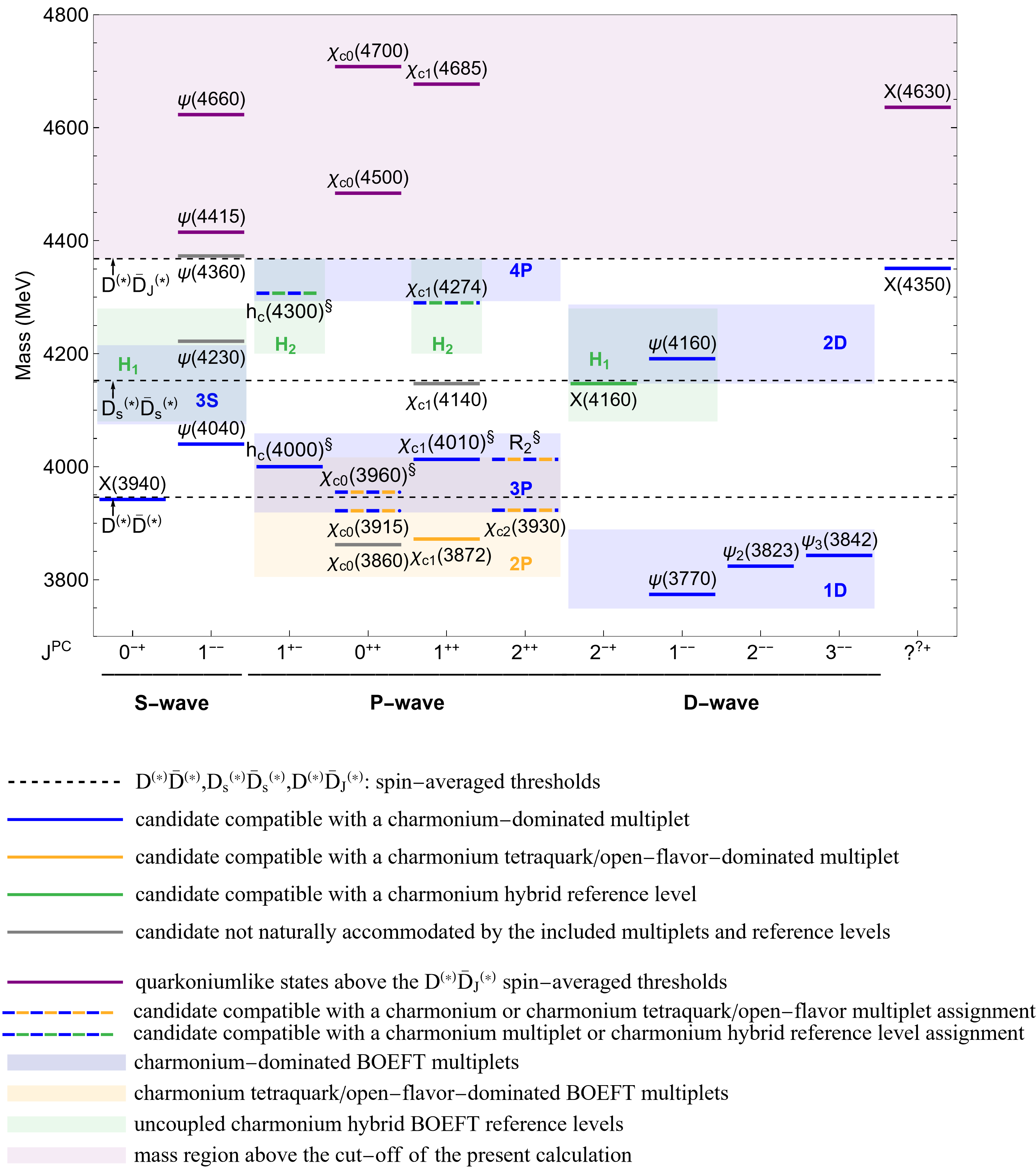}
\caption{\justifying{
Comparison between the charmoniumlike multiplets predicted in BOEFT (shaded boxes) by solving Eqs.~\eqref{eq:diabatic},~\eqref{eq:diabaticl0},~\eqref{eq:diffeq-hyb}, including also the charmonium hybrid levels predictions, and the experimental spectrum (continuous lines) above the $D \bar{D}$ threshold and up to the $D^{(\ast)} \bar{D}_J^{(\ast)}$ spin-averaged thresholds.
The charmoniumlike states listed in PDG and lying above the spin-averaged $D^{(\ast)} \bar{D}_J^{(\ast)}$ spin-isospin averaged thresholds are also displayed.
The description of these states will be addressed within the BOEFT formalism in future works, where the mixing between quarkonium and the $D^{(\ast)} \bar{D}_J^{(\ast)}$ meson-antimeson thresholds, as well as hybrid-quarkonium mixing, are going to be accounted for.
The hybrid spectrum shown here should be regarded as spectroscopic levels in the absence of mixing effects (see Section~\ref{sec:hybrids}), rather than as direct predictions for the physical resonances. 
Uncertainties on the masses are estimated from the relativistic corrections for charmonium-dominated multiplets ($135~{\rm MeV}$), from meson-antimeson spin-splitting corrections for charmonium tetraquark/open-flavor-dominated multiplets ($210~{\rm MeV}$). For the charmonium hybrid reference levels, the displayed band represents the common $\pm 100~\mathrm{MeV}$ uncertainty associated with matching the pure-gauge hybrid static energies to the additive heavy-light scheme used in the present work. Experimental states not listed in the PDG are marked by a ``§'' sign.
}}
\label{fig:cc sector summary}
\end{figure*}

The next systematic improvements should include spin-dependent
interactions and physical heavy-light threshold splittings, whose
BOEFT structure is known
\cite{Oncala:2017hop,Soto:2020xpm,Brambilla:2018pyn,
Brambilla:2019jfi,Soto:2023lbh}, as well as hidden-strange channels
and the tetraquark BO potentials approaching the $S+P$ heavy-meson
thresholds; see Fig.~\ref{fig:cc sector summary}. Their coupling to
quarkonium and hybrid configurations may generate additional
threshold-dominated states and modify nearby pole masses, widths, and
compositions. The most important nonperturbative lattice QCD inputs
identified by the present analysis are the lowest $1^{--}$
adjoint meson mass, the presently unknown $V_{\Pi_g}$ tetraquark
potential, and the mixing potentials connecting hybrid static energies
to the relevant $S+S$ and $S+P$ tetraquark/open-flavor channels.

The principal outcome of this work is that predominantly quarkonium
resonances and exceptional extended exotic states need not be
described through unrelated dynamical models. They arise from the same
QCD-constrained coupled equations, and are organized into a common
HQSS-multiplet structure. The shallow states are highly sensitive to
the adjoint meson mass and to their resulting proximity to the
critical condition for binding, whereas the organization and dominant
channel content of the higher predominantly quarkonium spectrum are considerably more robust.
BOEFT thereby turns the quarkoniumlike spectrum above threshold from a collection of individually interpreted candidates into a systematic
multiplet problem and identifies the additional BO sectors and mixing
potentials needed for a more complete description.

\section*{Acknowledgements}
We acknowledge the DFG cluster of excellence ORIGINS funded by the Deutsche Forschungsgemeinschaft under Germany’s Excellence Strategy-EXC-2094-390783311. N. B. acknowledges the ERC Advanced Grant ERC-2023-ADG-Project EFT-XYZ. R. B. and F.-Z. P. acknowledge sponsorship from the Alexander von Humboldt Foundation. A. M. acknowledges support from the Physical Research Laboratory,
Ahmedabad and the Department of Space (DOS), Government of India. We thank Christoph Hanhart, Jakob Hoffmann, Tetsuo Hyodo, Roberto Mussa, Marco Scodeggio, Christopher Thomas, and Antonio Vairo for helpful discussions. 

\appendix

\section{Unitary transformation to the BO diabatic basis}\label{app:changeofbasis}
After simplifying the angular part of the quarkonium and tetraquark wavefunctions as in~\cite{Berwein:2024ztx}, the coupled Schr\"odinger equation accounting for the mixing between the $V_{\Sigma_g^+}$, $V_{\Sigma_g^{+ \prime}}$, $V_{\Pi_g}$ static potentials discussed in Section~\ref{subsec:quantum_numbers} takes the form

\begin{align}\label{eq:coupledSchr2}
&\left[
-\frac{1}{m_Qr^2}\,\partial_rr^2\partial_r+\frac{1}{m_Qr^2}
{\begin{pmatrix}
l\left(l+1\right) & 0 & 0\\[6pt]
0                 & l(l+1)+2        & -2\sqrt{l(l+1)} \\[6pt]
0                 & -2\sqrt{l(l+1)} & l(l+1)
\end{pmatrix}}\right.
\nonumber\\
&\hspace{5.0 cm}\left.
+\begin{pmatrix} V_{\Sigma_{g}^{+}}(r) &  V_{\Sigma_{g}^{+}-\Sigma_{g}^{+\prime}}(r) & 0 \\[6pt]
    V_{\Sigma_{g}^{+}-\Sigma_{g}^{+\prime}}(r) & V_{\Sigma_{g}^{+\prime}}(r) & 0\\[6pt]
      0 & 0 & V_{\Pi_g}(r)\end{pmatrix}
      \right] 
      \hspace{-1pt}\begin{pmatrix} \psi_{\Sigma_g^+}^{l}(r) \\[6pt] \psi_{\Sigma_g^{+\prime}}^{l}(r) \\[6pt] \psi_{\Pi_g}^{l}(r)\end{pmatrix}={\mathcal{E}}_{l} \begin{pmatrix} \psi_{\Sigma_g^+}^{l}(r) \\[6pt] \psi_{\Sigma_g^{+\prime}}^{l}(r) \\[6pt] \psi_{\Pi_g}^{l}(r)\end{pmatrix},
\end{align}

and has already been presented in our previous studies~\cite{Brambilla:2024imu,Brambilla:2026zfw}.
The parametrization for the $V_{\Sigma_{g}^{+}}$, $V_{\Sigma_{g}^{+\prime}}$, $V_{\Sigma_{g}^{+}-\Sigma_{g}^{+\prime}}$ static potentials around the string-breaking region and in the basis of Eq.~\eqref{eq:coupledSchr2} have been given in the lattice QCD studies~\cite{Bulava:2019iut, Bulava:2024jpj}.
We now rewrite Eq.~\eqref{eq:coupledSchr2} in the BO diabatic basis, which is better suited for the calculations presented in Section~\ref{sec:spectrum}, as the tetraquark components are expressed in definite partial waves.
To this aim, we diagonalize the angular momentum matrix via the unitary transformation
\begin{eqnarray}
U=
\begin{pmatrix}
1 & 0 & 0\\
0 & \sqrt{\frac{l}{2l + 1}} & \sqrt{\frac{l+1}{2l+1}} \\
0 & \sqrt{\frac{l+1}{2l+1}} &  -\sqrt{\frac{l}{2l + 1}}
\end{pmatrix},
\label{eq:T-matrix elements}
\end{eqnarray}

so that the coupled Schr\"odinger equation takes the form\footnote{For $l=0$, Eq.~\eqref{eq:1++BOEFT} reduces to a two-channel equation consisting of the quarkonium and $\Sigma_g^{+\prime}$ tetraquark components, see Eq.~\eqref{eq:diabaticl0}, which is already written in the BO diabatic basis.}
\begin{align}\label{eq:1++BOEFT}
\hspace{+0cm}\left[
-\frac{1}{m_Qr^2}\,\partial_r r^2 \partial_r + \frac{1}{m_Qr^2}
{\begin{pmatrix}
l (l + 1) & 0 & 0 \\[4pt]
0 & (l - 1) l & 0 \\[4pt]
0 & 0 & (l + 1) (l  + 2) \\[4pt]
\end{pmatrix}}\right. \hspace{0cm} \left. \right. \nonumber \\
&\hspace{-8cm}\left. + \begin{pmatrix} 
V_{\Sigma_{g}^+} & \sqrt{\frac{l}{2l + 1}} V_{\Sigma_g^+ \text{-} \Sigma_g^{+\prime}} & \sqrt{\frac{l + 1}{2l + 1}} V_{\Sigma_g^+ \text{-} \Sigma_g^{+\prime}} \\[6pt]
\sqrt{\frac{l}{2l + 1}} V_{\Sigma_g^+ \text{-} \Sigma_g^{+\prime}} & \frac{l}{2l + 1} V_{\Sigma_g^{+'}} + \frac{l + 1}{2l + 1} V_{\Pi_g}  & \frac{\sqrt{l(l + 1)}}{2l + 1}(V_{\Sigma_g^{+'}} - V_{\Pi_g}) \\[6pt]
\sqrt{\frac{l + 1}{2l + 1}} V_{\Sigma_g^+ \text{-} \Sigma_g^{+\prime}} & \frac{\sqrt{l(l + 1)}}{2l + 1}(V_{\Sigma_g^{+'}} - V_{\Pi_g}) & \frac{l + 1}{2l + 1} V_{\Sigma_g^{+'}} + \frac{l}{2l + 1}V_{\Pi_g}  \\
\end{pmatrix} \right]
\begin{pmatrix} \psi_{Q \bar{Q}(l)}^{l} \\[4pt] \psi_{M\bar{M}(l-1)}^{l}\\[4pt] \psi_{M\bar{M}(l+1)}^{l} \end{pmatrix}  
=  \mathcal{E}_{l} \begin{pmatrix} \psi_{Q \bar{Q}(l)}^{l} \\[4pt] \psi_{M\bar{M}(l-1)}^{l} \\[4pt] \psi_{M\bar{M}(l+1)}^{l} \end{pmatrix},
\end{align}
which coincide with Eq.~\eqref{eq:diabatic} introduced in Section~\ref{subsec:coupledequation}.

\section{Resonances in the bottom-charmed sector}
\label{app:bottom-charm}

To derive the resonances in the bottom-charmed sector, the heavy-quark pair $Q\bar{Q}$ is taken to be either $b\bar{c}$ or $c\bar{b}$. The factors $1/m_Q$ in the Schr\"odinger equations \eqref{eq:diabatic} and \eqref{eq:diabaticl0} should then be replaced by $1/(2\mu)$, where the reduced mass $\mu$ is given by $\mu=m_b^{\rm HL} m_c^{\rm HL}/(m_b^{\rm HL}+m_c^{\rm HL})=1.439$~GeV. The threshold in the bottom-charmed sector is $m_\text{thresh}=m_b^{\rm HL}+m_c^{\rm HL}=7.286$~GeV.

Within the energy region between the $S+S$ and $S+P$ thresholds, four resonances, including the $2D$, $3P$, $4S$, and $3D$ states, are predicted in the bottom-charmed sector.
Note that with the current tuning of the adjoint meson mass, we obtain no tetraquark bound state below the $S+S$ charm-bottom meson-antimeson threshold.
This result is in contrast to the charmoniumlike and bottomoniumlike sectors, where tetraquark states close below the relevant threshold were found.

The masses and decay widths and pole couplings into $S+S$ open-flavor meson-antimeson pairs extracted from the $T$ matrix and $K$ matrix methods are listed in Table~\ref{Table: c b-bar masses and decay widths}. From the components $Z$ and $X_i$ in Table~\ref{Table: c b-bar table of compositeness} and the normalized composition measures in Table~\ref{Table: table of c b-bar probabilites}, we see that all states are dominated by the $c\bar{b}$ channel, with the $2D$ resonance having the largest admixture from the tetraquark channels.

\begin{table}
    \renewcommand{\arraystretch}{1.2}
    \centering
    \begin{tabular}{cccrrccrr}
        \toprule [1.5pt]
        \multirow{2}{*}{$nl$} & \multicolumn{4}{c}{$T$ matrix} & \multicolumn{4}{c}{$K$ matrix}  \\
         \cmidrule(lr){2-5} \cmidrule(lr){6-9} 
       & $M$ (MeV) & $\Gamma$ (MeV) & $g^2_{l-1}$ & $g^2_{l+1}$ & $M$ (MeV) &  $\Gamma$ (MeV) & $g^2_{l-1}$ & $g^2_{l+1}$ \\
     \midrule [1.5pt]
     $2D$ &  $7356$   & $33$ &  $84\%$ & $16\%$    & $7352$   & $43$  & $90\%$ & $10\%$  \\
        $3P$ & $7501$   &  $10$  &   $66\%$ & $34\%$   & $7499$ & $11$ & $66\%$ & $34\%$ \\
        $4S$ &  $7617$ & $7$  &     &  $100\%$ & $7615$ & $8$  & & $100\%$ \\
     $3D$ &  $7671$ & $4$  &  $7\%$ & $93\%$   & $7672$ & $4$ & $8\%$ & $92\%$ \\ 
\bottomrule  [1.5pt]
    \end{tabular}
   \caption{Masses $M$ and widths $\Gamma$ of the isoscalar bottom-charmed states obtained from the $T$ matrix and $K$ matrix poles. The normalized pole couplings $g^2_{l-1}$ and $g^2_{l+1}$ characterize the relative residue strengths in the $D^{(\ast)}\bar{B}^{(\ast)}$ open-flavor channels with orbital angular momentum $l-1$ and $l+1$, respectively. For narrow isolated resonances, they may approximate branching fraction ratios within this truncated channel
   space.}
     \label{Table: c b-bar masses and decay widths}
\end{table}

\begin{table}
    \renewcommand{\arraystretch}{1.2}
    \centering
    \begin{tabular}{crrr} 
        \toprule [1.5pt]
        \multirow{2}{*}{$nl$} & \multirow{2}{*}{Channel} & \multicolumn{2}{c}{$Z$ or $X_i$} \\
        \cmidrule(lr){3-4}
        & & \multicolumn{1}{c}{CSM}  & \multicolumn{1}{c}{$K$ matrix}  \\
        \midrule [1.5pt]
        \multirow{3}{*}{$2D$} & $c\bar{b}$ ($D$-wave) &  $1.18+0.15i$  & $1.29$  \\  
         & $D^{(\ast)}B^{(\ast)}\, \text{($P$-wave)}$ &  $-0.20-0.29i$ & $-0.26$ \\
         & $D^{(\ast)}B^{(\ast)}\, \text{($F$-wave)}$ &  $0.02+0.14i$ & $-0.03$ \\ 
        \hline
        \multirow{3}{*}{$3P$} & $c\bar{b}$ ($P$-wave) &  $0.92+0.13i$ & $0.97$ \\ 
         & $D^{(\ast)}B^{(\ast)} \, \text{($S$-wave)}$ & $-0.01-0.06i$ & $0.02$ \\
         & $D^{(\ast)}B^{(\ast)} \, \text{($D$-wave)}$ &  $0.09-0.07i$ & $0.01$ \\ 
         \hline
         \multirow{2}{*}{$4S$} & $c\bar{b}$ ($S$-wave)   &  $0.95+0.09i$ & $1.00$  \\
         & $D^{(\ast)}B^{(\ast)} \, \text{($P$-wave)}$ &  $0.05-0.09i$ & $0.00$  \\
        \hline
        \multirow{3}{*}{$3D$} & $c\bar{b}$ ($D$-wave) &  $0.96-0.03i$ & $0.96$ \\ 
         & $D^{(\ast)}B^{(\ast)} \, \text{($P$-wave)}$ &   $0.03-0.01i$  & $0.00$  \\ 
         & $D^{(\ast)}B^{(\ast)}\, \text{($F$-wave)}$ &   $0.01+0.04i$ & $0.04$  \\
        \bottomrule [1.5pt]
    \end{tabular}
    \caption{Quarkonium component $Z$ and tetraquark/open flavor components $X_i$ of the bottom-charmed resonances.
    The complex scaling method (CSM) entries are generally complex, whereas the $K$ matrix entries are real but may lie outside the interval $[0,1]$. The two columns correspond to different prescriptions and should not be identified numerically.}
     \label{Table: c b-bar table of compositeness}
\end{table}

\begin{table}
    \renewcommand{\arraystretch}{1.2}
    \centering
    \begin{tabular}{crrr} 
        \toprule [1.5pt]
        \multirow{2}{*}{$nl$} & \multirow{2}{*}{Channel} & \multicolumn{2}{c}{$P_i$} \\
        \cmidrule(lr){3-4}
        & & \multicolumn{1}{c}{CSM}  & \multicolumn{1}{c}{$K$ matrix}  \\
        \midrule  [1.5pt]
        \multirow{3}{*}{$2D$} & $c\bar{b}$ ($D$-wave)  &  $71\%$ & $82\%$   \\ 
         & $D^{(\ast)}B^{(\ast)}\, \text{($P$-wave)}$ &  $21\%$ & $16\%$ \\
         & $D^{(\ast)}B^{(\ast)}\, \text{($F$-wave)}$  &  $8\%$ & $2\%$  \\ 
        \hline
        \multirow{3}{*}{$3P$} & $c\bar{b}$ ($P$-wave) &  $84\%$ & $97\%$   \\ 
         & $D^{(\ast)}B^{(\ast)} \, \text{($S$-wave)}$  & $6\%$ & $2\%$  \\
         & $D^{(\ast)}B^{(\ast)} \, \text{($D$-wave)}$ &  $10\%$ & $1\%$ \\ 
        \hline
         \multirow{2}{*}{$4S$} & $c\bar{b}$ ($S$-wave)  &  $90\%$ &  $100\%$ \\
         & $D^{(\ast)}B^{(\ast)} \, \text{($P$-wave)}$  & $10\%$ & $0\%$  \\
        \hline
        \multirow{3}{*}{$3D$} & $c\bar{b}$ ($D$-wave)  &  $93\%$& $96\%$   \\ 
         & $D^{(\ast)}B^{(\ast)} \, \text{($P$-wave)}$  &  $3\%$ & $0\%$  \\ 
         & $D^{(\ast)}B^{(\ast)}\, \text{($F$-wave)}$   &  $4\%$ & $4\%$  \\
        \bottomrule [1.5pt]
    \end{tabular}
    \caption{Normalized composition measures $P_i$  of the predicted bottom-charmed resonances obtained with the 
    complex scaling method (CSM) and the K matrix prescription. For resonances, $P_i$ are prescription-dependent measures of the channel content and not channel probabilities.}
     \label{Table: table of c b-bar probabilites}
\end{table}

\section{Quarkonium Hybrids: Schr\"odinger equations and Spectrum}\label{app:Hybrids}

In this appendix, we summarize the results on the hybrid Schr\"odinger equations, and the potential parametrization
following Refs.~\cite{Berwein:2015vca, Brambilla:2022hhi}.
In the absence of hybrid-quarkonium and hybrid-tetraquark mixing as discussed in Section~\ref{sec:hybrids}, the coupled Schr\"odinger equations are given by~\cite{Berwein:2015vca}:
\begin{align}
  \hspace{0.4 pt}&\left[-\frac{1}{m_Qr^2}\,\partial_rr^2\partial_r+\frac{1}{m_Qr^2}\begin{pmatrix} l(l+1)+2 & 2\sqrt{l(l+1)} \\
      2\sqrt{l(l+1)} & l(l+1) \end{pmatrix}+\begin{pmatrix} V_{\Sigma_{u}^{-}} & 0 \\
      0 & V_{\Pi_{u}} \end{pmatrix}\right]\hspace{-4pt} \, \begin{pmatrix} \psi_\Sigma^{l} \\
     \psi_{-\Pi}^{l}\end{pmatrix} = \mathcal{E}_l \begin{pmatrix} \psi_\Sigma^{l} \\
    \psi_{-\Pi}^{l}\end{pmatrix}\,,\nonumber\\
 &\hspace{4.0 cm}\left[-\frac{1}{m_Qr^2}\,\partial_r\,r^2\,\partial_r+\frac{l(l+1)}{m_Qr^2}+V_{\Pi_{u}}\right]\psi_{+\Pi}^{l} = \mathcal{E}_{l} \,\psi_{+\Pi}^{l}\,,
\label{eq:diffeq-hyb}
\end{align}
where $l$ is the BO angular momentum, the quark mass is taken in the heavy-light scheme $m_Q=m_Q^{\rm HL}$ defined in
Eqs.~\eqref{eq:Dmeson} and \eqref{eq:Bmeson}.
The eigenvalue $\mathcal E_l$ is the hybrid energy in the
convention specified below, and the remaining quantum
numbers are defined in Section~\ref{subsec:coupledequation}.

 In Eq.~\eqref{eq:diffeq-hyb}, the radial wavefunctions $\psi^{l}_\Sigma$ and $\psi^{l}_{-\Pi}$ correspond to states with the same parity, while the radial wavefunction $\psi^{l}_{+\Pi}$ corresponds to a state with opposite parity. The off-diagonal terms mix the $\psi^{l}_\Sigma$ with $\psi^{l}_{-\Pi}$ and vice versa, but they do not affect the parity, and hence, the  $\psi^{l}_{+\Pi}$ decouples, and follows a single channel Schr\"odinger equation. The coupled equations in~\eqref{eq:diffeq-hyb} lead to degenerate spin multiplets due to HQSS\footnote{For $l=0$, the two-channel coupled equation in~\eqref{eq:diffeq-hyb} reduces to a single channel consisting exclusively of the $\Sigma_u^-$ component.}.
The $J^{PC}$ quantum numbers are  $\left\{l^{\pm\pm};(l-1)^{\pm\mp},l^{\pm\mp},(l+1)^{\pm\mp}\right\}$,
where the first entry corresponds to the spin-$0$ combination and the next three entries to the spin-$1$ combinations.
For $l=0$, there is only one spin-$1$ combination as well as only one parity or charge conjugation state.
In Table~\ref{tab:multiplets}, we show the first five degenerate multiplets.
The coupled equation in~\eqref{eq:diffeq-hyb} describes the hybrid multiplets $H_1$, $H_3$, and $H_4$, 
while the single-channel equation describes the hybrid multiplets $H_2$ and $H_5$. 
Because the hybrid potentials are placed in the same additive
heavy-light scheme as the potentials of
Section~\ref{subsec:parametrization}, the physical mass associated
with a hybrid eigenvalue $\mathcal E_{l}$ is
\begin{equation}
 M_{l}
 =
 \mathcal E_{l}
 +2m_Q^{\rm HL}
 -E_1,
 \label{eq:hybrid-mass-convention}
\end{equation}
where $E_1=5~\mathrm{MeV}$ is the asymptotic value of the
tetraquark/open-flavor potentials in Eq.~\eqref{eq:QQbar_V}.
Equation~\eqref{eq:hybrid-mass-convention} is the hybrid analogue
of Eq.~\eqref{eq:quarkoniumlike-masses}.

\begin{table}
\renewcommand{\arraystretch}{1.2}
    \centering
\begin{tabular}{ccccc}  
    \toprule [1.5pt]
   Multiplet &
  $\,\,\,\,\,l\,\,\,\,\,$ & $J^{PC}$       & $\Lambda^{\sigma}_{\eta}$           \\
        \midrule  [1.5pt]
   $H_1$ &
                                  $1$     & $1^{--},(0,1,2)^{-+}$ & $\Sigma_u^-$, $\Pi_u$ \\
   $H_2$ &
                   $1$     & $1^{++},(0,1,2)^{+-}$& $\Pi_u$               \\
   $H_3$ &
                   $0$     & $0^{++},1^{+-}$       & $\Sigma_u^-$          \\
   $H_4$ &
                                   $2$     & $2^{++},(1,2,3)^{+-}$& $\Sigma_u^-$, $\Pi_u$ \\
   $H_5$ &
                   $2$     & $2^{--},(1,2,3)^{-+}$ & $\Pi_u$               \\
        \bottomrule [1.5pt]
\end{tabular}
\caption{The low-lying hybrid multiplets coming from the $\Sigma_u^-$ and $\Pi_u$ hybrid static energies with $J^{PC}$ quantum numbers ($l\leq2$). The multiplets are ordered by increasing value of the heavy quark-antiquark pair orbital angular momentum.
In Ref.~\cite{Oncala:2017hop}, the multiplets $H_1$, $H_2$, $H_3$, $H_4$ and $H_5$ are named $(s/d)_1$, $p_1$, $p_0$, $(p/f)_2$ and $d_2$, respectively.}
 \label{tab:multiplets}
\end{table}

We take the shapes of the $\Pi_u$ and $\Sigma_u^-$ static energies
from the pure-SU(3) lattice parametrization of
Ref.~\cite{Alasiri:2024nue}. We denote the $\Pi_u$ potential in the
additive normalization of that reference by
$\widetilde V_{\Pi_u}^{\rm ABM}(r)$.

\begin{align}
\widetilde V_{\Pi_u}^{\rm ABM}(r)=
\begin{cases}
\frac{\kappa_8}{r}+E_{1^{+-}}^{\rm ABM}
+A_{\Pi_u}r^2
+B_{\Pi_u}r^4
+C_{\Pi_u}r^6,
 &r<r_{\Pi_u}
\\
V_1^{\rm str}(r)+E_0^{\rm ABM}
+\frac{D_{\Pi_u}}{r^4},
 &r>r_{\Pi_u}
\end{cases},
\label{eq:VLambdaeta}
\end{align}

with 
\begin{equation}
 V_N^{\rm str}(r)=
 \sqrt{\widetilde\sigma^{\,2}r^2
 +2\pi\left(N-\frac{1}{12}\right)\widetilde\sigma}.
\label{V-N}
\end{equation}
$\widetilde\sigma$ is the string tension, and $N$ is the quantum number for transverse vibrational excitations of the string~\cite{Juge:2002br}. For $\Sigma_u^-$ static energy, the potential $\tilde{V}^{\rm ABM}_{\Sigma_u^-}$ is defined from its difference with $\tilde{V}^{\rm ABM}_{\Pi_u}$, i.e., $\tilde{V}^{\rm ABM}_{\Sigma_u^-} = \tilde{V}^{\rm ABM}_{\Pi_u} + \Delta \tilde{V}_{\Sigma_u^-}$, with $\Delta \tilde{V}_{\Sigma_u^-}$ given by
\begin{align}
\Delta \tilde{V}_{\Sigma_u^-} = 
\begin{cases}
 \Delta A_{\Sigma_u^-} \,r^2  + \Delta B_{\Sigma_u^-} \,r^4 + \Delta C_{\Sigma_u^-} \,r^6, 
  &r < r_{\Sigma_u^-}
\\
V_3^{\rm str}(r) - V_1^{\rm str}(r)  +  \frac{\Delta D_{\Sigma_u^-}}{r^4},   
  &r > r_{\Sigma_u^-}
 \end{cases},
\end{align}
with $\Delta A_{\Sigma_u^-} = A_{\Sigma_u^-} -  A_{\Pi_u}$, $\Delta B_{\Sigma_u^-} = B_{\Sigma_u^-} -  B_{\Pi_u}$, and so on. 
The constant $E_{1^{+-}}^{\rm ABM}$ is the $1^{+-}$ gluelump
energy parameter in the additive lattice normalization of
Ref.~\cite{Alasiri:2024nue}. Like any absolute energy associated with a static color source, it depends on the common additive
normalization and therefore cannot be identified directly with the gluelump energy parameter in the heavy-light scheme used
in the present work. The relevant ABM input is instead the quarkonium--hybrid static-energy difference,
in which the common static-source self-energy cancels.

For the matching described below, we use the ABM $\Sigma_g^+$ static potential parametrization
\begin{align}
\widetilde V_{\Sigma_g^+}^{\rm ABM}(r)
=
\begin{cases}
\frac{\kappa_1}{r}+E_{0^{++}}^{\rm ABM}
+A_{\Sigma_g^+}r^2+B_{\Sigma_g^+}r^4
+C_{\Sigma_g^+}r^6,
&r<r_{\Sigma_g^+}
\\
V_0^{\rm str}(r)+E_0^{\rm ABM}
+\frac{D_{\Sigma_g^+}}{r^4},
&r>r_{\Sigma_g^+}
\end{cases}.
\label{eq:ABM-singlet}
\end{align}

The ABM analysis determines the ground-state $\Sigma_g^+$
static potential and the hybrid static potentials on a common energy axis.
In particular, the string tension and the large-distance offset
$E_0^{\rm ABM}$ extracted from the $\Sigma_g^+$ fit are used in
the $\Pi_u$ fit. The separation between quarkonium $\Sigma_g^+$ and hybrid static potentials is consequently independent of the common static-source self-energy.
The potential parameters are summarized in Table~\ref{tab:SigmagpPiu-params}.

The quarkonium potential used in our coupled calculation in Eqs.~\eqref{eq:diabatic},~\eqref{eq:diabaticl0}  is
\begin{equation}
 V_{\Sigma_g^+}(r)
 =
 V_0+\frac{\gamma}{r}+\sigma r.
\label{eq:our-singlet-for-hybrid-match}
\end{equation}

The ABM potentials in Eq.~\eqref{eq:ABM-singlet} and
Eq.~\eqref{eq:our-singlet-for-hybrid-match} have different
Coulomb coefficients and string tensions. They therefore cannot
be related exactly by a constant over all distances. We define
their difference by
\begin{equation}
 \Delta_\Sigma(r)
 =
 V_{\Sigma_g^+}(r)
 -
 \widetilde V_{\Sigma_g^+}^{\rm ABM}(r).
 \label{eq:singlet-matching-difference}
\end{equation}
This difference is nearly constant over the intermediate-distance region
\begin{equation}
 r_a=0.4~{\rm fm}
 \leq r\leq
 r_b=0.9~{\rm fm},
 \label{eq:interval}
\end{equation}
which represents the characteristic distance region sampled by
the low-lying hybrid states.
Working in this region, we define a common matching shift whose central parameters are 
given by 
\begin{equation}
 \delta_g=-1.232~{\rm GeV}.
 \label{eq:deltag_value}
\end{equation}
The chosen interval in Eq.~\eqref{eq:interval} also corresponds to a near-minimum of the variance of $\Delta_\Sigma(r)$ among windows of comparable width. The corresponding root-mean-square deviation of $\Delta_\Sigma(r)$ from the constant shift in Eq.~\eqref{eq:deltag_value} is only about $3~\mathrm{MeV}$, confirming that the extracted shift is largely insensitive to the residual $r$ dependence of the two parametrizations.

The matching is nevertheless approximate, since the two singlet parametrizations do not have identical $r$ dependence. We estimate the associated normalization uncertainty by comparing the matching shift extracted over the intermediate-distance region with that obtained from shorter-distance intervals within the above range. This yields the fully correlated matching uncertainty
\begin{equation}
 \delta_g=-1.232\pm0.100~{\rm GeV}.
 \label{eq:hybrid-matching-uncertainty}
\end{equation}
This uncertainty controls the absolute placement of the complete
uncoupled hybrid tower and does not affect its level splittings.
It does not include the mass shifts or widths generated by the
omitted hybrid--tetraquark and hybrid--quarkonium mixings.

The hybrid potentials entering
Eq.~\eqref{eq:diffeq-hyb} are therefore
\begin{equation}
 V_{\Lambda_\eta^\sigma}(r)
 =
 \widetilde V_{\Lambda_\eta^\sigma}^{\rm ABM}(r)
 +\delta_g,
 \qquad
 \Lambda_\eta^\sigma\in\{\Pi_u,\Sigma_u^-\}.
 \label{eq:hybrid-potential-HL}
\end{equation}
The same shift is used in the charmonium and bottomonium sectors,
since the additive normalization of a static energy is independent
of the heavy flavor.

Notice that this matching procedure does not require an absolute
color-octet-referenced gluelump mass. A direct perturbative
conversion of the renormalon-subtracted gluelump parameter of
Ref.~\cite{Herr:2023xzj} to an unsubtracted pole-like convention
is very much  limited by the leading $u=1/2$ renormalon 
and therefore does not immediately provide a sufficiently precise absolute
normalization for the hybrid potentials. By contrast, the
quarkonium--hybrid static-energy difference used in the present
matching is free of the common static-source self-energy and of
the corresponding leading renormalon.

With the central matching shift in
Eq.~\eqref{eq:deltag_value}, the $1^{+-}$ gluelump energy parameter in the heavy-light scheme is
\begin{equation}
 \bar\Lambda_{1^{+-}}^{\rm HL}
 =
 E_{1^{+-}}^{\rm ABM}+\delta_g
 =
 -0.054\pm0.100~{\rm GeV},
 \label{eq:gluelump-HL}
\end{equation}
where $E_{1^{+-}}^{\rm ABM}=1.178~{\rm GeV}$. In the
Cornell-normalized scheme used for the adjoint meson mass, this becomes
\begin{equation}
 \Lambda_{1^{+-}}^{\rm C}
 =
 \bar\Lambda_{1^{+-}}^{\rm HL}-V_0
 =
 1.088\pm0.100~{\rm GeV}.
 \label{eq:gluelump-Cornell}
\end{equation}
Using the calibrated adjoint-meson parameter in the heavy-light scheme
\begin{equation}
 \bar\Lambda_{1^{--}}^{\rm HL}=-0.113~{\rm GeV},
 \label{eq:cornell-Adj}
\end{equation}
we obtain
\begin{equation}
\bar\Lambda_{1^{+-}}^{\rm HL}-\bar\Lambda_{1^{--}}^{\rm HL} =
 0.059\pm0.100~{\rm GeV}.
 \label{eq:gluelump-adjoint-difference}
\end{equation}
The common additive-scheme dependence cancels in this difference.
Within the present matching construction, the resulting value is
compatible with Ref.~\cite{Foster:1998wu} within the quoted matching
uncertainty.

The central matching prescription, therefore, places the
$1^{+-}$ gluelump approximately $59~{\rm MeV}$ above the
$1^{--}$ adjoint meson. Matching the short-distance quarkonium potential constants in Eq.~\eqref{eq:ABM-singlet} and \eqref{eq:our-singlet-for-hybrid-match} instead gives a difference with respect to adjoint meson in Eq.~\eqref{eq:cornell-Adj} of approximately
$118~{\rm MeV}$. The spread between these prescriptions is the
dominant origin of the $\pm0.100~{\rm GeV}$ matching uncertainty
assigned above. This comparison should consequently be regarded
as a consistency check of the matching procedure rather than as
an independent precision determination of the
gluelump--adjoint splitting.

\begin{table*}[t!]
\label{tab:SigmagpPiu-params}
\renewcommand{\arraystretch}{1.2}
\centering
\scriptsize
\resizebox{\textwidth}{!}{
\begin{tabular}{cccccccccccc}
\toprule[1.5pt]
$k^{PC}$ &
$\Lambda_\eta^\sigma$ &
$N$ &
$r_0^2\widetilde\sigma$ &
$r_0E_0^{\rm ABM}$ &
$\kappa_{1,8}$ &
$r_0E_{k^{PC}}^{\rm ABM}$ &
$r_0^3A_{\Lambda_\eta^\sigma}$ &
$r_0^5B_{\Lambda_\eta^\sigma}$ &
$r_0^7C_{\Lambda_\eta^\sigma}$ &
$D_{\Lambda_\eta^\sigma}/r_0^3$ &
$r_{\Lambda_\eta^\sigma}/r_0$
\\
\midrule[1.5pt]
$0^{++}$ & $\Sigma_g^+$ & $0$ &
$1.384$ & $-0.057$ & $-0.240$ & $0.013$ &
$3.200$ & $-4.690$ & $3.495$ & $0.029$ & $0.713$
\\
$1^{+-}$ & $\Pi_u$ & $1$ &
$1.384$ & $-0.057$ & $+0.037$ & $2.984$ &
$-0.010$ & $0.209$ & $-0.053$ & $0.127$ & $1.298$
\\
$1^{+-}$ & $\Sigma_u^-$ & $3$ &
$1.384$ & $-0.057$ & $+0.037$ & $2.984$ &
$0.683$ & $0.009$ & $-0.032$ & $-16.968$ & $2.216$
\\
\bottomrule[1.5pt]
\end{tabular}
}
\caption{
Dimensionless parameters of the ABM $\Sigma_g^+$, $\Pi_u$, and
$\Sigma_u^-$ parametrizations in pure SU(3) gauge theory
\cite{Alasiri:2024nue}. The Sommer scale is
$r_0=0.5~{\rm fm}$. The entries are the unshifted ABM parameters.
The potentials entering Eq.~\eqref{eq:diffeq-hyb} are obtained by
adding the common shift in
Eq.~\eqref{eq:hybrid-potential-HL}. For the $\Sigma_u^-$ row, the
listed coefficients are the combined coefficients obtained after
adding the fitted $\Sigma_u^--\Pi_u$ difference to the $\Pi_u$
parametrization.
}
\end{table*}
Using $r_0^{-1}=0.3947~{\rm GeV}$, the unshifted ABM parameters from Table~\ref{tab:SigmagpPiu-params} are
\begin{equation}
E_{1^{+-}}^{\rm ABM}=1.178~{\rm GeV},
\qquad
E_0^{\rm ABM}=-0.023~{\rm GeV}.
\end{equation}
After the application of the central matching shift, these additive
parameters become
\begin{equation}
E_{1^{+-}}^{\rm HL}=-0.054~{\rm GeV},
\qquad
E_0^{\rm HL}=-1.256~{\rm GeV}.
\end{equation}
Only the common additive parameters are shifted. The Coulomb
coefficient, string tension, higher-order coefficients, matching
radii, and gluelump-energy splittings remain unchanged.
The resulting uncoupled hybrid reference masses are reported in
Tables~\ref{tab:ccbarhybspectrum} and
\ref{tab:bbbarhybspectrum}.

\bibliography{bibliography}
\end{document}